%% file: main.tex
\documentclass[pdflatex,sn-basic]{sn-jnl}

\usepackage{graphicx}
\usepackage{multirow}
\usepackage{amsmath,amssymb,amsfonts}
\usepackage{amsthm}
\usepackage{mathrsfs}
\usepackage[title]{appendix}
\usepackage[table,svgnames]{xcolor}
\usepackage{booktabs}
\usepackage{subcaption}
\usepackage{ulem}
\usepackage{bm}
\usepackage{mathtools}
\usepackage[export]{adjustbox}
\usepackage{pdflscape}

\input{commands}

\begin{document}

\title{Conditional pathways-based climate attribution}

\author*[1]{\fnm{Christopher} \spfx{R.} \sur{Wentland}}\email{crwentl@sandia.gov}
\author[2]{\fnm{Michael} \sur{Weylandt}}
\author[3]{\fnm{Laura} \spfx{P.} \sur{Swiler}}
\author[3]{\fnm{Diana} \spfx{L.} \sur{Bull}}

\affil*[1]{\orgname{Sandia National Laboratories}, \orgaddress{\city{Livermore}, \state{CA}, \country{USA}}}
\affil[2]{\orgdiv{Zicklin School of Business}, \orgname{Baruch College, CUNY}, \orgaddress{\city{New York}, \state{NY}, \country{USA}}}
\affil[3]{\orgname{Sandia National Laboratories}, \orgaddress{\city{Albuquerque}, \state{NM}, \country{USA}}}

\abstract{
Attribution of climate impacts to natural and anthropogenic source forcings is essential for understanding and addressing climate effects. While standard methods like optimal fingerprinting have been effective for long-term changes, they often struggle in low signal-to-noise regimes typical of short-term forcings or with climate variables loosely related to the forcing. Single-step approaches fail to leverage additional climate information to enhance attribution certainty. To overcome these limitations, we propose a formal statistical framework that incorporates hypothesized physical pathways linking source forcings to downstream impacts. By establishing relationships based on scalar features and simple forcing response models, we create a series of conditional probabilities that describe the likelihood of the final impact. This method captures both primary and secondary processes by which the downstream impact evolves. Through hypothesis testing in a likelihood ratio framework, we demonstrate improved attribution confidence for source magnitudes in low signal-to-noise scenarios. Using the 1991 eruption of Mt. Pinatubo as a case study, we show that incorporating near-surface temperature and stratospheric radiative flux measurements enhances attribution certainty compared to analyses based solely on temperature, even at seasonal and regional scales. This framework holds promise for improving climate attribution assessments for unknown source magnitudes and low signal-to-noise impacts, where traditional methods may falter. Additionally, the formal inclusion of pathways allows for a deeper exploration of complex, multivariate relationships influencing source attribution.
}


\maketitle

\section{Introduction}\label{sec:intro}
\input{intro}

\section{Case Study: Mount Pinatubo}\label{sec:mtpinatubo}
\input{mtpinatubo}

\section{Methodology}\label{sec:methods}
\input{methods}

\section{Results}\label{sec:results}
\input{results}

\section{Discussion}\label{sec:discussion}
\input{discussion}

\section{Conclusions}
\input{conclusions}

\backmatter

\bmhead{Acknowledgments}
\input{acknowledgements}

\section*{Statements and Declarations}
\input{declarations}

\begin{appendices}
\input{appendices}
\end{appendices}

\clearpage
\bibliography{climate}

\end{document}

%% file: commands.tex
\newcommand{\eqdef}{\ensuremath{:=}}

\newcommand{\bbeta}{\bm{\beta}}
\newcommand{\btheta}{\bm{\theta}}

\newcommand{\bepsilon}{\bm{\epsilon}}

\newcommand{\SOtwo}{SO$_2$}

\newcommand{\inROne}[1]{\ensuremath{\in \mathbb{R}^{#1}}}
\newcommand{\inRTwo}[2]{\ensuremath{\in \mathbb{R}^{#1 \times #2}}}
\newcommand{\cond}[2]{\ensuremath{ \left( #1 \ \middle\vert \ #2 \right)}}
\newcommand{\tsim}{\ensuremath{\sim}}

\newcommand{\varIdx}{\ensuremath{v}}
\newcommand{\forceIdx}{\ensuremath{f}}
\newcommand{\ensIdx}{\ensuremath{e}}

\newcommand{\timeIdx}{\ensuremath{t}}

\newcommand{\numTime}{\ensuremath{N_t}}
\newcommand{\numEns}{\ensuremath{N_e}}
\newcommand{\numVars}{\ensuremath{N_v}}
\newcommand{\numForce}{\ensuremath{N_f}}

\newcommand{\numParents}{\ensuremath{N_{p,\varIdx}}}

\newcommand{\stateVar}{\ensuremath{q}}
\newcommand{\stateVec}{\ensuremath{\bm{\stateVar}}}
\newcommand{\stateMat}{\ensuremath{\bm{Q}}}
\newcommand{\metricVar}{\ensuremath{k}}
\newcommand{\metricRandom}{\ensuremath{K}}
\newcommand{\metricVec}{\ensuremath{\bm{\metricVar}}}
\newcommand{\metricMat}{\ensuremath{\bm{K}}}

\newcommand{\linparamvec}{\ensuremath{\btheta}}
\newcommand{\linparamest}{\ensuremath{\widehat{\theta}}}
\newcommand{\linparamestvec}{\ensuremath{\widehat{\btheta}}}

\newcommand{\fingparamvec}{\ensuremath{\bbeta}}
\newcommand{\fingparamest}{\ensuremath{\widehat{\beta}}}
\newcommand{\fingparamestvec}{\ensuremath{\widehat{\bbeta}}}
\newcommand{\linvarest}{\ensuremath{\widehat{\sigma}^2}}

\newcommand{\forceVar}{\ensuremath{f}}
\newcommand{\forceRandom}{\ensuremath{F}}
\newcommand{\forceSet}{\ensuremath{\mathcal{F}}}
\newcommand{\forceVec}{\ensuremath{\mathbf{\forceVar}}}
\newcommand{\obsVar}{\ensuremath{O}}

\newcommand{\obsSet}{\ensuremath{\mathscr{O}}}
\newcommand{\dataSet}{\ensuremath{{\mathcal{D}}}}

\newcommand{\pdfVar}{\ensuremath{P}}
\newcommand{\likelihood}{\ensuremath{\mathcal{L}}}
\newcommand{\testStat}{\ensuremath{\lambda}}
\newcommand{\testDist}{\ensuremath{\Lambda}}
\newcommand{\parentFunc}[1]{\ensuremath{\mathcal{P}(#1)}}

\definecolor{vira}{RGB}{253, 231, 37}
\definecolor{virb}{RGB}{152, 216, 62}
\definecolor{virc}{RGB}{64, 189, 114}
\definecolor{vird}{RGB}{31, 153, 138}
\definecolor{vire}{RGB}{43, 116, 142}
\newcommand{\tca}{\cellcolor{vira}} % p < 0.001
\newcommand{\tcb}{\cellcolor{virb}} % 0.001 < p < 0.01
\newcommand{\tcc}{\cellcolor{virc}} % 0.01 < p < 0.05
\newcommand{\tcd}{\cellcolor{vird}} % 0.05 < p < 0.1
\newcommand{\tce}{\cellcolor{vire}} % p > 0.1
\newcommand{\phlt}{\phantom{$<$ }}

%% file: intro.tex
Detecting and attributing the effects of anthropogenic activity on the global climate is an important and ongoing subject of climate research~\citep{santer1993,Hasselmann1993, hegerl1997A, hegerl1997B, North1998, Berliner2000, IPCCar3ch12, IPCCar6}. While the relationship between anthropogenic activity and long-term changes in global mean surface temperature is now widely accepted as ``established fact''~\citep{IPCCar6}, there are still many challenges in determining pathways connecting forcings and responses within the climate system. In particular, the detection and attribution (D\&A) of responses that are short-lived and/or spatially localized, and hence possess significant variability, is an outstanding problem~\citep{Bindoff2013,Lehner2016}. Process-based attribution and extreme weather event (EWE) storylines, described more fully below, focus on regional and short-lived responses. However, unlike traditional D\&A which assumes separable forcings and unconditional probabilities~\citep{hegerl2010, hasselmann1997, ribes2013}, these process-based and storyline approaches employ \textit{conditional} probabilities which explicitly account for interacting and dependent factors which produce a climate response~\citep{LloydShepherd2023}.

In this paper, we propose a novel approach to construct and analyze conditional relationships, inspired by the process-based and EWE storyline attribution communities, that is especially well suited for D\&A of spatio-temporally localized responses to a forcing. Leveraging process knowledge to construct ``pathways'' that connect ``upstream'' and ``downstream'' variables, we factorize the total climate system forcing response into conditional relationships. This framework is well suited to (1) progressively interrogating complex, multivariate relationships in the climate system, (2) discriminating the forcing magnitude producing the chain of responses, and (3) deciphering forcing-driven responses on short-time scales and in confined regions. Using a comprehensive simulation study of the 1991 Mt.\ Pinatubo eruption, we characterize the statistical properties of our proposed approach and show that it has far greater power -- ability to correctly detect and attribute global and localized responses -- than similar unconditional approaches. 

Process-based attribution has arisen to determine the physical processes influencing the response of interest to external forcing and internal variability~\citep{IPCCar6ch10}. It infers the underlying mechanisms driving a response by conditioning on the degree of climate change~\citep{Wohland2022,Wu2020,Malchow2023}. Attribution can range from illustrating consistency with a proposed physical process ~\citep{Wohland2022}, to developing surrogate models composed of only the physical processes of interest in a particular region ~\citep{Wu2020}, to embedding process models in a robust statistical frameworks~\citep{Malchow2023}. Concepts from this D\&A approach have also been used to identify the 2019 Australian bushfires as a potential cause for the observed ``triple-dip'' La Ni\~na~\citep{Fasullo2023}. An attribution study in spirit, this research connected visual representations of spatio-temporal patterns of subtropical clouds and radiation to a lagged decrease in humidity and temperature, driving the intertropical convergence zone (ITCZ) northward, leading to a decrease in equatorial Pacific surface temperatures and the multiple La Ni\~nas~\citep{Fasullo2023}. With significant subject matter expertise, the causal chain from forcing (bushfires) to response (triple dip La Ni\~na) was supported by the timing and spatial structures of variables revealing the relationships~\citep{Fasullo2023}. Process-based attribution methods require bespoke mechanistic modeling approaches, thus limiting their broad applicability. By contrast, our proposed approach requires only minimal statistical modeling of quantities of interest and can be straightforwardly applied to a variety of climate responses to build frameworks for powerful pathways-based attribution.

EWE storylines condition on certain aspects of variability, like large scale dynamics, to understand the role of anthropogenic climate change (ACC) in transitory events~\citep{Cattiaux2010, Trenberth2015, Shepherd2016, Lackmann2015, ZappaShepherd2017, Mindlin2020}. These studies analyze the \textit{magnitude} of a given response as a function both the large scale dynamical state and the degree of ACC; by contrast, non-storyline approaches typically analyze the probability of an event as a function the degree of ACC, e.g., the probability of extreme weather events~\citep{Otto2017}. For instance, given the observed state of the atmosphere (e.g., goepotential heights, wind speeds), a storyline analysis found that Hurricane Sandy's intensity would be stronger and make landfall further north in a warmer future~\citep{Lackmann2015}. Predictions of regional and seasonal storm tracks in the Northern Hemisphere~\citep{ZappaShepherd2017} and Southern Hemisphere~\citep{Mindlin2020} have been shown to be dependent upon the polar vortex and midlatitude westerlies; both of these large scale dynamical phenomena have dependencies on the degree of ACC making traditional unconditional attribution infeasible. With a storyline approach, however, the polar vortex state could be conditionally attributed to storm track features.

Storyline-type analyses have also been successfully employed outside of the EWE community. \citet{Lehner2016} was able to demonstrate D\&A consistency between observations and the naturally forced response to volcanic eruptions in the Coupled Model Intercomparison Project Phase 5 (CMIP5) multi-model ensemble by only evaluating simulated output for which the El Ni\~no Southern Oscillation (ENSO) was in an El Ni\~no phase in the first boreal winter following three volcanic eruption, by conditioning on the ENSO phase. Without this conditioning, the natural forced response over a 16 year period in the CMIP5 ensemble was inconsistent with observations, highlighting the importance of conditioning on major modes of internal variability when analyzing responses possessing similar time scales as those modes. Our proposed approach extends this use of conditional analysis beyond major climate modes to arbitrary upstream quantities, e.g., the processes defining a pathway from an unknown source magnitude to downstream impact.

We propose a flexible framework for \textit{conditional} analyses that directly incorporates process-based knowledge of expected climate responses through an explicit pathway model, combining strengths of both process-based and storyline-based approaches. By using knowledge of climate processes to develop a set of interrelated conditional analyses, we significantly improve the statistical power of D\&A and are able to make confident attribution statements at seasonal and regional scales. Formally, for a known set of climate quantities $F \to Y_1 \to Y_2 \to \dots \to Y_K$, where $F$ is the forcing of interest and $Y_1, Y_2, \dots, Y_K$ are observed climate quantities, we factorize the joint multivariate probability of $(Y_1, \dots, Y_K)$ into a series of univariate conditional relationships:
\begin{equation*}
    \pdfVar (Y_1, \dots, Y_K | F) = \pdfVar (Y_K | Y_1, \dots, \dots, Y_{K-1}, F) * \pdfVar (Y_{K-1} | Y_1, \dots, Y_{K-2}, F) * \dots * \pdfVar (Y_1 | F).
\end{equation*}
Here, the climate pathway is used principally to guide factorization of the joint probability into a series of conditional probabilities; these univariate relationships are, in turn, far easier to model, facilitating application of our approach to complex climate systems. These conditional distributions are determined through models of how each $Y_k$, inclusive of internal variability, is impacted by both the forcing magnitude ($F$) and by ``upstream'' variables ($Y_1, \dots, Y_{k-1}$). In the present work, these relationships are estimated in a data-driven manner, through regression modeling of scalar features, though any probabilistic model of conditional dependence could be used. Each model progressively increases in dimension as additional downstream variables in the pathway are included.

Once these conditional models are estimated and combined to compute the joint probability $\pdfVar (Y_1, \dots, Y_K | F)$, the resulting probability statements are embedded into a hypothesis testing framework to support or reject attribution statements. Hypothesis tests on the forcing magnitude are developed under a flexible likelihood ratio (LR) framework, in a similar fashion to the proposal of~\citet{Ribes2017}. Specifically, we propose a null hypothesis ($H_0$) that the true forcing lies within a set of forcings $\forceSet_0$ and an alternative hypothesis ($H_1$) that the observed forcing instead falls in a suitable set of alternatives $\forceSet_1$ ($\forceSet_0 \cap \forceSet_1 = \emptyset$). Rejection of the null hypothesis indicates \textit{attribution} of the observed impacts $(Y_1, \dots, Y_K)$ to the set of alternatives $\forceSet_1$. The LR framework we employ is flexible and statistically powerful, but should be distinguished from the risk ratios or probability ratios used in extreme weather attribution~\citep{NAP21852,Swain2020,Paciorek2018,Chiang2021}; risk ratios capture the increased rate of adverse events in different scenarios, while LRs assess whether $\forceSet_0$ or $\forceSet_1$ is more consistent with an impact already observed. 

Our use of a conditional multi-step decomposition stands in direct contrast to the multi-step methodology considered by~\citet{hegerl2010}. Specifically, they claim that the strength of a multivariate (pathway) approach is limited by the weakest step in the pathway. They argue that this follows from conditional decompositions of multivariate probabilities: if the chain is $F \to Y_1 \to Y_2$, then $\pdfVar (Y_2|F) = \pdfVar (Y_2|Y_1) * \pdfVar (Y_1|F) \leq \max\{\pdfVar (Y_2|Y_1), \ \pdfVar (Y_1|F)\}$. This presupposes that $Y_2$ arises only from the influence of $Y_1$, and is conditionally independent of $F$ or any other intermediate effects. However, in the climate this assumption is unlikely to hold; temperature, for instance, is not solely dictated by incoming radiative flux, it is also influenced by, for example, surface albedo or the degree of water vapor in the air. It is possible that $F$ could influence more than just one factor influencing temperature, i.e., that $F$ could be the direct parent of each downstream variable. Through another lens, the impact of $F$ on $Y_2$ could be larger than what we would expect from $Y_1$ alone; by controlling for the effect of $Y_1$ on $Y_2$ in a regression model, we are able to isolate the remaining variability in $Y_2$ and determine whether $F$ is correlated with that remaining variability. If it is, then we can improve our attribution of the joint effect $(Y_1, Y_2)$ to the forcing. Thus, \textit{contra}~\citet{hegerl2010}, we find that adding additional variables \textit{increases} attribution certainty.

In this paper we demonstrate that traditional unconditional fingerprinting approaches struggle to distinguish between the global near-surface temperature responses over 3 years due to varied emissions from Mt. Pinatubo, but a conditional approach (evaluating the likelihood of the near-surface temperature response conditioned on the forcing magnitude and intermediary variables) enables attribution of a 10 Tg eruption with high discrimination. This framework is further applied to show seasonal and regional downstream responses can still be attributed to the correct forcing magnitude, within a range, using the conditional pathway. A conditional attribution method of this nature is well suited to a variety of applications where the magnitude of some forcings are highly uncertain, as is the case for aerosols~\citep{Watson2020,Kahn2023}, volcanic eruptions~\citep{Zanchettin2019,Ukhov2023}, particulate matter or black carbon from wildfires~\citep{Li2021Aus}, and even natural methane emissions~\citep{Saunois2024}. Additionally, the mediating drivers of a response could be unknown, as in the studies of~\citet{Wu2020}, \citet{Wohland2022}, and~\citet{Malchow2023}. By hypothesizing and testing powers of distinguishability between multiple driver options, one could then infer with more confidence which mediating mechanisms are important to a response.

The remainder of this paper is organized as follows: Section 2 describes the case study of Mount Pinatubo, Section 3 presents the methodology including the formulation of pathways, conditional likelihoods, and the likelihood ratio test used, Section 4 provides results, and Section 5 and 6 present the discussion and conclusions, respectively. 

%% file: mtpinatubo.tex
This section provides an overview of the eruption of Mt.\ Pinatubo and its impacts, and outlines the set of simulations used to demonstrate multi-step attribution. Mt.\ Pinatubo is an attractive case study because its impacts have been well studied, allowing us to focus on demonstrating the proposed novel attribution framework and rely on known properties of the eruption response.

\subsection{Impacts from Mt.\ Pinatubo}

Large volcanic eruptions (e.g., from Mt.\ Tambora, Krakatoa, Mt.\ Pinatubo) are a significant source of aerosol forcing in the stratosphere. The resultant impacts from aerosol forcings in the stratosphere due to explosive volcanic eruptions are as wide ranging as surface temperature decreases~\citep{parker1996,soden2002}, lower stratosphere temperature increases~\citep{labitzke1992}, reduction in global precipitation~\citep{gillett2004}, lowering of global sea-level~\citep{church2005}, and increased diffusivity of incoming radiation~\citep{Robock:2000,proctor2018} with resultant impacts on net primary productivity of plants~\citep{gu2003,proctor2018,greenwald2006}. The magnitude of these impacts, which strongly influences detectability and attribution, is dependent upon the magnitude of the eruption~\citep{marshall2019} as well as the state of the climate at the time of the eruption~\citep{Zanchettin2022,Lehner2016,Mcgraw2016}.

In this manuscript, we are concerned with the primary radiative and temperature impacts from Mt.\ Pinatubo. Mt.\ Pinatubo released 18-19 Tg of \SOtwo\ into the atmosphere~\citep{guo2004} with only \tsim 10 Tg remaining in the stratosphere for further microphysical and chemical evolution into sulfate aerosols~\citep{Kremser:2016}. These sulfate aerosols modified radiative forcing by scattering incoming shortwave radiation and absorbing longwave and near-infrared radiation~\citep{Robock:2000}. Incoming shortwave radiation was partially backscattered into space by the aerosols, reducing the amount of energy incident to Earth as confirmed by the Earth Radiation Budget Satellite~\citep{Minnis1993}. The net reduction of radiative forcing cooled the troposphere~\citep{Santer2014,Kremser:2016}, achieving a global maximum surface cooling of \tsim 0.4 K between June 1992 and October 1992~\citep{ramachandran2000}.

\subsection{Simulations and data preparation}\label{subsec:simulations}

The above impacts of Mt.\ Pinatubo were simulated using the U.S. Department of Energy's Energy Exascale Earth System Model, version 2 (E3SMv2)~\citep{Golaz:2022}. These runs utilized recent aerosol modeling capabilities, referred to as ``stratospheric prognostic aerosols'' (SPA)~\citep{hbrown2024}, which simulate sulfate aerosol formation and evolution in the stratosphere from the injection of volcanic \SOtwo, ensuring dynamical consistency between atmospheric transport and aerosol evolution. The complete implementation of E3SMv2-SPA is described by~\citet{hbrown2024}, which details changes to the 4-mode Modal Aerosol Module microphysics~\citep{liu2012,liu2016} and validates its performance against observations.

A simulation campaign employing E3SMv2-SPA was launched on the ne30pg2 mesh, with \tsim 110 km horizontal resolution and 72 vertical layers up to \tsim 0.1 hPa; the campaign is detailed by~\citet{ehrmann2024}. Internal climate variability and the initial conditions at time of eruption significantly affect the short-term responses to a volcanic forcing, like temperature~\citep{Zanchettin2022,Lehner2016,Mcgraw2016}. Thus our simulations are initialized with modes of variability (ENSO and QBO) in historically accurate states (the ``limited variability'' simulations of~\citet{ehrmann2024}). Fifteen fully-coupled freely-running ensemble members were generated by randomly perturbing the initial temperature field by values near machine precision, which diverge according to their own synoptic dynamics. These limited variability ensembles were run from June 1, 1991 to December 31, 1998 under several stratospheric \SOtwo\ injections scenarios, ranging from no eruption (0 Tg \SOtwo, the ``counterfactual'') to 15 Tg \SOtwo\ (\tsim 50\% greater than the estimated historical eruption of \tsim 10 Tg of stratospheric \SOtwo). Monthly averages of field data are saved over this period.

As discussed by~\citet{ehrmann2024}, the ``limited variability'' initialization reduces variability between ensemble members for roughly the first year of the simulation, after which the ensemble members can be considered independent with variabilities expected from standard initializations. The inter-ensemble variability, as measured by the degree of radiative flux and near-surface temperature variance, is later incorporated into our statistical framework. We do not explicitly treat internal variability as in traditional attribution methods, which incorporate it through a linear regression covariance term. Instead, our framework incorporates the variability represented by the ensemble spread in an impact metric across forcing levels, inclusive of the unforced case. This is calculated as the residual variance from a forcing response model; further details are presented in Section~\ref{subsec:linregress}.

Several steps were taken to prepare the data for the multi-step attribution process detailed in Section~\ref{sec:methods}. First, all fields of interest were remapped to a $1^{\circ} {\times} 1^{\circ}$ latitude-longitude grid, and clipped between latitudes 66S--66N. This latter operation is performed to exclude missing radiation data during polar winter. Before making any further data reductions, the ensemble mean of the counterfactual simulations is computed, and this mean space-time field is subtracted from all ensemble members (including the counterfactual runs themselves). This has the effect of transforming the data from raw measurements to ``impacts''; that is, the centered fields isolate the impact of Mt.\ Pinatubo's eruption relative to a scenario without an eruption. Next, latitude-weighted averages of global, Northern Hemisphere (NH, 0-66N, 180W-180E), and North American (NA, 25N-66N, 170W-60W) regions are computed from the two-dimensional impact fields. Finally, the time series data is trimmed for multiple time periods: a 3-year period from June of 1991 to June of 1994, the first winter post eruption as defined by January, February, and March 1992, and the first full summer post eruption as defined by June, July, and August 1992. The 3-year cutoff was chosen by determining the point at which all forced (1 Tg and greater eruptions) ensemble mean global time series permanently returned to within two standard deviations of the counterfactual global time series ensemble. This ultimately produces a scalar time series for each region, variable of interest, and ensemble member for the attribution studies presented below.

%% file: methods.tex
Before detailing the proposed conditional attribution methodology, we begin by describing the standard fingerprinting approach and note several challenges that arise when attempting to determine the magnitude of a climate forcing rather than the more common task of distinguishing between different classes of forcings (e.g. greenhouse gases (GHGs) vs. aerosols).

\subsection{Classical detection and attribution via fingerprinting}\label{subsec:fingerprinting}

Traditional detection and attribution by fingerprinting is typically formulated to distinguish between fundamentally different climate forcing types, e.g., anthropogenic GHGs and anthropogenic aerosols~\citep{santer1993, hasselmann1997, hegerl1997A, North1998, IPCCar3ch12, IPCCar6ch3}. This process generally begins by simulating the climate system with only one of these forcing types at a time, and another ``counterfactual'' climate system with none of the forcings (i.e., natural variability only). After these simulations are complete, space-time data are extracted for the climate variable(s) of interest (e.g., sea surface temperature, precipitation), and reduced to a one-dimensional time series. Conventionally, this reduction is performed by taking an (area-weighted) average over a spatial region of specific interest, e.g., globally, across North America, or over the Sahel~\citep{Marvel2020, santer2011}. Alternatively, the ``optimal fingerprinting'' strategy computes the projection of each variable onto a small number of empirical orthogonal functions (EOFs). This EOF-based projection both identifies major spatial patterns of climate variability and maximally captures the variance of the original data on a low-dimensional linear subspace~\citep{Hasselmann1993, hegerl1997B, Allen2003, ribes2013, Weylandt:2024-BeyondPCA}.

In this manuscript, we present fingerprinting analyses for globally averaged time series data under a ``perfect model'' assumption. This format assumes that the simulation model (here, E3SMv2-SPA) accurately represents real climate dynamics, and pseudo-observational data is extracted in a ``leave-one-out'' fashion. That is, the climate system is simulated $\numEns$ times under various forcing types. Then, the attribution analysis is repeated $\numEns$ times, each time using one ensemble member as the observational data and using the remaining simulations as the simulated realizations. Attribution over all $\numEns$ analyses is considered in the aggregate (e.g., 75\% achieved successful attribution) to give an estimate of the power and reliability of the different approaches. We note that the perfect model assumption is certainly false and that results of a perfect model study are best understood as an upper bound on expected attribution performance when actual observational or reanalysis data are used. Use of the perfect model structure focuses analysis on comparing alternative attribution strategies and avoids any additional complexities arising from climate model biases.

To formally describe fingerprinting D\&A, we first denote (spatially-averaged) time series fingerprints as $\stateVec_{\varIdx,\forceIdx,\ensIdx} \eqdef [\stateVar_{\varIdx,\forceIdx,\ensIdx,1}, \ \hdots, \ \stateVar_{\varIdx,\forceIdx,\ensIdx,\numTime}] \inROne{\numTime}$, where $\numTime$ is the number of observations, $\varIdx$ is the variable of interest, $\forceIdx$ is the candidate forcing, and $\ensIdx$ indexes the simulation ensemble member. In the multivariate context, where $\numVars$ different variables are analyzed jointly, the separate time series 
are ``stacked'' as $\stateVec_{\forceIdx,\ensIdx} \eqdef [\stateVec_{1,\forceIdx,\ensIdx}^{\top}, \ \hdots, \ \stateVec_{\numVars,\forceIdx,\ensIdx}^{\top}]^{\top} \inROne{\numTime \numVars}$. We denote the associated ensemble mean over all $\numEns$ simulations by $\stateVec_{\forceIdx} \eqdef \numEns^{-1} \sum_{\ensIdx}^{\numEns} \stateVec_{\forceIdx,\ensIdx} \inROne{\numTime\numVars}$. Finally, we concatenate these impact vectors across $\numForce$ different forcings into a single matrix $\stateMat \eqdef [\stateVec_{f_1}, \ \hdots \ , \ \stateVec_{f_{\numForce}}] \inRTwo{\numTime \numVars}{\numForce}$. 

These simulation results, $\stateMat$, are then regressed against a comparable time series derived from observational data, $\stateVec_o \inROne{\numTime\numVars}$, using a linear model of the form
\begin{equation}\label{eq:finger-regress}
    \stateVec_o = \stateMat \fingparamvec + \bm{\epsilon}.
\end{equation}
Here, $\fingparamvec \inROne{\numForce}$ are regression coefficients to be estimated and $\bm{\epsilon} \inROne{\numTime \numVars}$ is a vector of errors, or unexplained variability, associated with each observation. Eq.~\ref{eq:finger-regress} may be supplemented with an intercept term. If we assume the elements of $\bm{\epsilon}$ are IID Gaussian errors, the resulting estimate of $\fingparamvec$ is given by the ordinary least squares solution 
\begin{equation}
    \fingparamestvec = \left[ \stateMat^{\top} \stateMat \right]^{-1} \stateMat^{\top} \stateVec_o.
\end{equation}
If the elements of $\bm{\epsilon}$ are not IID Gaussian, weighted or generalized least squares may be used instead~\citep{Allen1999}.

Confidence intervals on each regression parameter in $\fingparamestvec$ are then assessed under a fixed confidence level. If the confidence interval for the parameter associated with a given forcing type $\forceIdx$ does not contain zero, then that forcing's effect on the observed climate impact is said to be ``detected'' at the given confidence level. If the confidence interval instead contains unity, then the observed climate impact is said to be ``attributed'' to that forcing type at that confidence level~\citep{Lee2005}.

This classical attribution strategy has demonstrated enormous success for long-term, categorical climate forcings. For short-term forcings, such as volcanic eruptions, the high natural variability of the climate system on small time scales often precludes successful attribution~\citep{Bindoff2013,Lehner2016}; specifically, on short time scales, the low signal-to-noise ratio of most forcings results in exceptionally wide confidence intervals, making detection difficult. Further, traditional fingerprinting is not designed to attribute between different \textit{magnitudes} of the \textit{same} forcing, as it assumes separability of the forcing impacts. This may be a reasonable assumption when considering the impacts of GHGs versus aerosols, but is certainly untrue when distinguishing, for example, between 7, 10, and 13 Tg \SOtwo\ volcanic eruptions.

\subsection{Pathway selection}\label{subsec:pathway}

As described previously, traditional climate fingerprinting links a climate phenomenon to a variety of possible climate forcings (e.g., anthropogenic aerosols, anthropogenic GHG emissions). This ultimately seeks a direct statistical relationship from source to impact. In contrast to this one-to-one relationship, our multi-step attribution process begins by proposing a pathway of arbitrary length and connectedness by which the climate impact may have occurred. The pathway is posited \textit{a priori} using expert understanding of the climate system.

\begin{figure}
    \centering
    \includegraphics[width=0.7\linewidth]{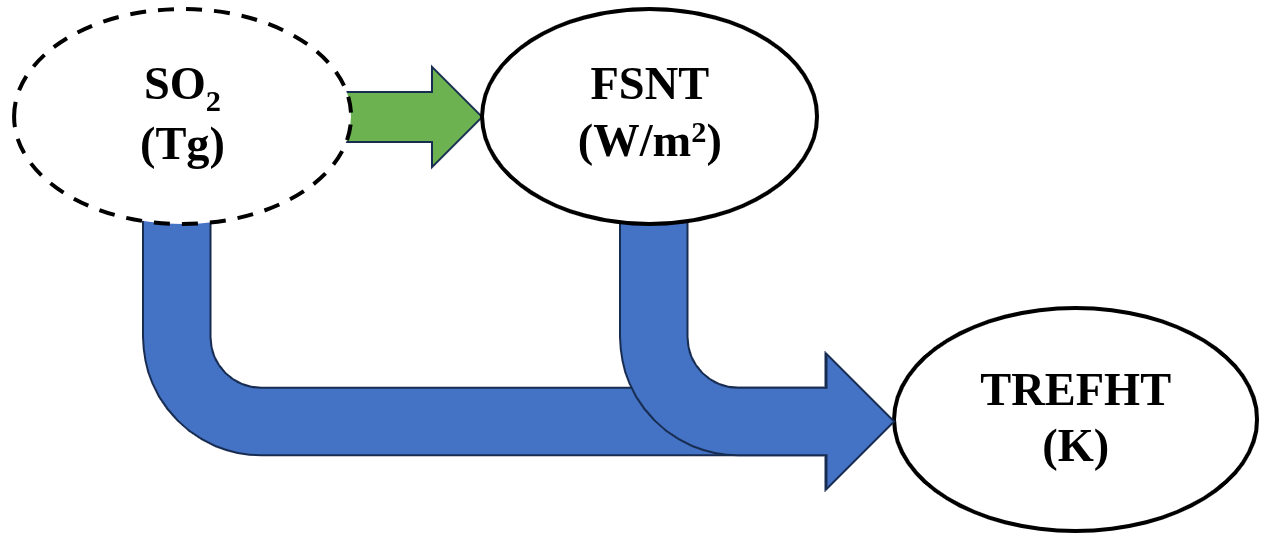}
    \caption{\label{fig:pathgraph}Graph representing the proposed climate impact pathway, with arrows indicating a direction of influence. The source forcing (\SOtwo) is marked by a dashed oval, while solid ovals indicate downstream forcing response variables. Each arrow color represents a separate regression as computed according to Section~\ref{subsec:linregress}.}
\end{figure}

In this paper, we propose a fairly simple pathway by which the eruption of Mt.\ Pinatubo affected the climate. This ``surface cooling'' pathway supposes that the injected aerosols (\SOtwo, measured in teragrams mass) reflect incoming shortwave radiation, resulting in a lower net shortwave radiative flux at the top of atmosphere (FSNT, in W/m$^2$), creating a net cooling effect of the temperature at a reference height of two meters (TREFHT, in K). This pathway architecture is illustrated in Figure~\ref{fig:pathgraph}. More complex climate phenomena, such as the eruption's effect on agricultural productivity, will no doubt require commensurately more complex pathways. The following procedures can be generalized to such scenarios and will be the subject of future work. 

While the arrows connecting \SOtwo\ to FSNT and FSNT to TREFHT capture the primary mechanisms of surface cooling, the arrow connecting \SOtwo\ to TREFHT is an important element of our proposal and merits additional discussion. The \SOtwo\ to TREFHT arrow serves to capture secondary impacts of atmospheric \SOtwo\ injection that are not strictly mediated by changes in FSNT, such as changes in surface albedo or the degree of airborne water vapor. These secondary impacts may be in the same or opposite direction as the primary impact. The use of \SOtwo\ as a direct upstream variable (parent) of downstream terms (TREFHT) allows our model to capture additional variance in the downstream response which is not correlated with mediating variables (FSNT), but is still correlated with the forcing of interest through unspecified secondary mechanisms. Because this capture of secondary effects is key to effective multivariate attribution, we recommend that the forcing variable be specified as a parent to \textit{all} other variables in the pathway. 

\subsection{Scalar metric analysis}

After a pathway has been proposed, average impact time series data are computed from fully-coupled climate simulations according to the process outlined in Section~\ref{subsec:simulations}. These simulations must necessarily include at least two different forcing magnitudes, though a greater number of simulated forcings ensures that the attribution assessment considers a variety of forcing levels. In this paper, we simulate responses to stratospheric \SOtwo\ injections of 0, 1, 3, 5, 7, 10, 13, and 15 Tg. Varying the forcing will be used later to approximate a functional form of the climate system with respect to the forcing magnitude. The historical forcing (10 Tg \SOtwo) provides a proxy for true observational data and is not used to approximate a functional form of the climate system. The no-eruption counterfactual scenario (0 Tg \SOtwo) provides a baseline against which any forcing magnitude may be compared. 

We first define the impact time series containing $\numTime$ time samples as $\stateVec_{\varIdx,\forceIdx,\ensIdx} \eqdef [\stateVar_{\varIdx,\forceIdx,\ensIdx,1}, \ \hdots, \ \stateVar_{\varIdx,\forceIdx,\ensIdx,\numTime}] \inROne{\numTime}$ for the $\varIdx$\textsuperscript{th} variable in the pathway, $\forceIdx$\textsuperscript{th} forcing magnitude, and $\ensIdx$\textsuperscript{th} ensemble member. Examples of the time series data for the global, NH, and NA regions can be found in Figure~\ref{fig:FSNT_TS}. Further, we do not consider the source forcing magnitude as a time series, but rather as a single scalar value which represents the amount of \SOtwo\ injected during June of 1991.

Note that in many cases, the characteristic time scales of a forcing's impact on different climate fields may vary significantly. In the case of the Mt.\ Pinatubo eruption, a decreased reference height temperature is sustained for much longer than the shortwave radiative flux decrease, as displayed in Figure~\ref{fig:FSNT_TS}. Thus, we extract the \textit{average} (detrended) impact from each time series, computed as an arithmetic mean over each spatial region's time series, denoted as 
\begin{equation}
    \metricVar_{\varIdx,\forceIdx,\ensIdx} \eqdef \frac{1}{\numTime} \sum_{\timeIdx=1}^{\numTime} \stateVar_{\varIdx,\forceIdx,\ensIdx,\timeIdx}.
\end{equation}
We associate the scalar forcing with a scalar summarization of the observed climate impacts to link variables in the proposed pathway. Any analysis of relationships between variables over the entire time series (as is standard in optimal fingerprinting) introduces significant additional complexity and we leave this extension for future work.

This procedure determines the average deviation of the impact time series from counterfactual conditions, quantifying the effect of the forcing as a single scalar. While this somewhat limits the generalization of this approach by eliminating temporal variations in the analysis, there is practical value in understanding the aggregate climate impact. Further, as will be shown later, the temporal averaging can subset to seasonal averages to establish trends on shorter time-scales. Depending on the desired analysis, other time-independent scalar values such as maxima or integrated quantities may also serve as metrics. This analysis follows other literature using scalar features~\citep{Wohland2022,Wu2020}. Additionally, scalar metrics are widely used by the community to represent complex climate states~\citep{Reed2022}, such as total precipitation, ITCZ location, gross primary productivity, and number/persistence of atmospheric rivers.

In the following formulations, the scalar metrics from all ensemble members for a given variable and forcing magnitude are collected in the vector denoted by
\begin{equation}
    \metricVec_{\varIdx,\forceIdx} \eqdef [\metricVar_{\varIdx,\forceIdx,1}, \hdots , \metricVar_{\varIdx,\forceIdx,\numEns}] \inROne{\numEns}.
\end{equation}
The inter-ensemble variability of the scalar metrics encodes the internal variability of the system. The ensemble members capture and represent different climate states evolving over time, making their variability a good representation of internal variability. 

\subsection{Forcing response model}\label{subsec:linregress}

From these time-averaged summaries, we now seek a relationship between the scalar measure of a downstream impact and those of any variables which are immediately upstream in the proposed pathway, as a result of varying the forcing magnitude. To begin, we introduce some notation to formalize the relationships between variables as dictated by the proposed pathway. We collect the scalar metrics of a given variable from $\numForce$ forcing levels into the vector,
\begin{equation}
    \metricVec_{\varIdx} \eqdef [\metricVec_{\varIdx,1}^\top, \; \hdots \;, \; \metricVec_{\varIdx,\numForce}^\top]^\top \inROne{\numEns \numForce}.
\end{equation}
To separately represent the forcing magnitude, given for the $\forceIdx$\textsuperscript{th} forcing level as $F_{\forceIdx}$, we create a vector with repeated instances of this scalar as $\forceVec_{\forceIdx} \eqdef [F_{\forceIdx}, \ \hdots \ , \ F_{\forceIdx}]^\top \inROne{\numEns}$. Then, these are assembled for $\numForce$ forcing levels as
\begin{equation}
    \forceVec \eqdef [\forceVec_{1}^\top, \ \hdots \ , \ \forceVec_{\numForce}^\top]^\top \inROne{\numEns \numForce}.
\end{equation}

Under a given pathway, each variable (except for the source forcing) will have variables which are immediately upstream in the pathway. Under the surface cooling pathway in Figure~\ref{fig:pathgraph}, only \SOtwo\ is immediately upstream of FSNT, while \SOtwo\ and FSNT are immediately upstream of TREFHT. We define the \textit{parent set} of a given variable, written as $\parentFunc{\metricVec_{\varIdx}}$, as the variables directly upstream of a given variable. Under the surface cooling pathway, we would thus have $\parentFunc{\metricVec_{\text{FSNT}}} = \{ \forceVec \}$, and $\parentFunc{\metricVec_{\text{TREFHT}}} = \{ \forceVec, \ \metricVec_{\text{FSNT}} \}$. The number of parents for a given variable is given as $\numParents \eqdef |\parentFunc{\metricVec_{\varIdx}}|$. As discussed previously, by including $\forceVec$ in the parent set of each variable, we allow the inference to capture physical relationships not explicitly specified in our probabilistic model.

Next, a model form must be proposed to relate downstream average impacts as the forcing magnitude varies. While the following steps in the proposed pathways-based attribution method readily generalize to more complex model forms (e.g., polynomial or logarithmic), we have found that a strong linear relationship exists between the average impacts of the pathways studied in Section~\ref{sec:results}. If we concatenate the forcing magnitude and average impacts of parent variables into the matrix $\metricMat_{\varIdx} \eqdef [\parentFunc{\metricVec_{\varIdx}}] \inRTwo{\numEns \numForce}{\numParents}$, the linear relationship in predicting downstream average impacts can be written as,
\begin{equation}\label{eq:linear-model}
    \metricVec_{\varIdx} = \metricMat_{\varIdx} \linparamvec_{\varIdx} + \bepsilon_{\varIdx},
\end{equation}
where the terms $\linparamvec_{\varIdx} \inROne{\numParents}$ are the regression parameters to be estimated, and $\bepsilon_{\varIdx} \inROne{\numEns \numForce}$ are the errors for this model. We emphasize that, unlike Eq.~\ref{eq:finger-regress}, the quantities on the left and right hand sides of Eq.~\ref{eq:linear-model} are not the same variable. As such, the resulting coefficients $(\linparamvec_{\varIdx})$ are a measure of the forcing response, not the consistency of simulated and observed data, and is not generally near unity in magnitude. For the simple pathway we propose in this paper, the first step in a pathway (predicting FSNT average impacts from \SOtwo\ magnitude) is a univariate model and the associated $\linparamvec$ has units of $\text{W}/\text{m}^2\text{-Tg}$, unlike the dimensionless regression coefficients of traditional D\&A analyses. Later steps (TREFHT average impact predictions from FSNT and \SOtwo) are multivariate models. Because Eq.~\ref{eq:linear-model} relates different quantities, we suggest that an intercept term be included, though we omit it here for brevity.

As with the basic fingerprinting linear regression described in Section~\ref{subsec:fingerprinting}, we make the assumption that the errors $\bepsilon_{\varIdx}$ are uncorrelated, have equal variance, and are distributed normally, Thus, the maximum likelihood estimator of the model parameters $\linparamvec_{\varIdx}$ is the ordinary least squares estimator, given by
\begin{equation}\label{eq:beta-hat}
    \linparamestvec_{\varIdx} = \left[\metricMat_{\varIdx}^\top \metricMat_{\varIdx}\right]^{-1} \metricMat_{\varIdx}^\top \metricVec_{\varIdx}.
\end{equation}
Unlike the construction in Section~\ref{subsec:fingerprinting}, however, the assumption of uncorrelated $\bepsilon_{\varIdx}$ is an assumption of independence across simulations, not independence across time. The individual $\bepsilon_{\varIdx}$ terms arise from inter-ensemble variability and their standard deviation can be used as a proxy for the internal variability not explicitly represented in the posited pathway. 

We take care to note that when estimating model parameters using only simulation data, it is good practice to exclude any data associated with the forcing level that will be considered the ``true'' forcing level for which we wish to attribute a climate response. This prevents ``data leakage'' where the model is trained using the ``testing'' dataset, unduly improving the resulting attribution purely by construction~\citep{Kapoor2023}. As will be noted later, in our analyses the 10 Tg pseudo-observational dataset is thus not included when evaluating Eq.~\ref{eq:beta-hat}.

\subsection{Joint probability model of observed quantities}\label{subsec:likelihood}

Having estimated functional relationships among each step in our causal pathway, we are now ready to apply these models to the task of attributing observed impacts to specific forcing levels. As introduced in Section~\ref{sec:intro}, we seek a statistical measure of attribution strength that combines a variety of downstream impacts to identify and characterize upstream drivers. As we will show below, such a measure can be constructed by combining the functional forms estimated previously into a single joint likelihood, which we can then use to develop rigorous statistical tests. 

Before proceeding, we again introduce useful notation. We consider both the unknown forcing and the downstream impacts as random variables, respectively denoted $\forceRandom$ and $\metricRandom_{\varIdx}$. In order to assess the effectiveness of our approach, we select 10 Tg as the ``correct'' forcing level and use the associated values of the downstream climate variables as our ``pseudo-observations''. We emphasize that our pseudo-observations are simulation outputs for which the forcing level is exactly known, and not actual observational data or reanalysis product. The set of pseudo-observations derived from the 10 Tg simulations is denoted as $\obsSet \eqdef \{ \metricVar_1^{\obsVar}, \ \hdots \ , \ \metricVar_{\numVars}^{\obsVar} \}$. The forcing magnitude ($\forceRandom$) is not included in $\obsSet$ as it is our target of inference and not an observed quantity. In this paper, only FSNT and TREFHT are thus treated as observed variables (cf. Figure~\ref{fig:pathgraph}).

Following this notation, the joint probability of all variables in the given observational dataset resulting from a particular forcing level can be written as,
\begin{equation}\label{eq:like-simple}
    \pdfVar_{\forceVar} (\obsSet) \eqdef \pdfVar \cond{\obsSet}{\forceRandom = \forceVar},
\end{equation}
where $\pdfVar (\cdot)$ denotes a probability density function. For a set of $\numVars$ pathway variables, this can be expanded as
\begin{align}
    \pdfVar_{\forceVar}(\obsSet) &= \pdfVar \cond{\metricRandom_1 = \metricVar_1^{\obsVar}, \ \hdots \ , \ \metricRandom_{\numVars} = \metricVar_{\numVars}^{\obsVar}}{\forceRandom = \forceVar} \nonumber \\
    &= \pdfVar \cond{\metricRandom_1 = \metricVar_1^{\obsVar}}{\forceRandom = \forceVar} \prod_{\varIdx = 2}^{\numVars} \pdfVar \cond{\metricRandom_{\varIdx} = \metricVar_{\varIdx}^{\obsVar}}{\forceRandom = \forceVar, \ \metricRandom_1 = \metricVar_1^{\obsVar}, \ \hdots \ , \ \metricRandom_{\varIdx-1} = \metricVar_{\varIdx-1}^{\obsVar} } \nonumber \\
    &= \prod_{\varIdx = 1}^{\numVars} \pdfVar \cond{\metricRandom_{\varIdx} = \metricVar_{\varIdx}^{\obsVar}}{\forceRandom = \forceVar, \ \metricRandom_1 = \metricVar_1^{\obsVar}, \ \hdots \ , \ \metricRandom_{\varIdx-1} = \metricVar_{\varIdx-1}^{\obsVar} } \label{eq:like-prod} 
\end{align}
as a result of the chain rule of probability. (For the $v = 1$ term, the only conditioning is on $F$.) Recalling the parent set $\parentFunc{\cdot}$, if a variable is not a parent of a given variable, then it does not influence the associated term in the joint density. Thus, Eq.~\ref{eq:like-prod} can be written concisely as
\begin{equation}\label{eq:like-compact}
    \pdfVar_{\forceIdx} (\obsSet) = \prod_{\varIdx = 1}^{\numVars} \pdfVar \cond{\metricRandom_{\varIdx} = \metricVar_{\varIdx}^{\obsVar}}{\forceRandom = \forceVar, \ \parentFunc{\metricRandom_\varIdx} = \parentFunc{\metricVar_{\varIdx}^{\obsVar}} }, 
\end{equation}
Here, we have computed linear models for each downstream variable with regression parameters given by Eq.~\ref{eq:beta-hat}. Under the assumption that the errors are distributed normally, the conditional average impact predictions have the following Gaussian distribution 
\begin{equation}
    \metricRandom_{\varIdx} \ | \ \forceRandom, \ \parentFunc{\metricRandom_\varIdx} = \parentFunc{\metricVar_{\varIdx}^{\obsVar}} \sim \mathcal{N} \left( \linparamest_{\text{F} \rightarrow \varIdx} \forceRandom + \sum_{j \in \parentFunc{\metricRandom_{\varIdx}}} \linparamest_{j \rightarrow \varIdx} \metricRandom_{j} , \ \linvarest_{\varIdx} \right).
\end{equation}
\noindent Specifically, we take $\metricRandom_{\varIdx}$ to be (conditionally) Gaussian with conditional mean given by the regression model defined above in Equations~\ref{eq:linear-model} and~\ref{eq:beta-hat} and variance given by the the associated estimate of error variance, $\linvarest_{\varIdx}$.

Under this distribution, the probability density terms of each observed variable, $\varIdx$, may be analytically computed as
\begin{equation}\label{eq:normal-like}
    \begin{split}
        \pdfVar &\cond{\metricRandom_{\varIdx} = \metricVar_{\varIdx}^{\obsVar}}{\forceRandom = \forceVar, \ \parentFunc{\metricRandom_\varIdx} = \parentFunc{\metricVar_{\varIdx}^{\obsVar}}} \\
        &= \frac{1}{\sqrt{2 \pi \linvarest_{\varIdx}}} \exp \left( - \frac{1}{2\linvarest_{\varIdx}} \left( \metricVar_{\varIdx}^{\obsVar} - \left(\linparamest_{\text{F} \rightarrow \varIdx} \forceVar + \sum_{j \in \parentFunc{\metricRandom_{\varIdx}}} \linparamest_{j \rightarrow \varIdx} \metricVar_{j}^{\obsVar} \right) \right)^2 \right).
    \end{split}
\end{equation}
Computing the full joint probability density, $\pdfVar_{\forceIdx}(\obsSet)$, is a straightforward combination of Equations~\ref{eq:like-compact} and~\ref{eq:normal-like}, but can require somewhat cumbersome bookkeeping; the interested reader may find practical details in the provided code. 

Recall that the term $\linvarest_{\varIdx}$ is the sample variance of the residuals of the linear model which predicts the $\varIdx$\textsuperscript{th} variable. This encodes our modeling uncertainty and reflects our measure of internal variability. A particular strength of this framework is the potential for multi-step pathways to improve this inference, as computing the joint densities and resulting likelihoods with \textit{more} information tends to decrease uncertainty and to improve our ability to successfully attribute a climate response to the correct forcing. Specifically, including more predictors tends to reduce the residual variance $\linvarest_{\varIdx}$, but the set of predictors used in the pathway must be chosen with some care to avoid ``overfitting.''

Again, in the ``perfect model'' study we present here, the ``observational'' dataset is not derived from historical measurements. Rather, pseudo-observational data is constructed as the ensemble average scalar measurement for each downstream variable of interest. That is, 
\begin{equation}\label{eq:obs-mean}
    \metricVar_{\varIdx}^{\obsVar} = \frac{1}{\numEns} \sum_{e = 1}^{\numEns} \metricVar_{\varIdx,\forceIdx_{\obsVar},e}.
\end{equation}
where $\forceIdx_{\obsVar} = 10$ is the ``true'' forcing level we have selected. We emphasize that this choice is made only for demonstration purposes to be consistent with the actual Pinatubo eruption. Note also that this ``true'' forcing level is the same set which is deliberately excluded from estimating the forced model response for the sake of preventing data leakage and assessing the robustness of this multi-step attribution procedure, as noted in Section~\ref{subsec:linregress}.

\subsection{Likelihood ratio test}\label{subsec:ratios}

The final step to assess attribution involves comparing the likelihoods for each forcing, as computed from Eq.~\ref{eq:like-compact}. We begin by defining two sets of forcing magnitudes: a single fixed forcing magnitude $\forceVar_1$ for which we wish to assess attribution, and a set $\forceSet_0$ which \textit{excludes} $\forceVar_1$. We propose a \textit{series} of null hypotheses $H_{0,\forceVar}$ that the true forcing of the observational data is some $\forceVar_0 \in \forceSet_0$, and an alternative hypothesis $H_1$ that the true forcing is instead $\forceVar_1$. Rejection of each null hypothesis under some statistical test thus indicates \textit{attribution} of the observed impacts to the alternative hypothesis forcing. Following~\citet{Allen1999} and many others, our test is fundamentally a \textit{model consistency} test which is used to reject incompatible forcings. While many such consistency tests exist, we adopt a likelihood ratio testing framework, as it provides a flexible and intuitive approach to building powerful statistical tests under arbitrarily complex multivariate pathways. 

Given the functional forms of the joint probability density functions $\pdfVar_{\forceVar}(\cdot)$ as defined in Eq.~\ref{eq:like-compact}, and some data $\dataSet$ (not necessarily the observational data), the likelihood ratio test statistic is defined as
\begin{equation}\label{eq:test-stat}
    \testStat_{\forceVar_0, \forceVar_1} (\dataSet) \eqdef \log \left(\frac{ \likelihood_{\dataSet}(\forceVar_1)}{\likelihood_{\dataSet}(\forceVar_0)} \right) = \log\left(\frac{ \pdfVar_{\forceVar_1} (\dataSet)}{\pdfVar_{\forceVar_0} (\dataSet)} \right).
\end{equation}
where $\likelihood_{\dataSet}(\forceVar_1)$ is the likelihood of forcing $\forceVar_1$ associated with data $\dataSet$. Recall that the likelihood and probability density functions are generally numerically equal and differ principally in which quantities are considered fixed.
A larger test statistic indicates a greater likelihood of the alternative hypothesis, given the data $\dataSet$. We further define $\testDist_{\forceVar_0,\forceVar_1}$ as the distribution of the test statistic $\testStat_{\forceVar_0,\forceVar_1}$ when the null forcing magnitude is assumed to be true. 

In this work, we utilize Monte Carlo sampling to simulate each distribution $\testDist_{\forceVar_0,\forceVar_1}$ for each forcing magnitude $\forceVar_0 \in \forceSet_0$. From Eqs.~\ref{eq:like-compact}-\ref{eq:normal-like}, the likelihoods are normally distributed with means according to the linear regression parameters computed from Eq.~\ref{eq:beta-hat} and variance computed from the sample residuals. Thus, using a random number generator, we can draw random samples $\dataSet_{\forceVar_0,i}$ from the normal distributions $\likelihood_{\forceVar_0} (\cdot)$, compute the test statistic $\testStat_{\forceVar_0,\forceVar_1} (\dataSet_{\forceVar_0,i})$ for each random sample, and approximate $\testDist_{\forceVar_0,\forceVar_1}$ empirically from a large number of random samples. Given the observational test statistic $\testStat_{\forceVar_0,\forceVar_1} (\obsSet)$, we can compute the $p$-value for each forcing magnitude in the null set as
\begin{equation}\label{eq:lr-pval}
    p_{\forceVar_0} \eqdef \text{Pr} ( \testDist_{\forceVar_0,\forceVar_1} \ge \testStat_{\forceVar_0,\forceVar_1} (\obsSet) )
\end{equation}
This approximately computes the probability that we measure a test statistic at least as extreme as that which we observe, assuming that the null hypothesis is true. If this $p$-value is very low, it is highly unlikely that the null forcing level $\forceVar_0$ is capable of producing a test statistic greater than $\testStat_{\forceVar_0,\forceVar_1} (\obsSet)$, and we may reject that null hypothesis with a confidence level equal to $1 - p_{\forceVar_0}$.

To make the above description more concrete, in the following section the alternative forcing level $\forceVar_1$ will be the supposed 10 Tg eruption magnitude which we wish to attribute. The series of null hypothesis forcing levels may, in theory, be any forcing level other than 10 Tg. For the sake of simplicity, we will instead restrict the null forcing levels to be those which were simulated to generate the regression models. The above process is repeated for each of the null forcing magnitudes investigated, and a $p$-value reported in each case. If all null hypotheses can be rejected with a certain level of confidence, it amounts to attribution of the observed climate response to the alternative forcing.

%% file: results.tex
We now apply the methodology detailed above to the 1991 eruption of Mt.\ Pinatubo and the purported resulting surface cooling. We begin with a demonstration of the basic fingerprinting approach described in Section~\ref{subsec:fingerprinting} to illustrate the challenges in applying this traditional method to a short-term forcing of continuously-varying magnitude. We then apply the proposed multi-step conditional attribution method to analyze its ability to address the shortcomings of fingerprinting.

\begin{figure}
    \centering
    {\begin{minipage}{0.45\linewidth}
        \includegraphics[width=0.99\linewidth]{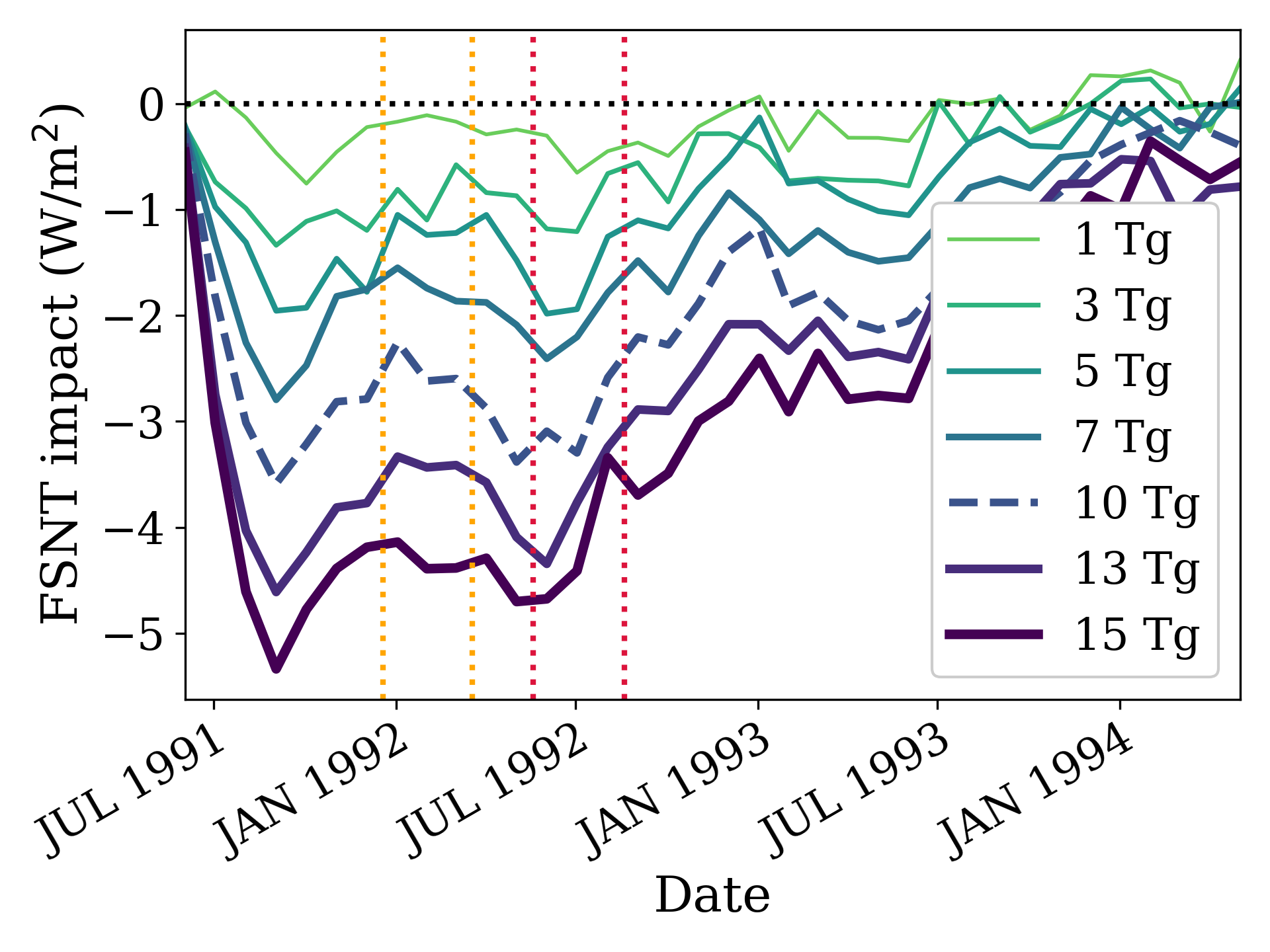}
    \end{minipage}
    \begin{minipage}{0.45\linewidth}
        \includegraphics[width=0.99\linewidth]{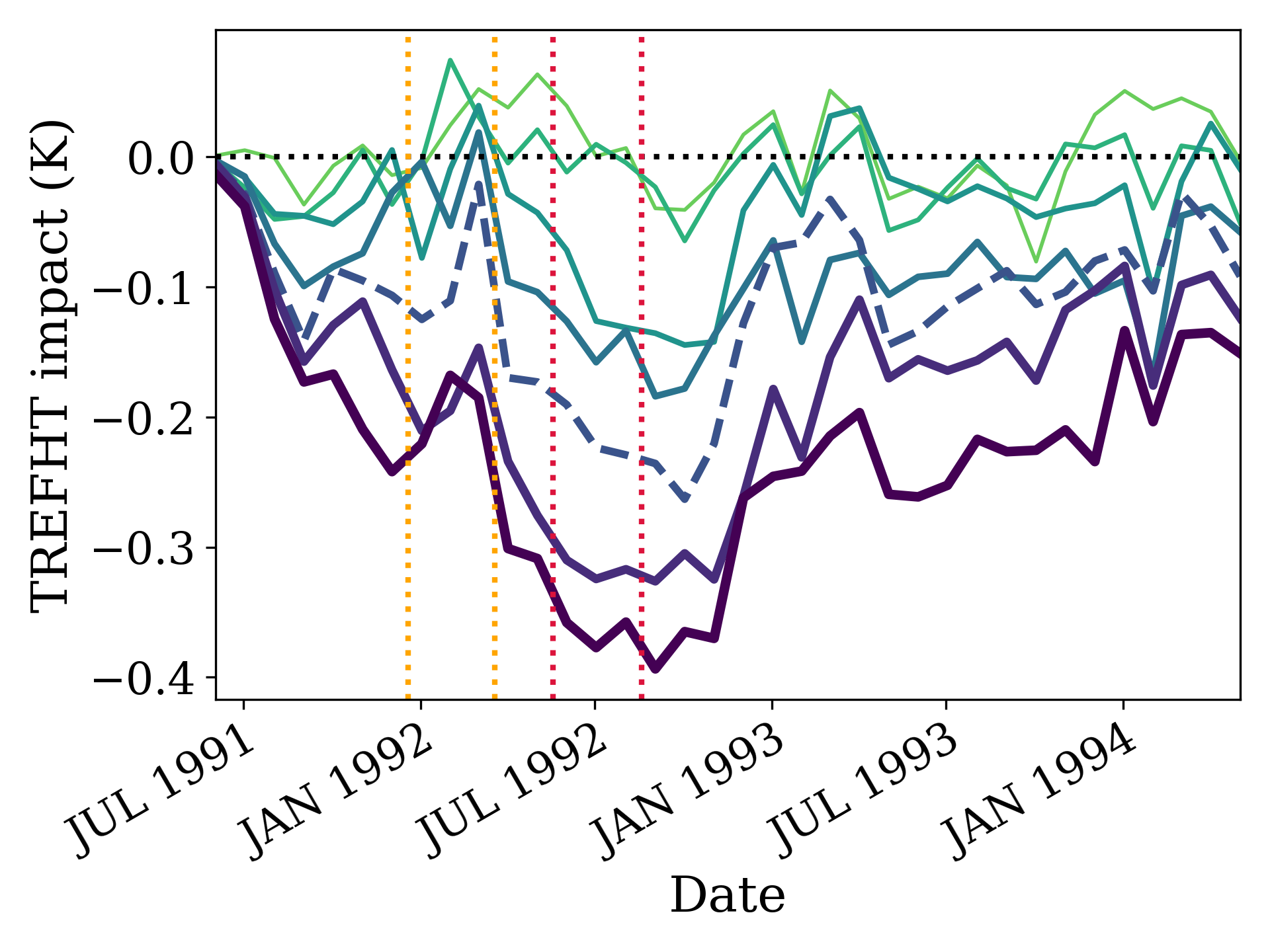}
    \end{minipage}
    \subcaption{Global}}
    
    {\begin{minipage}{0.45\linewidth}
        \includegraphics[width=0.99\linewidth]{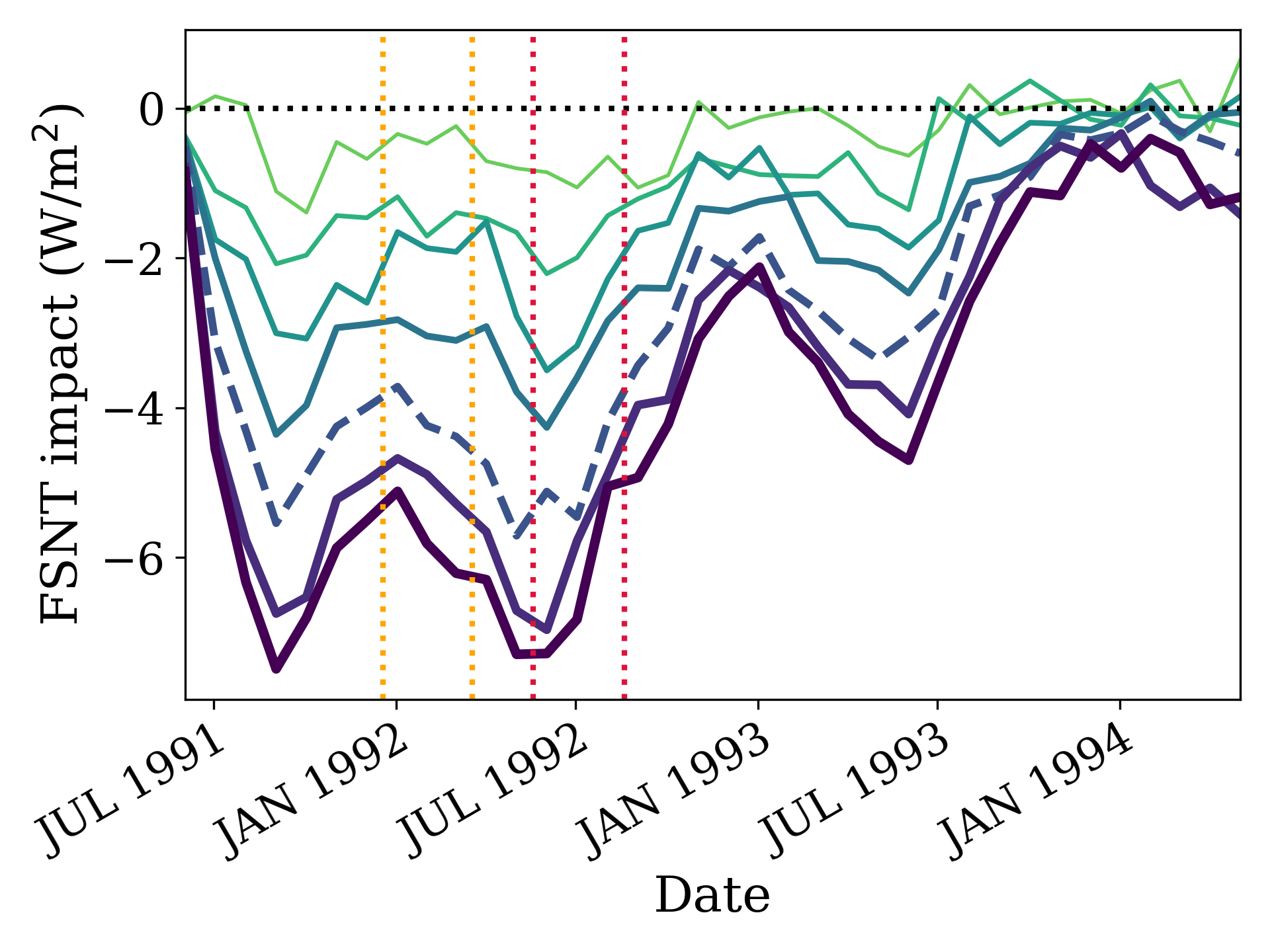}
    \end{minipage}
    \begin{minipage}{0.45\linewidth}
        \includegraphics[width=0.99\linewidth]{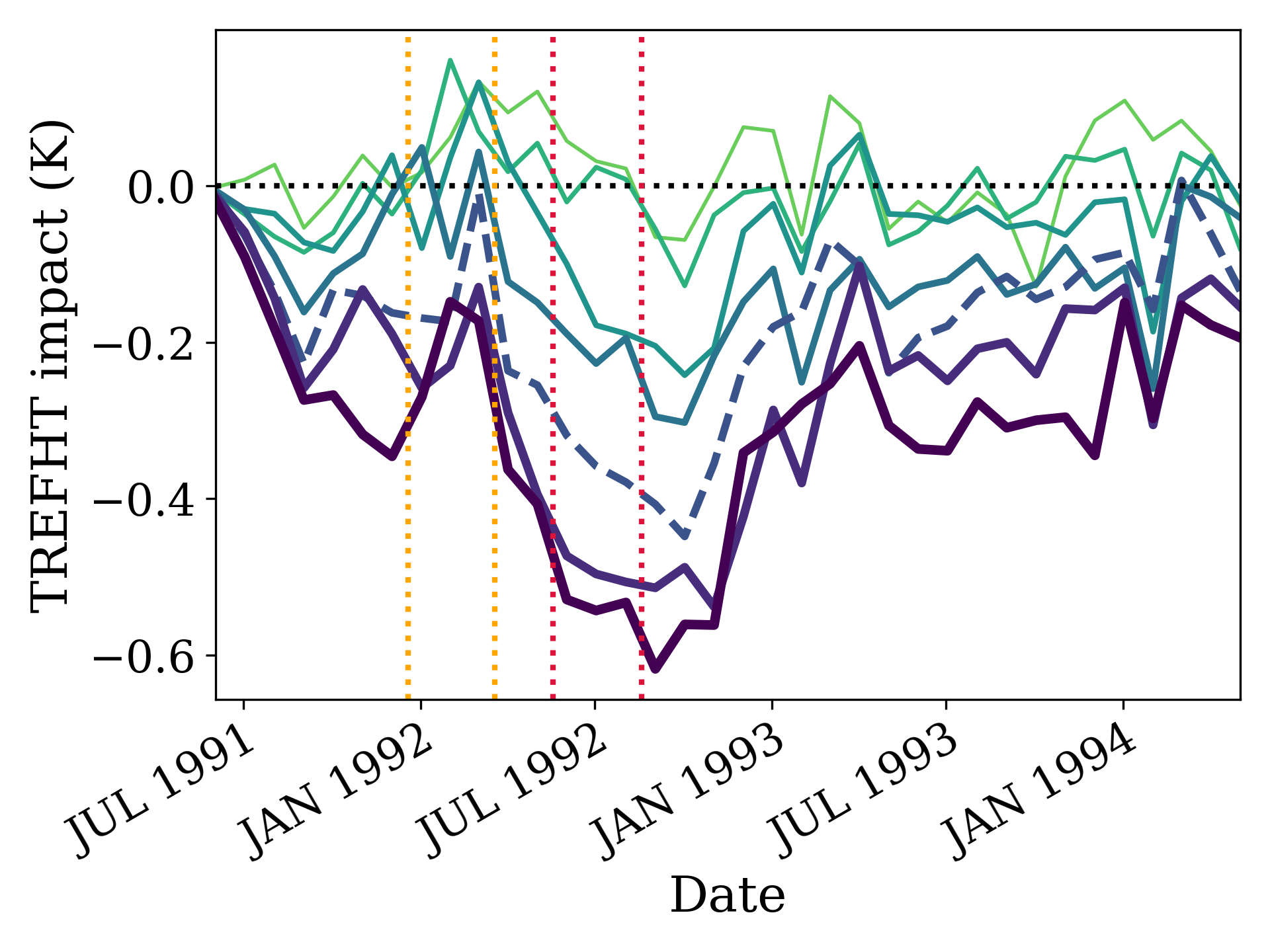}
    \end{minipage}
    \subcaption{Northern Hemisphere}}

    {\begin{minipage}{0.45\linewidth}
        \includegraphics[width=0.99\linewidth]{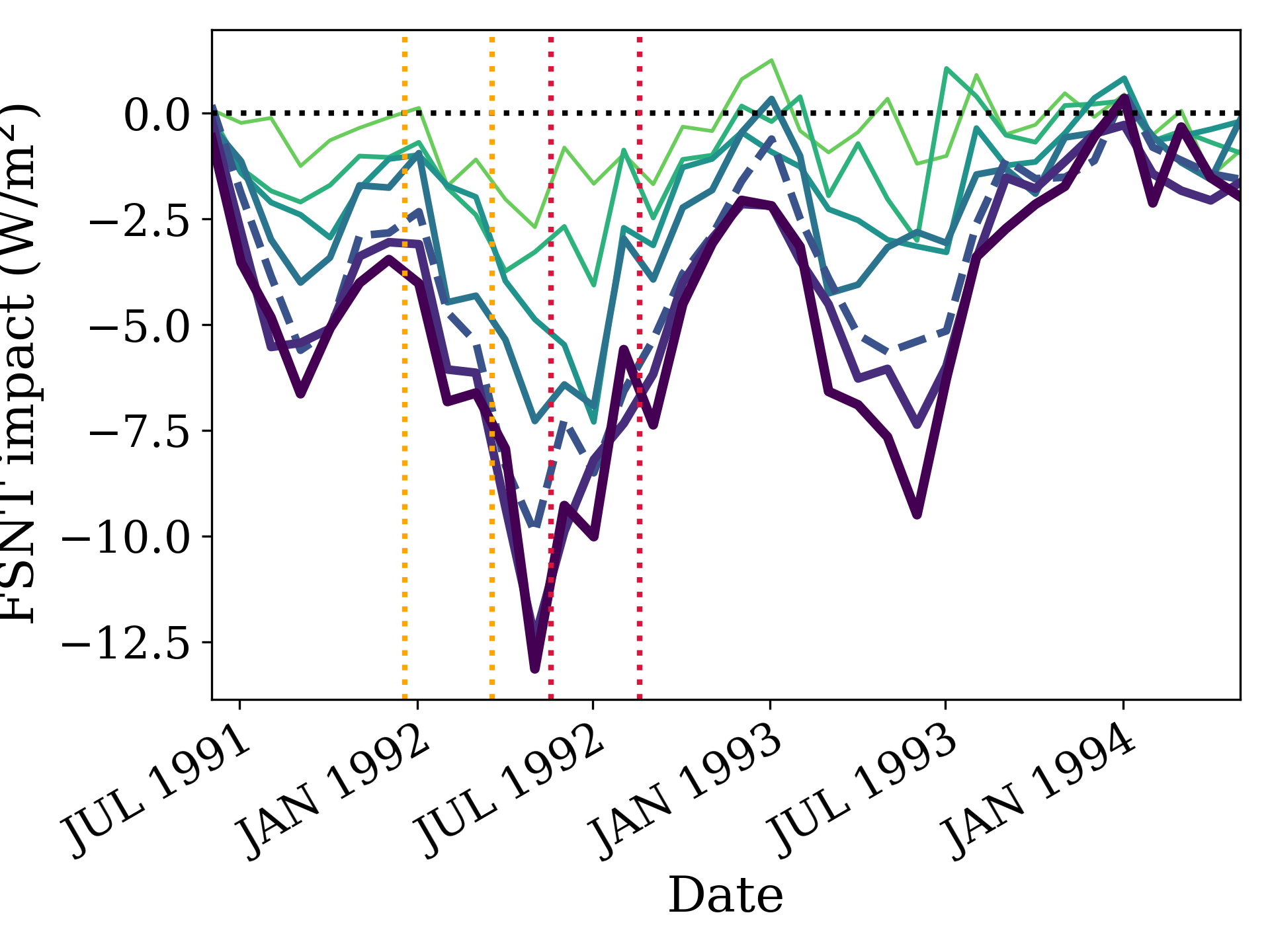}
    \end{minipage}
    \begin{minipage}{0.45\linewidth}
        \includegraphics[width=0.99\linewidth]{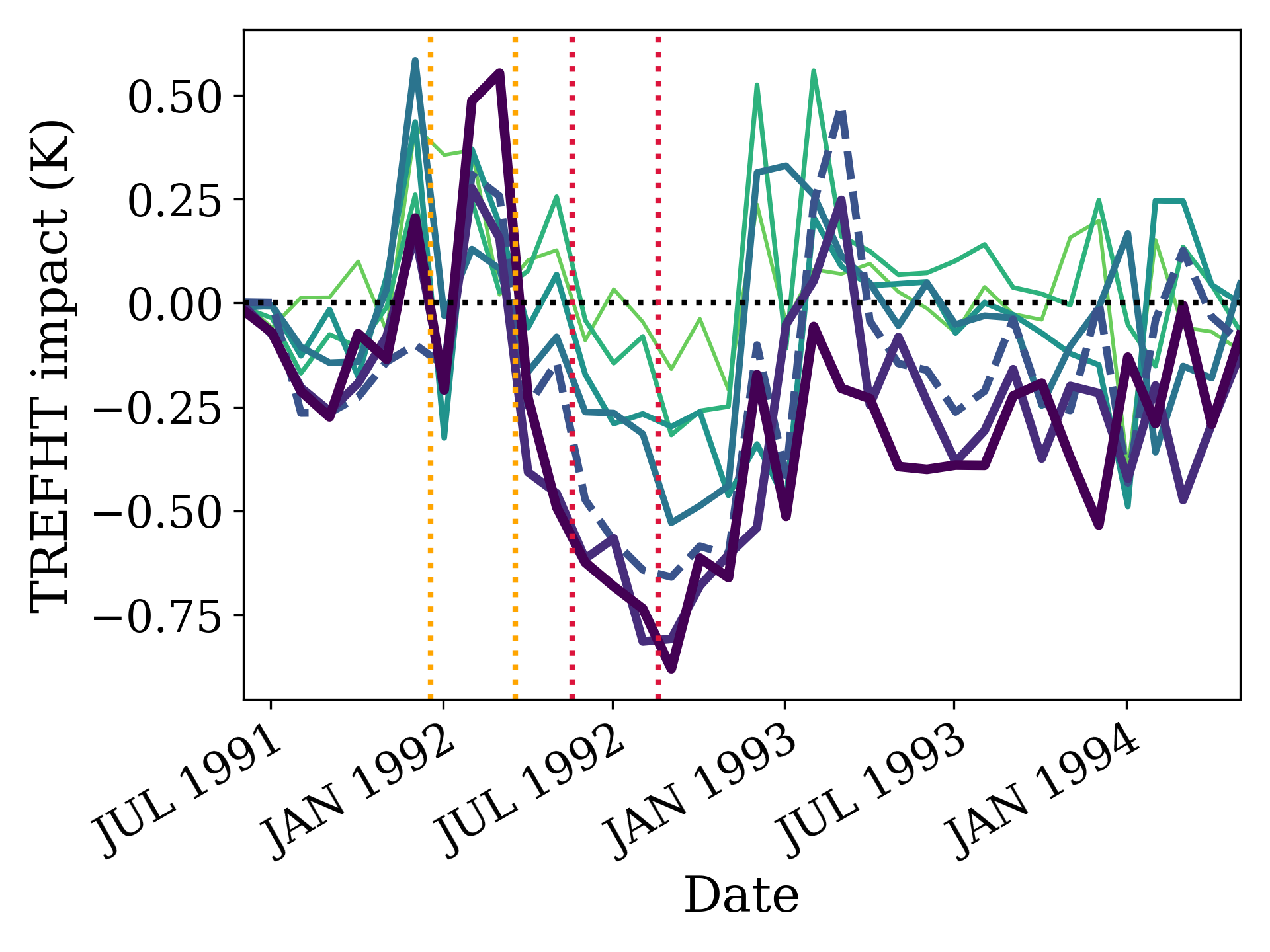}
    \end{minipage}
    \subcaption{North America}}
    
    \caption{\label{fig:FSNT_TS}Latitude-weighted average ensemble mean impact for FSNT (left) and TREFHT (right) under various forcing magnitudes over distinct spaial regions. The pseudo-observational 10 Tg impact is marked by a dashed line, while the zero impact line is marked with a dotted black line. The 1992 JFM and JJA time periods are bounded by vertical orange and red dotted lines, respectively. Note that the vertical axis limits differ for each spatial region.}
\end{figure}

As outlined in Section~\ref{subsec:simulations}, we simulate the eruption and the following three years under $\numForce = 8$ different forcing scenarios, each characterized by the mass of \SOtwo\ injected into the stratosphere by the eruption: 0 (no eruption, counterfactual), 1, 3, 5, 7, 10 (pseudo-observational forcing level), 13, and 15 Tg \SOtwo. We simulate $\numEns = 15$ ensemble members for each forcing level, for a total of 120 simulations. Latitude-weighted averages of the pathway variables are computed, and the ensemble means are plotted in Figure~\ref{fig:FSNT_TS}. Each row in Figure~\ref{fig:FSNT_TS} displays ensemble averages over spatial regions of decreasing area: global (66S-66N, 180W-180E), the Northern Hemisphere (NH, 0-66N, 180W-180E), and North America (NA, 25N-66N, 170W-60W) regions. The short time periods of additional interest, 1992 JFM and JJA, are bounded by vertical lines (orange and red respectively).

In addition to the primary response to aerosol lifetime, each region exhibits significant seasonal patterns in both FSNT and TREFHT impacts. As revealed by~\citet{hbrown2024} using the clear-sky (no influence from clouds) analog of FSNT, there is no seasonal tempering of FSNT impact in the winters (1991/1992, 1992/1993, and shown here, but not by~\citet{hbrown2024}, 1993/1994) implying the tempering is cloud-related. The temperature responds not only to the radiative changes from the presence of aerosols, but also cloud cover and ensuing dynamically driven changes. These seasonal cloud-related signatures are also present in the TREFHT histories. This pattern is enhanced in the Northern Hemisphere (and regions contained within the Northern Hemisphere) with the summer possessing more optically dense clouds~\citep{rossow1999} thus increasing the FSNT and TREFHT impacts to greater magnitudes. In contrast, in the winter with less optically dense clouds, there is a tempered response in the FSNT and TREFHT impact. This seasonal pattern is important to note as it will arise in our seasonally-focused and pathways-based conditional attribution results. 

Additionally, the temperature response in the winters after the eruption in NA exhibit positive impacts as opposed to the expected overall cooling. This is a well-researched secondary impact of the Mt.\ Pinatubo eruption with warm surface anomalies present over the northern continental landmasses in the winter(s) following the eruption~\citep{RobockMao1992,parker1996,kirchner1999}. This so-called Northern Hemisphere winter warming response to tropical volcanic eruptions like Mt.\ Pinatubo has been the subject of much research~\citep{Polvani:2019,Zanchettin2019,weierbach2023,dogar2024}, and is explored more fully in the dataset by~\citet{ehrmann2025}.

\subsection{Fingerprinting}\label{subsec:finger-results}

We begin the fingerprinting demonstration with a very simple case: distinguishing a 10 Tg eruption from the counterfactual climate in which no eruption occurs. Here, we only consider global average time series data over a three year time period, as it was found that projecting the dataset onto any number of leading EOFs only worsened attribution success (not shown) and any further restriction on the data (regional or temporal) further worsens the results (again, not shown). Recall that this three year time period is significantly reduced from the 16-year window used by~\citet{Lehner2016} on CMIP5 simulations and observations. This decrease in signal-to-noise increases the difficulty of the attribution beyond that of~\citet{Lehner2016}, though we employ a perfect model analysis and restrict ENSO at initiation instead of during the first boreal winter

Following the perfect model, leave-one-out (LOO) procedure detailed in Section~\ref{subsec:fingerprinting}, this process takes one member of the 10 Tg simulation global average time series ensemble as the pseudo-observation $\stateVec_o$, and computes the data matrix $\stateMat$ from the ensemble means of the remaining 10 Tg and counterfactual ensemble members. We consider both the univariate fingerprinting case using only TREFHT time series data, along with a multivariate fingerprinting scenario using both FSNT and TREFHT data. Given the disparate units of these quantities, we normalize each variable separately within the range $[-1, 1]$, using the minimum and maximum values over the entire dataset under the considered forcing levels.

\begin{figure}
    \centering
    \begin{minipage}{0.45\linewidth}
        \includegraphics[width=0.99\linewidth]{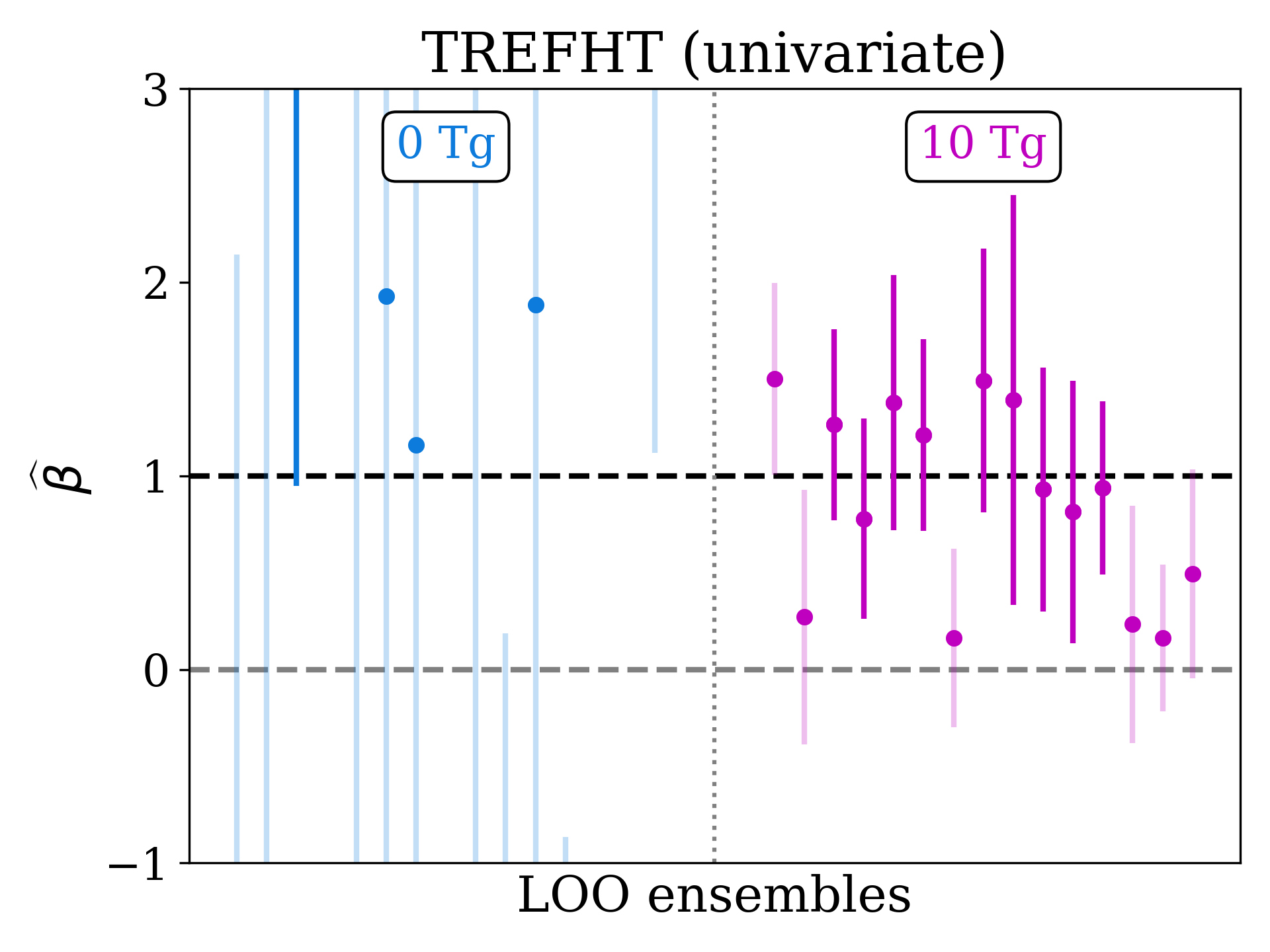}
    \end{minipage}
    \begin{minipage}{0.45\linewidth}
        \includegraphics[width=0.99\linewidth]{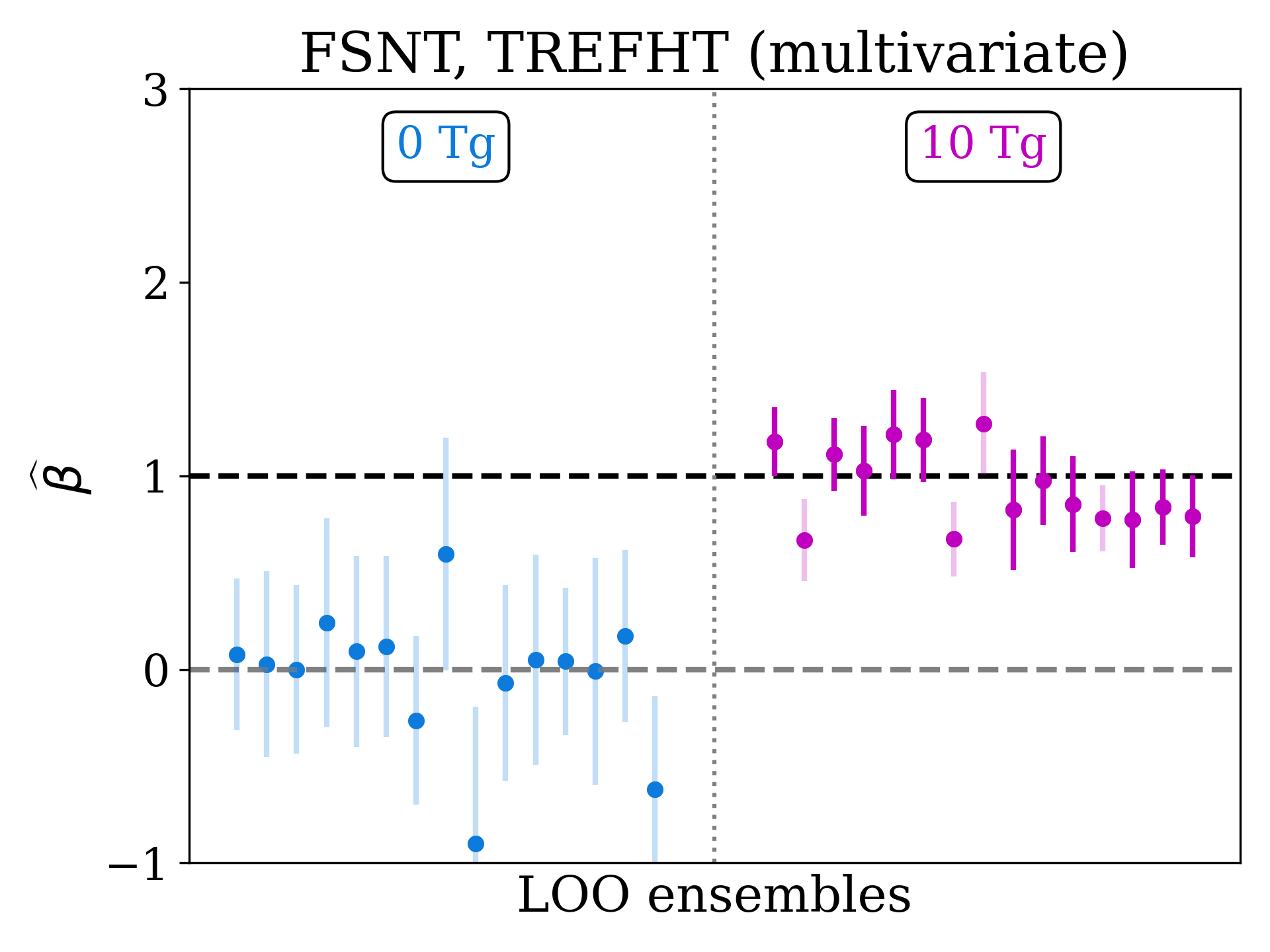}
    \end{minipage}
    \caption{\label{fig:betas-two}Leave-one-out perfect model fingerprinting of 10 Tg forced response vs. counterfactual unforced response, for TREFHT-only time series (left) and FSNT-TREFHT multivariate analysis (right). Dots indicate the inferred $\fingparamest$ value for either the 10 Tg or counterfactual response, for each LOO index. Error bars indicate the associated 95\% confidence interval. Translucent error bars indicate failed D\&A, while bold error bars indicate successful D\&A. The $\fingparamest = 0,\, 1$ levels are marked by horizontal dashed gray and black lines, respectively.}
\end{figure}

Successful detection and attribution is indicated by a confidence interval on the regression coefficient $\fingparamest$ associated with the 10 Tg time series which includes unity, but does not include zero. These coefficient values for each ensemble member, along with their 95\% confidence intervals, are displayed in Figure~\ref{fig:betas-two}. In both the univariate and multivariate analyses, successful attribution of the 10 Tg forcing is achieved in approximately 75\% of cases. The addition of FSNT in the multivariate case somewhat improves the value of the 10 Tg $\fingparamest$ value (closer to unity with a smaller confidence interval). In this case, traditional fingerprinting performs fairly well despite the noisy short-term climate response, but is only distinguishing between a very large volcanic eruption and the lack thereof.

\begin{figure}
    \centering
    \begin{minipage}{0.45\linewidth}
        \includegraphics[width=0.99\linewidth]{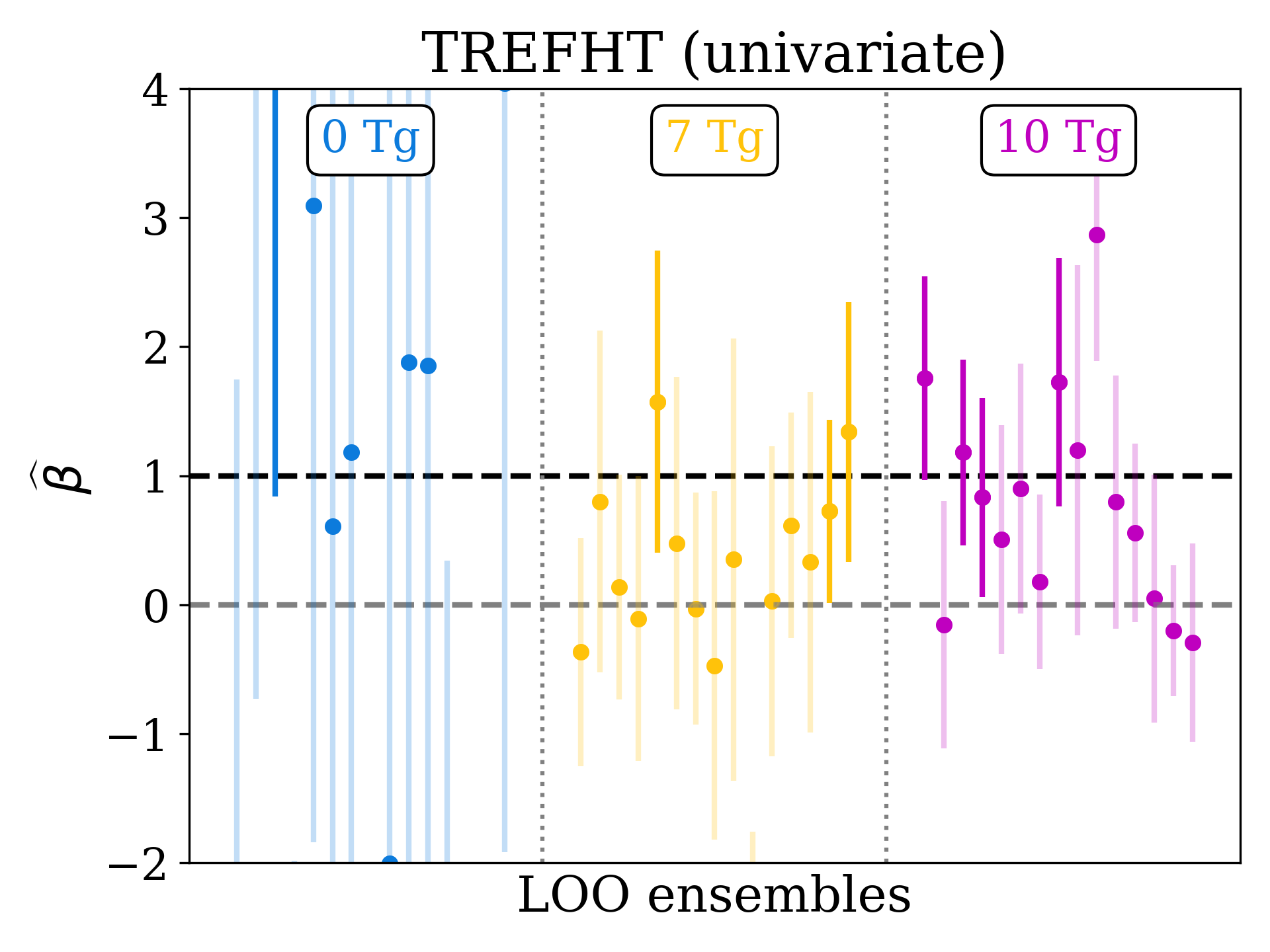}
    \end{minipage}
    \begin{minipage}{0.45\linewidth}
        \includegraphics[width=0.99\linewidth]{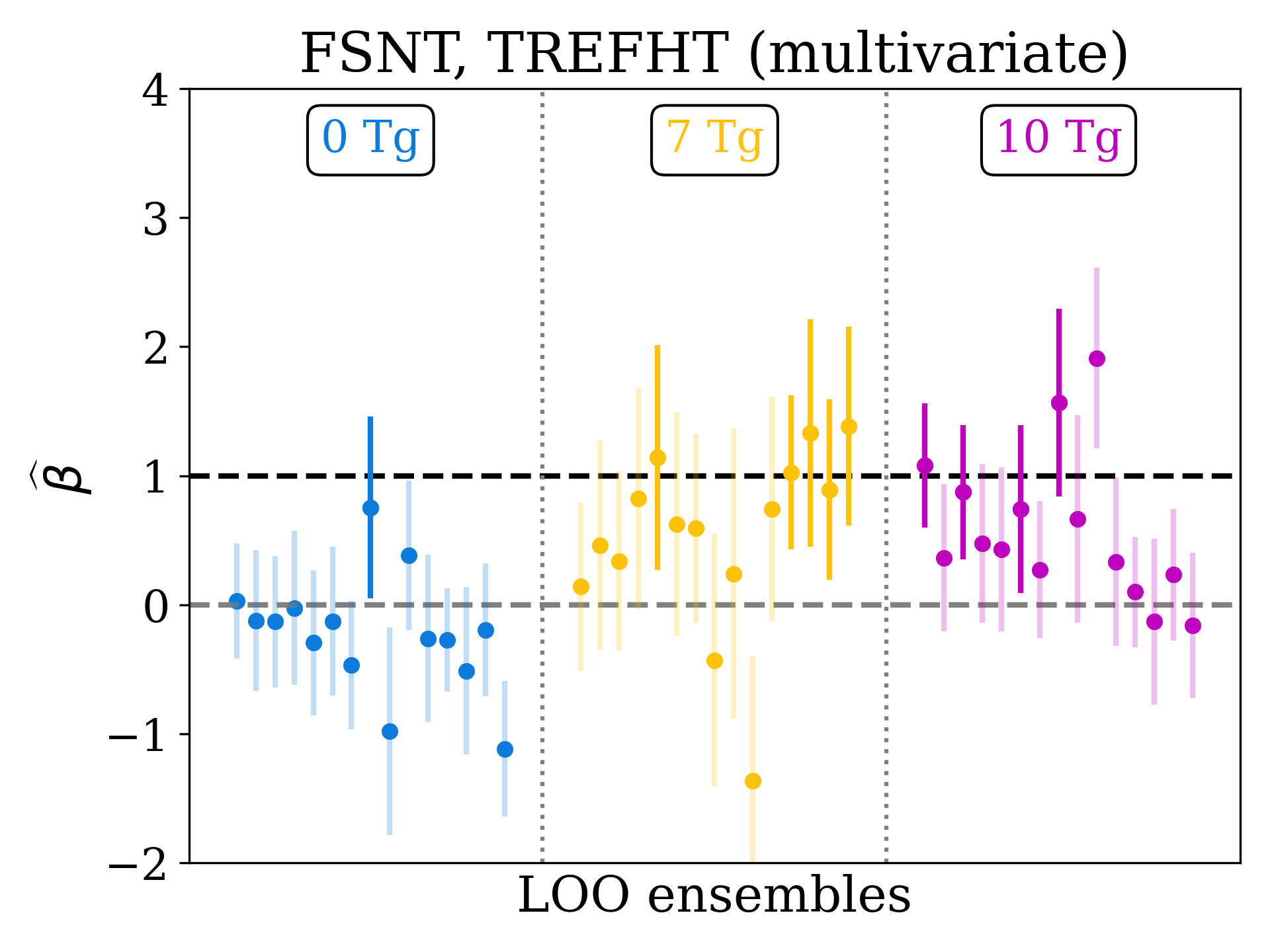}
    \end{minipage}
    \caption{\label{fig:betas-three}Same as Fig.~\ref{fig:betas-two}, but for analysis including three forcing levels (0, 7, and 10 Tg eruptions), additionally including the $\fingparamest$ values and confidence intervals for the 7 Tg response.}
\end{figure}

To illustrate the difficulty in applying fingerprinting to continuous (rather than categorical) climate forcings, we repeat the previous analysis, but additionally include time series data for the 7 Tg eruption. Thus, this attribution attempts to distinguish between three forcing levels, where one is relatively close to the 10 Tg ``true'' forcing. The resulting fingerprinting results are shown in Figure~\ref{fig:betas-three}. The addition of the 7 Tg data significantly worsens the ability to attribute the 10 Tg forcing, greatly decreasing the number of successful attributions for both the univariate and multivariate cases to less than 25\% of LOO cases. In the multivariate case, the 7 Tg forcing is incorrectly attributed in more instances than the 10 Tg forcing. This illustrates the effect of violating the assumption of forcing separability as is standard in typical fingerprinting approaches. Undoubtedly, adding data from simulations at more forcing levels will only exacerbate this issue further.

\subsection{Conditional pathways-based attribution}\label{subsec:multistep-results}

We now turn to the proposed conditional attribution described in Sections~\ref{subsec:pathway}-\ref{subsec:ratios}, following the pathway established in Section~\ref{subsec:pathway}. To motivate the use of intermediate physical effects (i.e., FSNT) in such multi-step pathways throughout this section we will make comparisons against the related single-step pathway; that is, relating a temperature impact directly from the \SOtwo\ forcing magnitude. Comparisons between the single-step and multi-step pathways demonstrate that the proposed attribution framework benefits from additional information, while traditional D\&A methods may not necessarily benefit from multivariate analyses, as seen in Section~\ref{subsec:finger-results}. We will further demonstrate this framework over very short time scales (winter and summer seasons in 1992) and in progressively smaller spatial regions (global, NH, and NA regions). 

\begin{figure}
    \centering
    \begin{minipage}{0.45\linewidth}
        \includegraphics[width=0.99\linewidth]{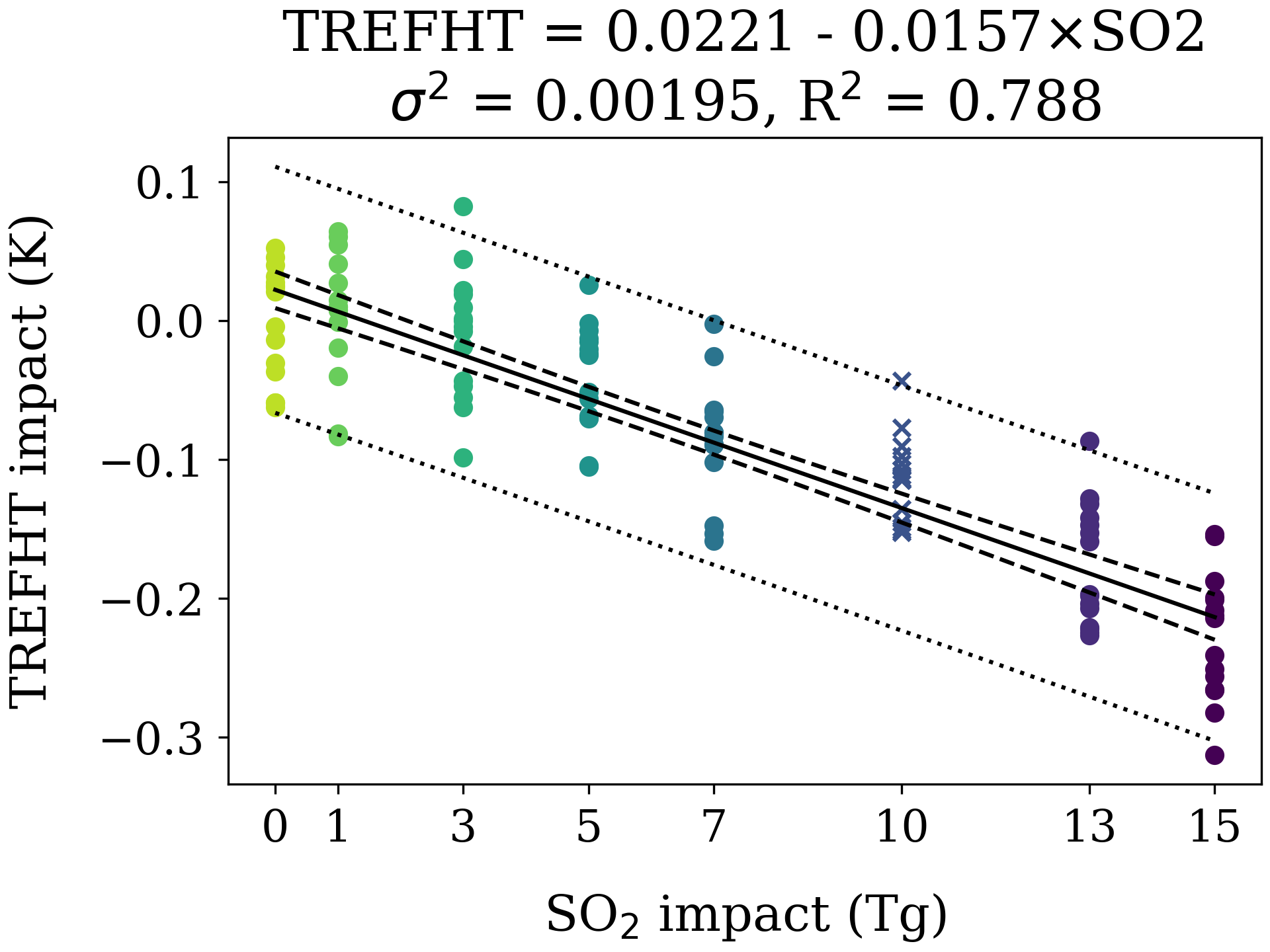}
    \end{minipage}
    
    \vspace{1em}
    \begin{minipage}{0.45\linewidth}
        \includegraphics[width=0.99\linewidth]{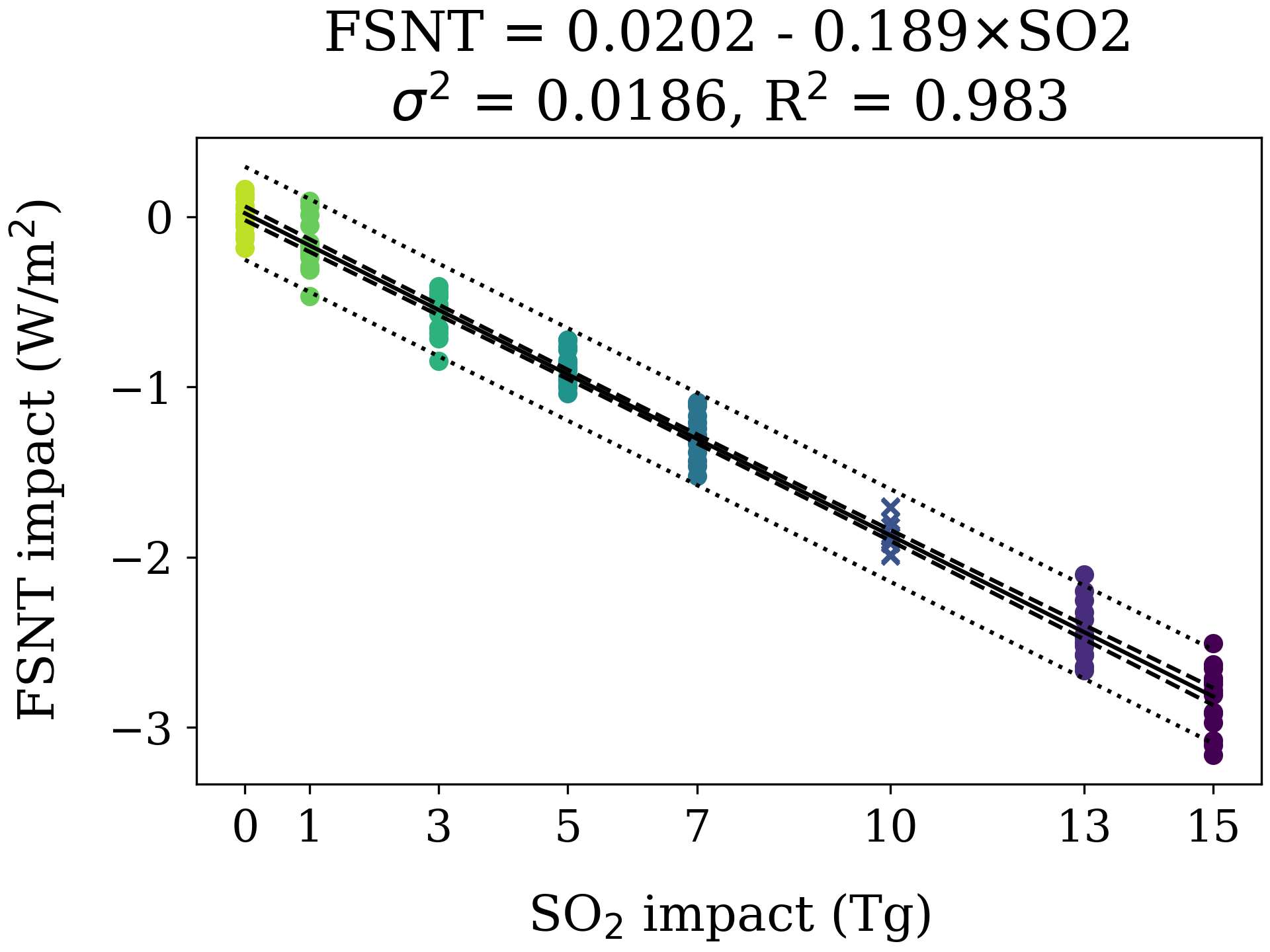}
    \end{minipage}
    \begin{minipage}{0.49\linewidth}
        \includegraphics[width=0.90\linewidth,trim={4em 3em 3em 1em},clip]{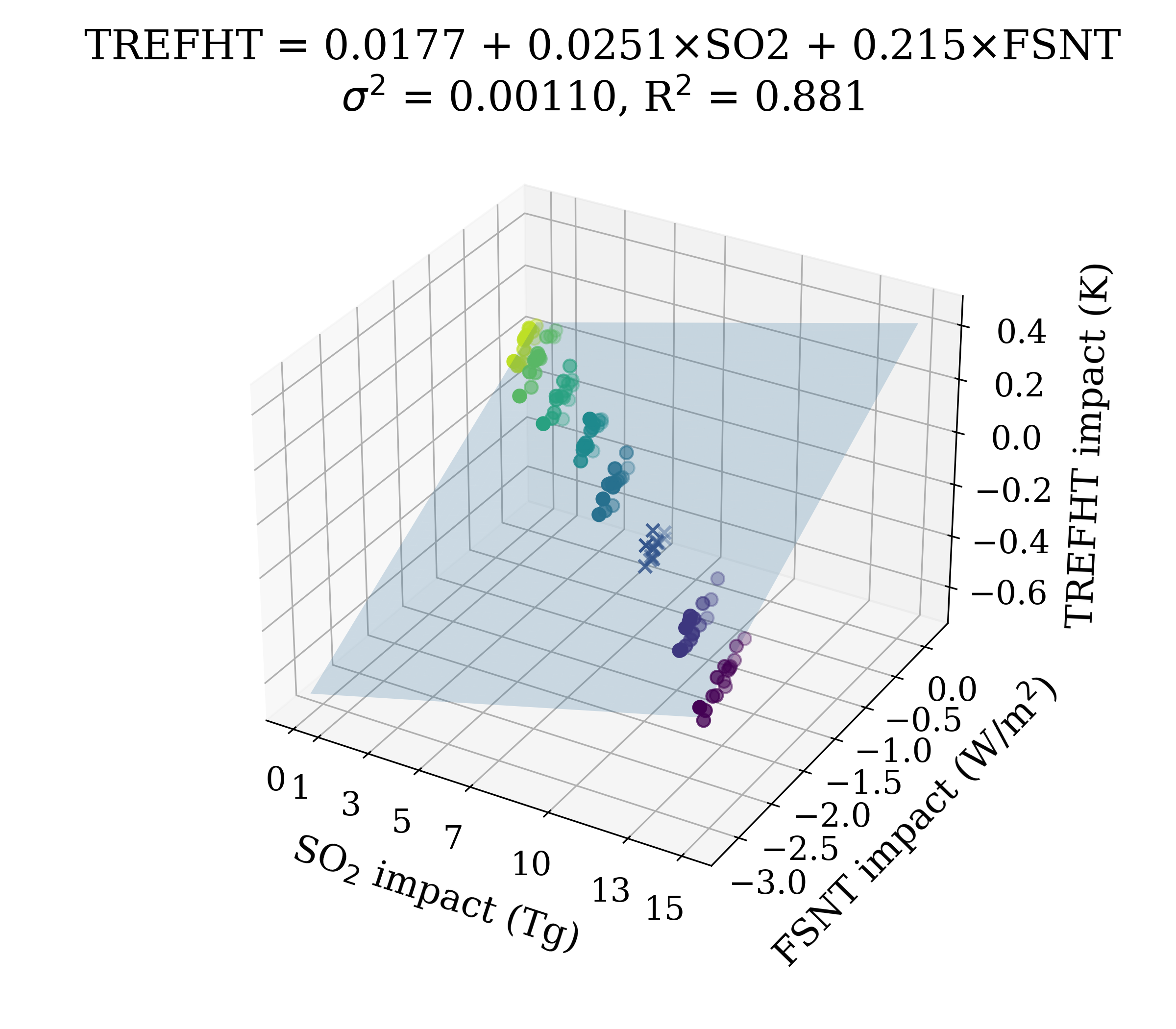}
    \end{minipage}
    \caption{\label{fig:regressions-glob}Global average impact linear regression fits for surface cooling pathway over three years (June 1991 -- June 1994); top row shows the required single-step regression and the bottom row the required multi-step pathway regressions. Univariate model plots include the OLS fit (solid line), 95\% confidence interval (dashed line), and 95\% prediction interval (dotted line). Multivariate model plots display the best fit plane (shaded gray). Model parameters, variance, and R$^2$ values are noted in figure titles. The 10 Tg data are not included in computing the regression, and are marked separately with X symbols on the plots. }
\end{figure}

Linear regression models for the average impact response are computed for each step in the proposed pathways according to Eq.~\ref{eq:beta-hat}. As mentioned in Section~\ref{subsec:linregress}, we deliberately remove any 10 Tg simulation data from this calculation, as this represents the pseudo-observational data whose response we wish to attribute. Figure~\ref{fig:regressions-glob} illustrates the average impact data and resulting linear models over the entire three year period for the single-step (top row) and multi-step pathway (bottom row) attribution framework. These fits exhibit strong linear relationships, lending some credence to the use of scalar metrics in our model: while the time-varying relationships between forcing level, radiative flux, and temperature are clearly highly non-linear from Figure~\ref{fig:FSNT_TS}, there exists a plausible linear relationship in average impact space. The corresponding linear regressions are similarly computed for the restricted spatial regions (NH and NA) and time periods (1992 JFM and JJA).

Table~\ref{tab:regressions} summarizes the estimated regression coefficients and Table~\ref{tab:r2} associated $R^2$ values. The single-step regression coefficients are given in un-normalized (actual) units, and estimate sensitivities of average temperature (K) to eruption size (Tg) across various temporal and spatial windows. For the single-step regressions, we note that all of the $\linparamest_{SO_2}$ are negative except for the last row representing the NA 1992 winter. The negative coefficients indicate that as the forcing level of the eruption (in Tg of \SOtwo) increases, the reference height temperature decreases. Specifically, we see that the global three year average change in TREFHT for a 10 Tg eruption is $10 \times -0.0157$ which represents an average decrease of -0.157 K globally over the three year period. We note that the summer 1992 months show a stronger decrease in average TREFHT temperature, especially regionally in the NH, and specifically in NA, where a decrease of nearly -0.5 $K$ is observed for a 10 Tg eruption. This enhanced impact, as discussed above, is partially the result of optically dense summer clouds. Further, the positive response in NA 1992 JFM is the result of the winter warming discussed previously and explored in detail by~\citet{ehrmann2025}. We also note that the 1992 JFM coefficients, in all regions over all time windows, indicate the smallest relative effect with the poorest regression fits.

The multi-step regression coefficients in Table~\ref{tab:regressions} are computed from data normalized to the range [-1, 1] such that the relative magnitude of the coefficients can be compared without units, giving a sense of relative importance of different predictors. In comparing the normalized contributions of $\linparamest_{SO_2}$ and $\linparamest_{FSNT}$ in the last two columns of Table~\ref{tab:regressions}, it is clear that both are important in defining the plane of best fit for TREFHT. In other words, by controlling for the effect of FSNT on TREFHT in a multivariate regression model, we are able to determine that a significant portion of downstream variability not fully explained by FSNT alone is correlated with \SOtwo. Comparing the magnitude of these coefficients allows us to quantify how much the secondary effects of the eruption contribute to changes in TREFHT, independently of how the eruption alters FSNT. As expected, the primary modulation of TREFHT is through FSNT, as indicated by the larger coefficients of $\linparamest_{FSNT}$, but the secondary effects captured by $\linparamest_{SO_2}$ are non-negligible. In particular, we see that the magnitude of $\linparamest_{SO_2}$ is typically one quarter to one half of the magnitude of the primary impact, $\linparamest_{FSNT}$. On closer inspection, we also note that both the magnitude and direction of these secondary effects ($\linparamest_{SO_2}$) vary seasonally, consistent with different seasonal cloud-based signatures (see Figure \ref{fig:FSNT_TS} and related discussion).

\begin{table}
    \caption{\label{tab:regressions} Estimated linear regression coefficients for TREFHT predictions. The middle column indicates coefficients for the single-step, un-normalized predictions of TREFHT from \SOtwo\ . The rightmost two columns indicate coefficients for the multi-step, normalized predictions of TREFHT from \SOtwo\ and FSNT jointly.}
    \begin{tabular}{llrcrr} 
    \toprule
    \multirow{2}*{Region} & \multirow{2}{*}{Time Window} & Single-step, dimensional & & \multicolumn{2}{c}{Multi-step, normalized} \\
    \cmidrule{3-3} \cmidrule{5-6}
     & & $\linparamest_{SO_2}$ (K/Tg) & & $\linparamest_{SO_2}$ (unitless) & $\linparamest_{FSNT}$ (unitless) \\
    \midrule
             & 3 years  & -0.0157 & &  0.953 & 1.810 \\
    Global   & 1992 JJA & -0.0265 & & -0.381 & 0.330 \\
             & 1992 JFM & -0.0155 & &  0.135 & 0.688 \\
    \midrule
             & 3 years  & -0.0223 & &  0.450 & 1.310 \\
    N. Hem   & 1992 JJA & -0.0403 & & -0.353 & 0.364 \\
             & 1992 JFM & -0.0188 & &  0.344 & 0.908 \\
    \midrule
             & 3 years  & -0.0212 & &  0.244 & 0.842 \\
    N. Amer. & 1992 JJA & -0.0487 & & -0.094 & 0.771 \\
             & 1992 JFM &  0.0007 & &  0.306 & 0.433 \\
    \bottomrule
    \end{tabular}
\end{table}

While not a major focus of our analysis, the residual variance $\linvarest_{\varIdx}$ associated with each regression step acts as a representation of the climate's internal variability, and is presented in Appendix~\ref{app:variances}. This unexplained variability is a significant driver of attribution certainty: pathways with high internal variability will have rather ``flat'' likelihoods that make attribution to a particular forcing level difficult. Intuitively, if the internal variability dominates the estimated impact ($R^2 \ll 1$), meaningful attribution becomes all but impossible. While it is clear that inclusion of additional variables will generally reduce $\linvarest_{\varIdx}$ and increase $R^2$, extension of the pathway without solid scientific support may lead to overfitting. Robust thresholds for selecting $R^2$ and $\linvarest_{\varIdx}$ are hard to determine, however, and we leave this as a question for future work.

\begin{table}
    \caption{\label{tab:r2}Linear regression $R^2$ values for single-step, intermediary, and multi-step regressions.}
    \begin{tabular}{llrrr} 
     \toprule
     Region & Time & \SOtwo $\rightarrow$ TREFHT & \SOtwo $\rightarrow$ FSNT & \SOtwo, FSNT $\rightarrow$ TREFHT \\
     \midrule
             & 3 years & 0.788 & 0.983 & 0.881 \\
    Global   & 1992 JJA & 0.796 & 0.922 & 0.806 \\
             & 1992 JFM & 0.421 & 0.910 & 0.498 \\
    \midrule
             & 3 years & 0.797 & 0.976 & 0.855 \\
    N. Hem   & 1992 JJA & 0.798 & 0.899 & 0.813 \\
             & 1992 JFM & 0.261 & 0.925 & 0.352 \\
    \midrule
             & 3 years & 0.406 & 0.897 & 0.543 \\
    N. Amer. & 1992 JJA & 0.629 & 0.680 & 0.824 \\
             & 1992 JFM & 0.005 & 0.747 & 0.075 \\
    \bottomrule
    \end{tabular}
\end{table}

\begin{figure}
    \begin{minipage}{0.32\linewidth}
        \includegraphics[width=0.99\linewidth,trim={15em 11em 15em 11em},clip]{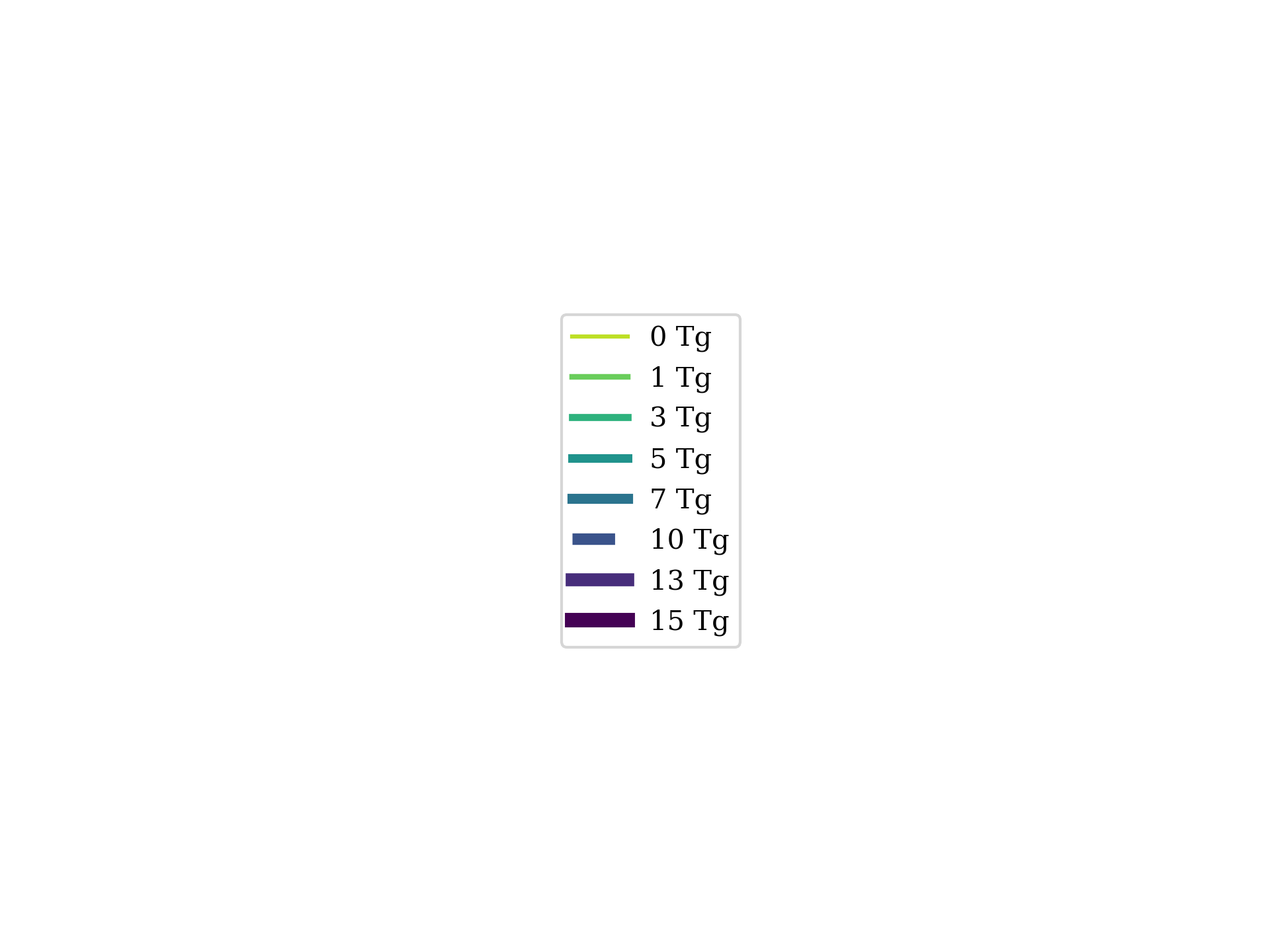}
    \end{minipage}
    \begin{minipage}{0.32\linewidth}
        \centering
        \vspace{3em} Single step $\longrightarrow$ \\ \vspace{3em} Multi-step
    \end{minipage}
    \begin{minipage}{0.32\linewidth}
        \includegraphics[width=0.99\linewidth]{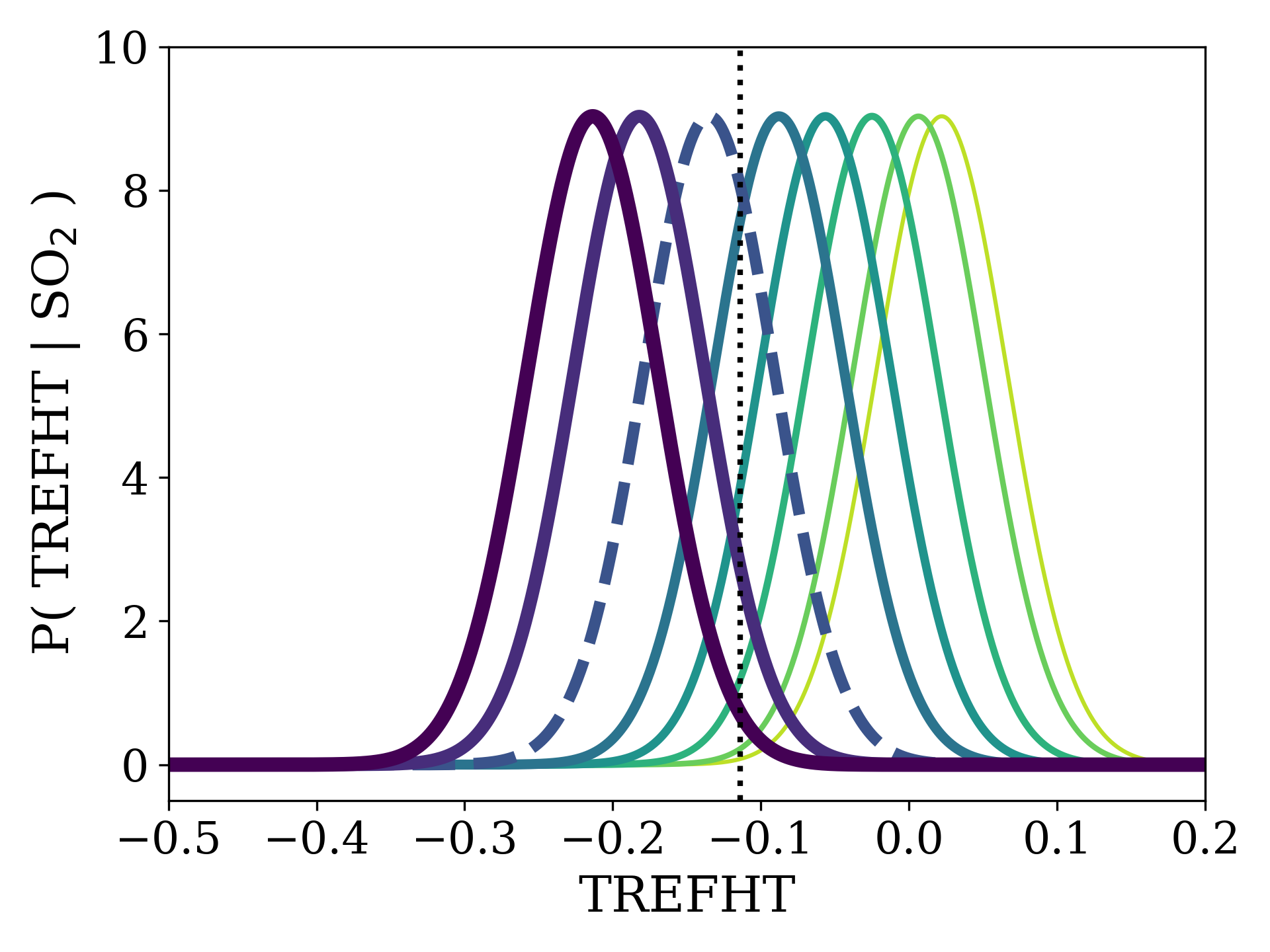}
    \end{minipage}
    \hspace{0.5em}

    \begin{center}
        $\overbrace{\qquad \qquad \qquad \qquad \qquad \qquad \qquad \qquad \qquad \qquad \qquad \qquad \qquad \qquad \qquad \qquad \qquad \qquad}$
    \end{center}
    
    \begin{minipage}{0.32\linewidth}
        \includegraphics[width=0.99\linewidth]{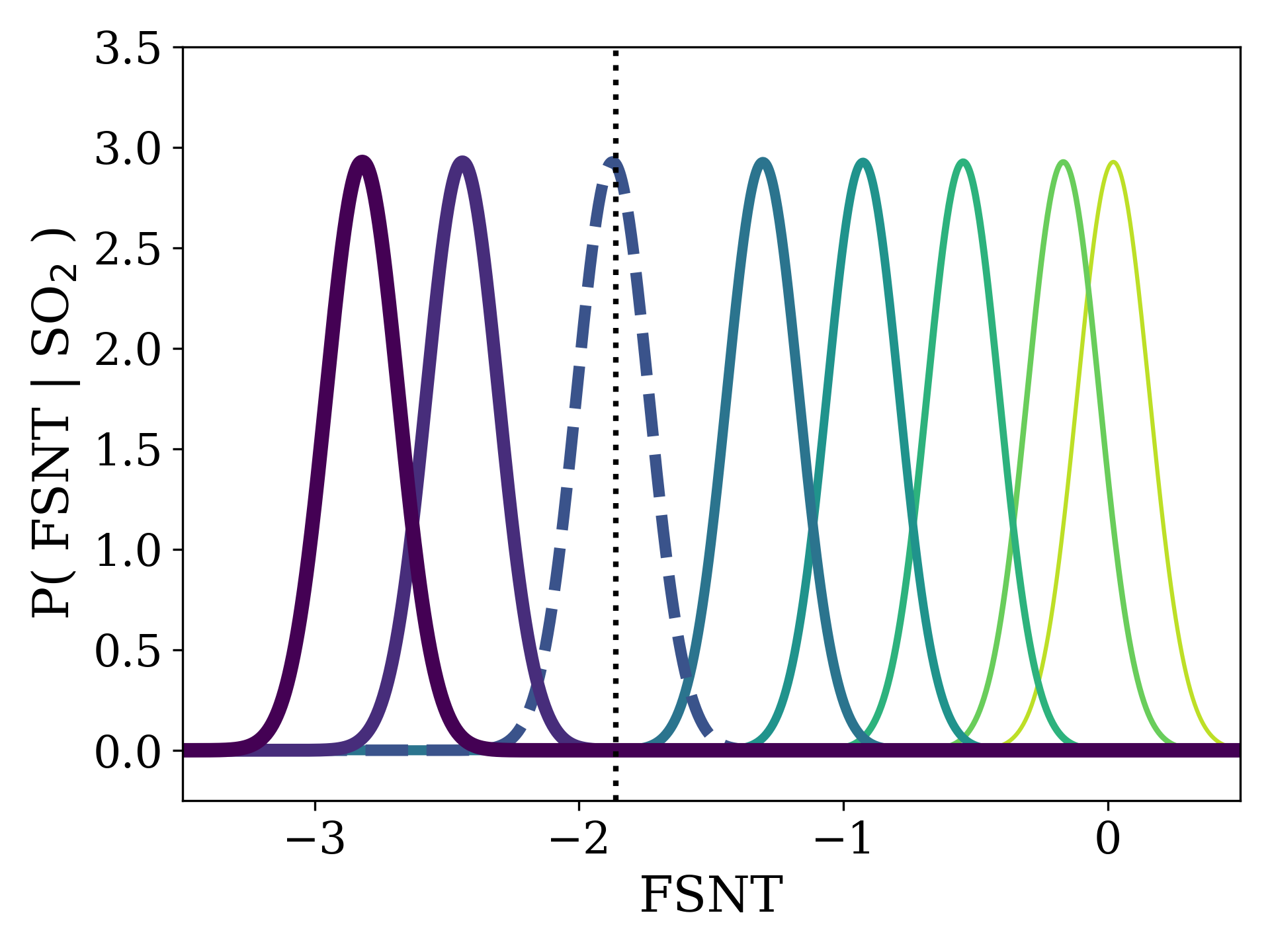}
    \end{minipage}
    \begin{minipage}{0.32\linewidth}
        \includegraphics[width=0.99\linewidth]{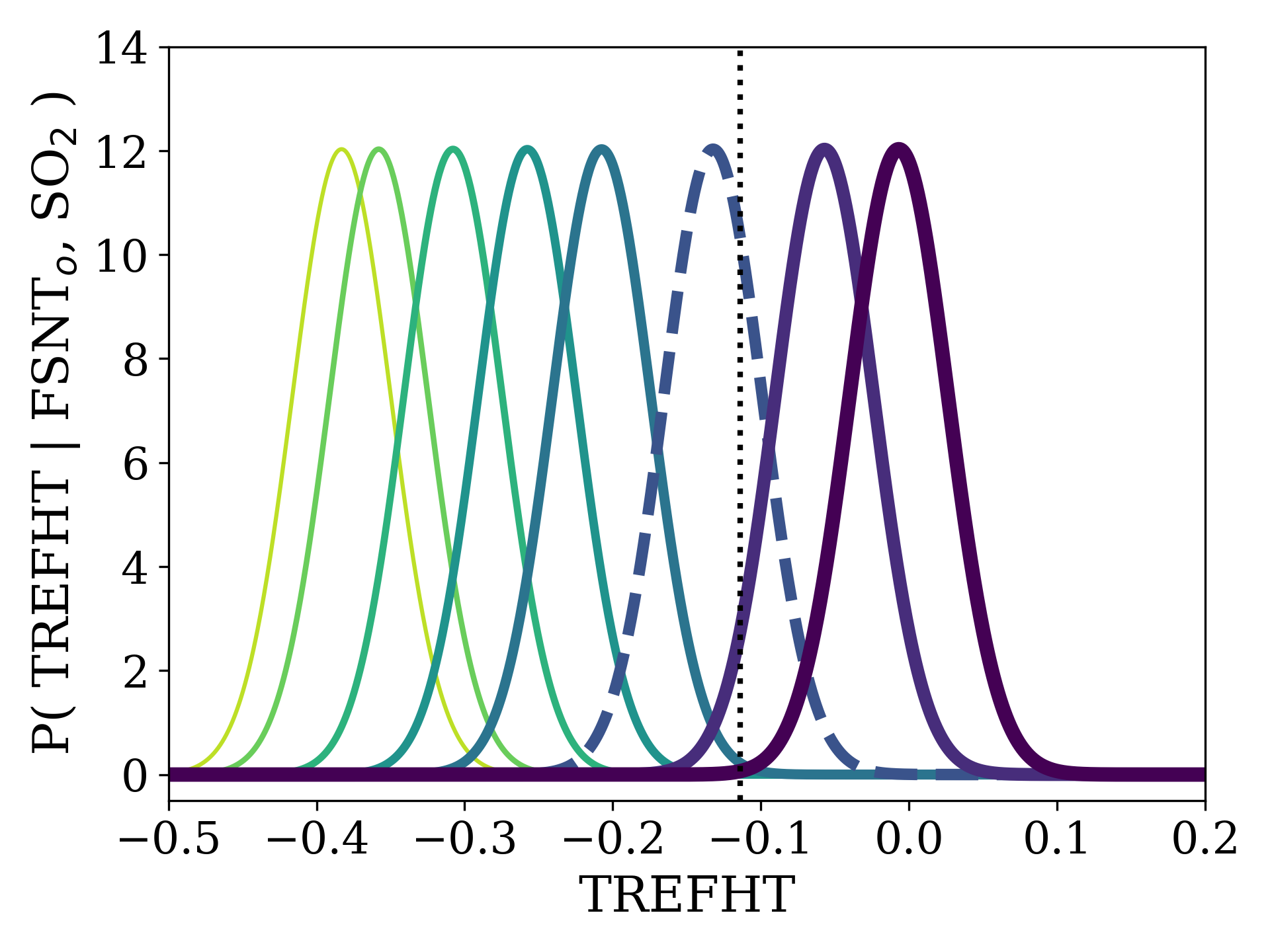}
    \end{minipage}
    \begin{minipage}{0.32\linewidth}
        \includegraphics[width=0.99\linewidth]{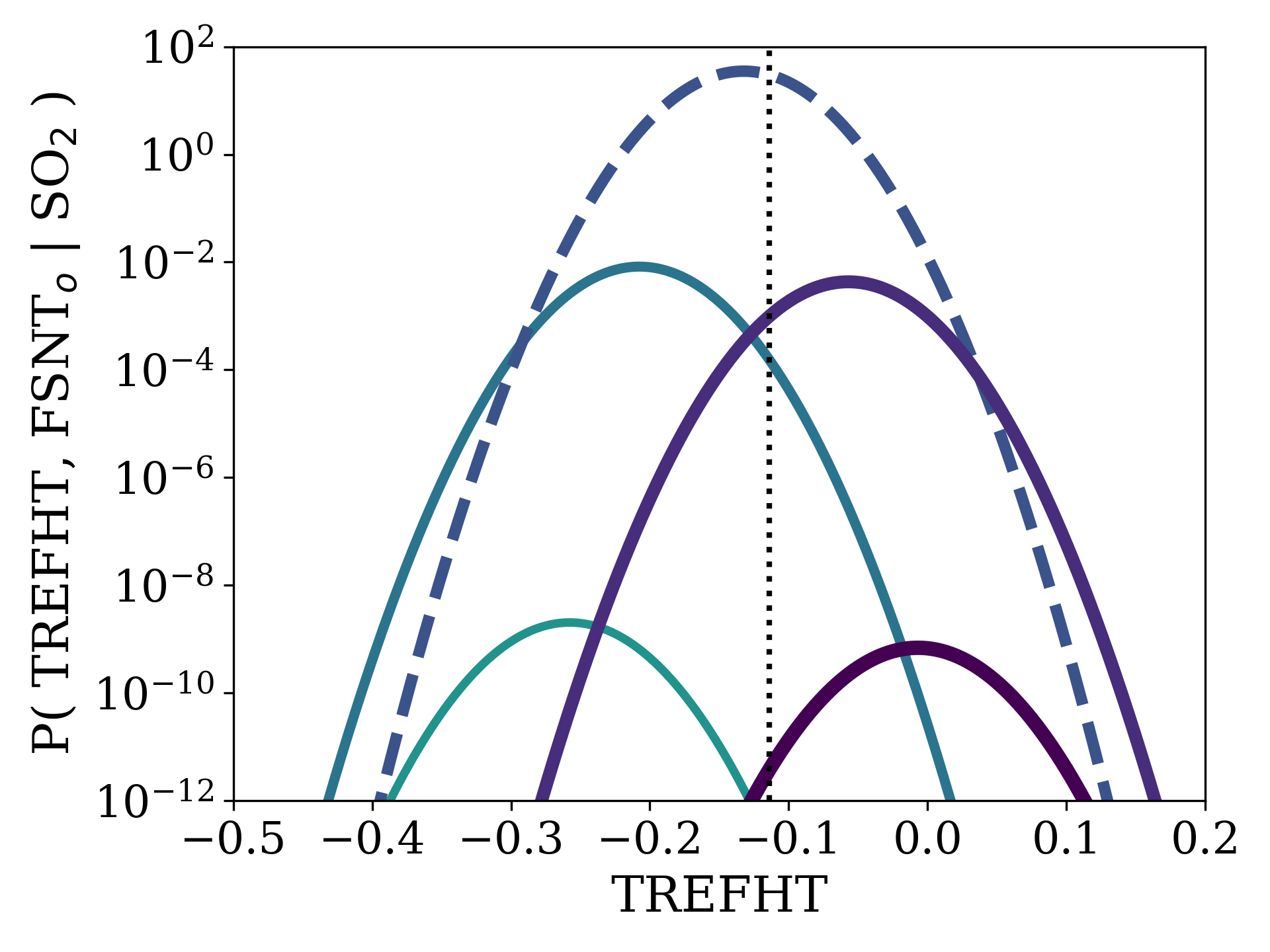}
    \end{minipage}
    \caption{\label{fig:like-dists} Global three year averaged likelihood probability density functions for the single-step TREFHT (top right), intermediate FSNT response (bottom left), the TREFHT response conditioned on the FSNT pseudo-observation (bottom center), and the joint multi-step TREFHT likelihood (bottom right) for all forcing levels. The vertical dotted line indicates the pseudo-observation value of the associated variable.}
\end{figure}

It is useful to emphasize the effect that multi-step conditioning has on the likelihood functions arising from the above regression models, as described in Section~\ref{subsec:likelihood}. Several component likelihood probability density functions are plotted in Figure~\ref{fig:like-dists}, and the relevant pseudo-observation values are marked with a vertical dashed line. The top right plot displays the single-step pathway likelihood functions, and it is immediately clear that the likelihood of the observed value is extremely similar under the 10 Tg and 7 Tg distributions. This hints that these eruption magnitudes may be very difficult to distinguish solely from reference height temperature data. On the other hand, the bottom left plot shows the likelihood distributions for FSNT, for which the likelihood at the observed value is overwhelmingly greater under the 10 Tg eruption than for any other eruption magnitude. Using this intermediate step pseudo-observation to condition the multi-step TREFHT likelihood distribution has a drastic effect, as seen in the bottom center plot where the conditioned TREFHT distributions exhibit far greater separability than in the unconditioned (top right) distributions. The final multi-step joint likelihoods are shown in the bottom right plot, where the 10 Tg likelihood is now orders of magnitude greater than that under any other eruption magnitude at the observed TREFHT value. Inclusion of intermediate variables thus greatly improves attribution strength by increasing the contrast between the different likelihood curves, and hence the ability to distinguish different forcing scenarios. 

Finally, we assess the ability of the proposed framework to achieve successful attribution, and whether incorporating additional information via a conditional pathway improves attribution. Following the procedure outlined in Section~\ref{subsec:ratios}, we consider various null hypotheses from the set $\forceSet_0 = \{0, 1, 3, 5, 7, 13, 15\}$ Tg, and compare against the alternative hypothesis forcing level $\forceIdx_1 = 10$ Tg. For simplicity, we compare individual elements of $\forceSet_0$ with $\forceIdx$ separately, resulting in a ``simple'' (non-composite) likelihood ratio test; extension to composite likelihoods (testing all elements of $\forceSet_0$ simultaneously) is straightforward but adds minimal value to our analysis. Using one million Monte Carlo samples for each forcing level $\forceVar_0$, with likelihood functions $\likelihood_{\dataSet}(\cdot)$ as defined by the linear regressions computed previously, we compute approximations of the likelihood ratio distributions $\testDist_{\forceVar_0,\forceVar_1}$. We emphasize that these Monte Carlo samples are drawn from the likelihood (regression) model developed above and can be computed quite cheaply; we do not require millions of expensive climate model simulations. Using the pseudo-observational data as computed by Eq.~\ref{eq:obs-mean} to compute the test statistic $\testStat_{\forceVar_0,\forceVar_1} (\obsSet)$, the resulting $p$-values are calculated according to Eq.~\ref{eq:lr-pval}. These values are reported for each spatial region and temporal period in Table~\ref{tab:pValue}. Recall that these $p$-values represent the probability of observing a test statistic greater than $\testStat_{\forceVar_0,\forceVar_1} (\obsSet)$ under the null hypothesis. When this value is small, it indicates that the observations are more consistent with the alternative hypothesis ($F=\forceVar_1$) than with the null hypothesis ($F=\forceVar_0$). Intuitively, we expect to obtain smaller $p$-values (higher confidence attribution) when the contrast between the null and alternative hypotheses is large (e.g., 0 vs 10 Tg) and we expect larger $p$-values (lower confidence attribution) when the contrast between the null and alternative hypotheses is smaller (e.g., 7 Tg vs 10 Tg). 

\begin{landscape}

\begin{table}
    \begin{minipage}{0.89\linewidth}
        \caption{Likelihood ratio test $p$-values as defined by Eq.~\ref{eq:lr-pval}, approximating the probability of measuring the observed test statistic under various null hypothesis forcing magnitude assumptions, with a 10 Tg alternative hypothesis.}
        \label{tab:pValue}
        \begin{tabular}{lllrrrrrrr}
            \toprule
            & Region & Time & 0 Tg & 1 Tg & 3 Tg & 5 Tg & 7 Tg & 13 Tg & 15 Tg \\
            \midrule
    
             &        & 3 years  & \tcb 1.06e-3 & \tcb 3.24e-3 & \tcc 2.21e-2 & \tcd 9.69e-2 & \phlt \tce 2.78e-1 & \phlt \tcd 6.09e-2 & \tcc 1.20e-2 \\
             & Global & 1992 JJA & \tca 2.83e-4 & \tcb 1.08e-3 & \tcb 9.45e-3 & \tcd 5.31e-2 & \tce 1.88e-1 & \tcd 9.72e-2 & \tcc 2.12e-2 \\
             &        & 1992 JFM & \tcd 8.77e-2 & \tce 1.16e-1 & \tce 1.89e-1 & \tce 2.85e-1 & \tce 4.01e-1 & \tce 2.44e-1 & \tce 1.58e-1 \\
    
            \cmidrule{2-10}
    
            &           & 3 years  & \tca 2.12e-4 & \tca 8.50e-4 & \tcb 8.08e-3 & \tcc 4.70e-2 & \tce 1.73e-1 & \tce 1.05e-1 & \tcc 2.35e-2 \\
            Single & NH & 1992 JJA & \tca 8.60e-5 & \tca 3.27e-4 & \tcb 3.87e-3 & \tcc 2.71e-2 & \tce 1.17e-1 & \tce 1.57e-1 & \tcc 4.08e-2 \\
            &           & 1992 JFM & \tce 1.17e-1 & \tce 1.40e-1 & \tce 1.95e-1 & \tce 2.61e-1 & \tce 3.36e-1 & \tce 4.08e-1 & \tce 3.25e-1 \\
    
            \cmidrule{2-10}
    
            &    & 3 years  & \tcd 5.63e-2 & \tcd 7.57e-2 & \tce 1.30e-1 & \tce 2.05e-1 & \tce 3.02e-1 & \tce 3.46e-1 & \tce 2.41e-1 \\
            & NA & 1992 JJA & \tcb 2.03e-3 & \tcb 4.21e-3 & \tcc 1.55e-2 & \tcc 4.61e-2 & \tce 1.16e-1 & \tce 4.04e-1 & \tce 2.35e-1 \\
            &    & 1992 JFM & \tce 4.45e-1 & \tce 4.50e-1 & \tce 4.61e-1 & \tce 4.71e-1 & \tce 4.81e-1 & \tce 4.85e-1 & \tce 4.74e-1 \\
    
            \addlinespace[1em]
            \midrule
            \addlinespace[1em]
    
            &        & 3 years  & \tca $<$ 1.00e-6 & \tca $<$ 1.00e-6 & \tca $<$ 1.00e-6 & \tca $<$ 1.00e-6 & \tca 1.40e-5 & \tca 1.00e-6 & \tca $<$ 1.00e-6 \\
            & Global & 1992 JJA & \tca $<$ 1.00e-6 & \tca $<$ 1.00e-6 & \tca 1.00e-6 & \tca 1.13e-4 & \tcc 1.20e-2 & \tcc 1.51e-2 & \tca 1.66e-4 \\
            &        & 1992 JFM & \tca $<$ 1.00e-6 & \tca 1.00e-6 & \tca 7.70e-5 & \tcb 4.73e-3 & \tcd 8.34e-2 & \tcc 1.14e-2 & \tca 1.64e-4 \\
    
            \cmidrule{2-10}
    
            &          & 3 years  & \tca $<$ 1.00e-6 & \tca $<$ 1.00e-6 & \tca $<$ 1.00e-6 & \tca $<$ 1.00e-6 & \tca 2.60e-5 & \tca 4.32e-4 & \tca $<$ 1.00e-6 \\
            Multi & NH & 1992 JJA & \tca $<$ 1.00e-6 & \tca $<$ 1.00e-6 & \tca $<$ 1.00e-6 & \tca 1.20e-4 & \tcb 9.20e-3 & \tcd 5.40e-2 & \tcb 2.04e-3 \\
            &          & 1992 JFM & \tca $<$ 1.00e-6 & \tca $<$ 1.00e-6 & \tca $<$ 1.00e-6 & \tca 8.30e-5 & \tcb 8.85e-3 & \tcd 6.25e-2 & \tcb 2.20e-3 \\
    
            \cmidrule{2-10}
    
            &    & 3 years  & \tca $<$ 1.00e-6 & \tca $<$ 1.00e-6 & \tca 2.00e-6 & \tca 3.94e-4 & \tcc 1.40e-2 & \tce 1.23e-1 & \tcc 1.28e-2 \\
            & NA & 1992 JJA & \tca $<$ 1.00e-6 & \tca $<$ 1.00e-6 & \tca 3.70e-5 & \tcb 6.87e-3 & \tcc 2.97e-2 & \tce 2.74e-1 & \tce 1.49e-1 \\
            &    & 1992 JFM & \tca 4.76e-4 & \tcb 1.31e-3 & \tcb 9.07e-3 & \tcc 4.23e-2 & \tce 1.37e-1 & \tce 2.06e-1 & \tcd 7.28e-2 \\
            
            \bottomrule
        \end{tabular}
    \end{minipage}
    \hspace{1em}
    \begin{minipage}{0.075\linewidth}
        \vspace{7.5em}
        \includegraphics[width=0.99\linewidth,trim={1em 0em 14em 0em},clip]{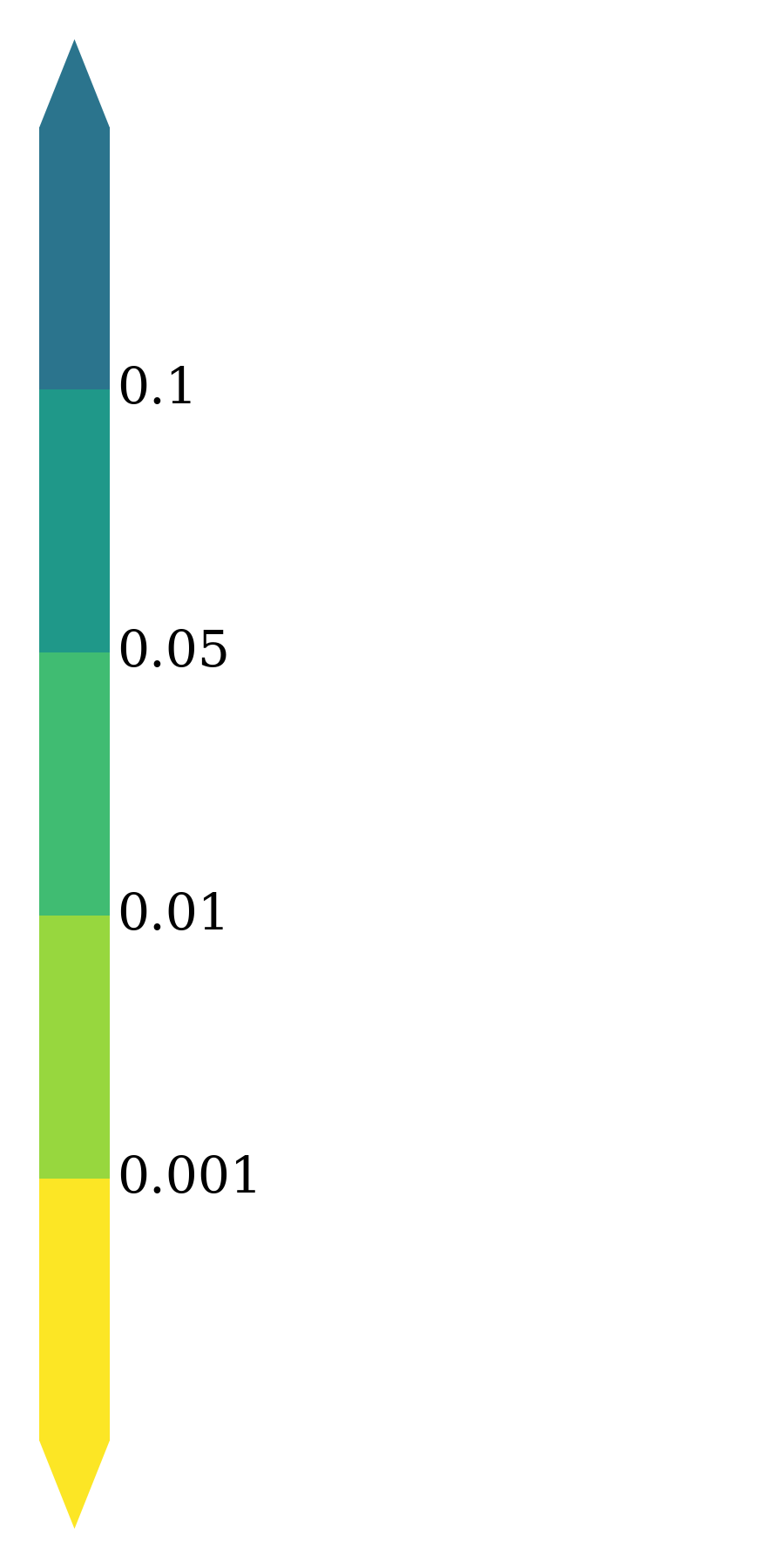}
    \end{minipage}
\end{table}

\end{landscape}

For the single-step analysis (upper half of Table~\ref{tab:pValue}), the $p$-values for the null hypothesis 0 Tg and 1 Tg forcings fall below 0.05 for the three year windows, but are slightly larger ($< 0.1$) for the shorter time frames. Comparing this to the corresponding multi-step analyses (lower half), we observe much smaller $p$-values for all analyses, indicating that the multi-step analysis provides much stronger evidence against a 0 Tg or 1 Tg eruption. In particular, we see that we can reject the null of a 0 Tg eruption at 99.9\% confidence for all analyses, even NA 1992 JFM, using the multi-step approach while the single-step approach is only able to reject the 0 Tg null at 91\% confidence for the global 1992 JFM analysis. 

As we consider larger eruptions, the contrast between hypotheses, e.g., a 7 Tg null hypothesis eruption and a 10 Tg alternative hypothesis eruption, is less clear, making it harder to distinguish scenarios and weakening attribution statements. The $p$-values for the single-step analysis increase materially and it becomes impossible to reject the null at even moderate confidence levels: for example, taking $\forceVar_0 = 7$ Tg for a global analysis over a three year window, we obtain a $p$-value of 0.278 and cannot even reject the null of a 7 Tg eruption at a 75\% confidence level. By contrast, $p$-values from the multi-step approach remain small ($p < 5\times 10^{-4}$) for the global three year analysis, enabling us to continue making strong attribution statements using the multi-step analysis that are not possible with single-step analysis. 

Comparing the two sections of Table \ref{tab:pValue}, it is clear that the conditional pathways-based attribution (lower half) provides systematic improvements over single-step attribution (upper half), as indicated by the higher prevalence of lighter colors. These improvements are perhaps most useful as we consider null hypothesis that are less easily distinguished ($\forceVar_0 = 7, \ 13$) or as we consider smaller spatial or temporal windows. The multi-step approach is not a panacea, however, and confident attribution for certain analyses, particularly NA with $\forceVar_0 = 7$ or 13, remains somewhat difficult given the high levels of climate variability and the low signal strength. Even in these challenging regimes, however, the multi-step approach provides meaningful improvements, allowing, e.g., 90\% confidence ($p < 0.1$) attribution of the 1992 JFM impact globally and in the NH against all alternatives.

%% file: discussion.tex
The results presented above demonstrate that the proposed conditional pathways-based attribution approach is able to distinguish the source magnitude giving rise to the chain of responses in the climate system, even on short-times scales and in confined regions. This discriminatory power is facilitated by evaluating the likelihood of the reference height temperature response conditioned on the forcing magnitude \textit{and} intermediary pathway variables. This conditional pathways-based attribution has wide applicability in the climate field as a method to determine the magnitude of forcing, attribute low signal-to-noise downstream impacts, and quantitatively evaluate mediating mechanisms of a downstream impact.

Discriminating between source forcings is quite important when the forcing magnitudes are uncertain. This is the case, for instance, for the 1815 Tambora eruption in which the $\pm2\sigma$ spread of uncertainty in the \SOtwo\ eruption mass was reported by~\citet{tooheyPersonalCorrespondence} to be approximately 8.98 Tg for a 28.08 Tg eruption~\citep{Zanchettin2019}. In this case, \citet{Zanchettin2019} simulated Tambora's low and high ($\pm2\sigma$) estimates to study the relative importance of forcing magnitude versus initial conditions on the surface temperature response. They showed strong distinguishability between volcanic forcings of different levels and internal variability for summertime global, NH, and NA surface temperature responses. However, wintertime temperature responses in the NH, and particularly NA, exhibited significant overlap between forcings and could be mistaken for deviations possible from internal variability alone. Overall, these results point to the dominant role of internal variability in the downstream temperature impact from a volcanic eruption. Given the size of Tambora, one might assume that the significant signal-to-noise ratio would overcome internal variability to exhibit clear temperature responses. However, over short periods and confined regions, the internal variability at mid and high latitudes was still dominant.

\begin{figure}
    \centering
    \begin{minipage}{0.49\linewidth}
        \includegraphics[width=0.99\linewidth]{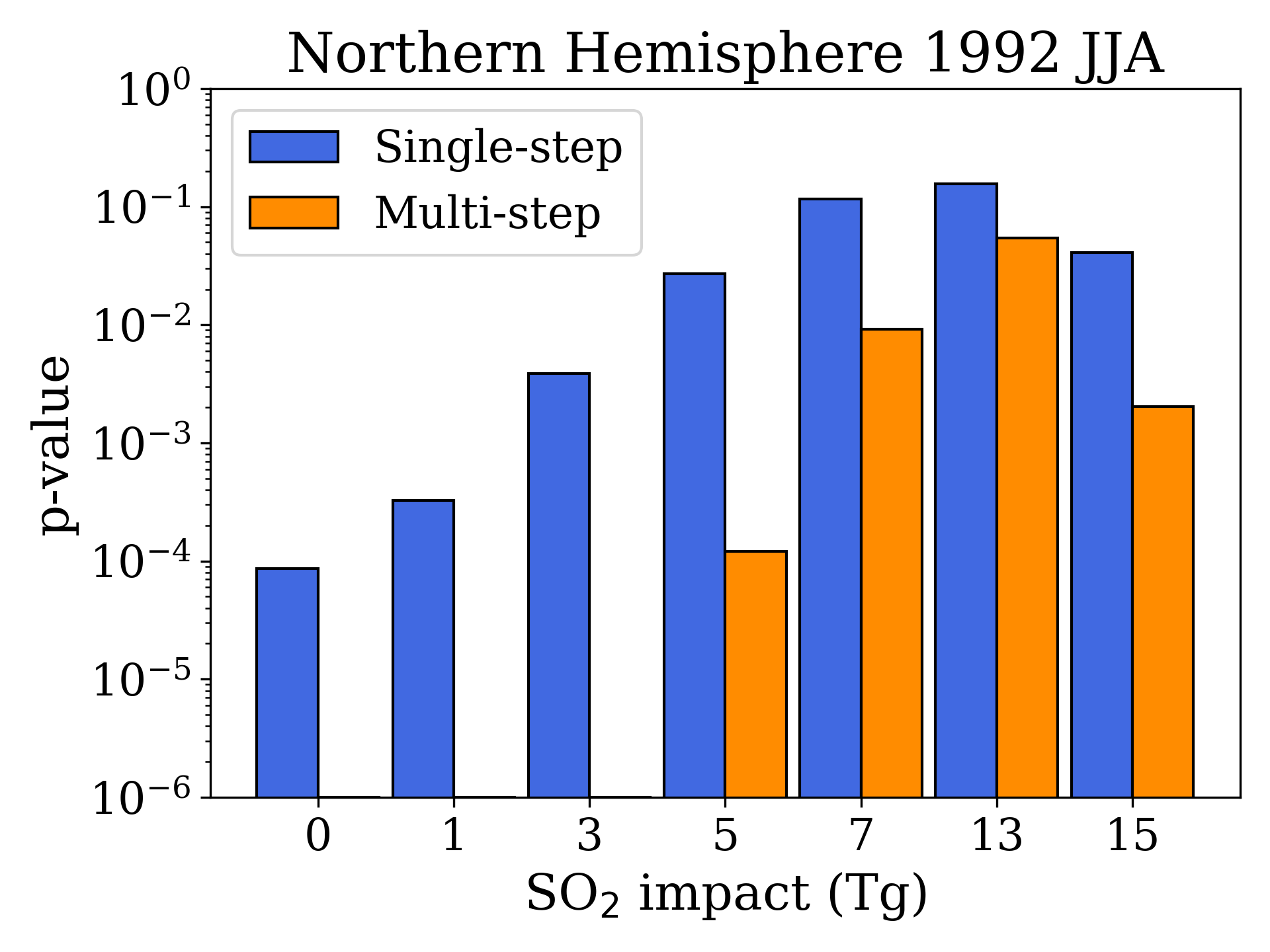}
    \end{minipage}
    \begin{minipage}{0.49\linewidth}
        \includegraphics[width=0.99\linewidth]{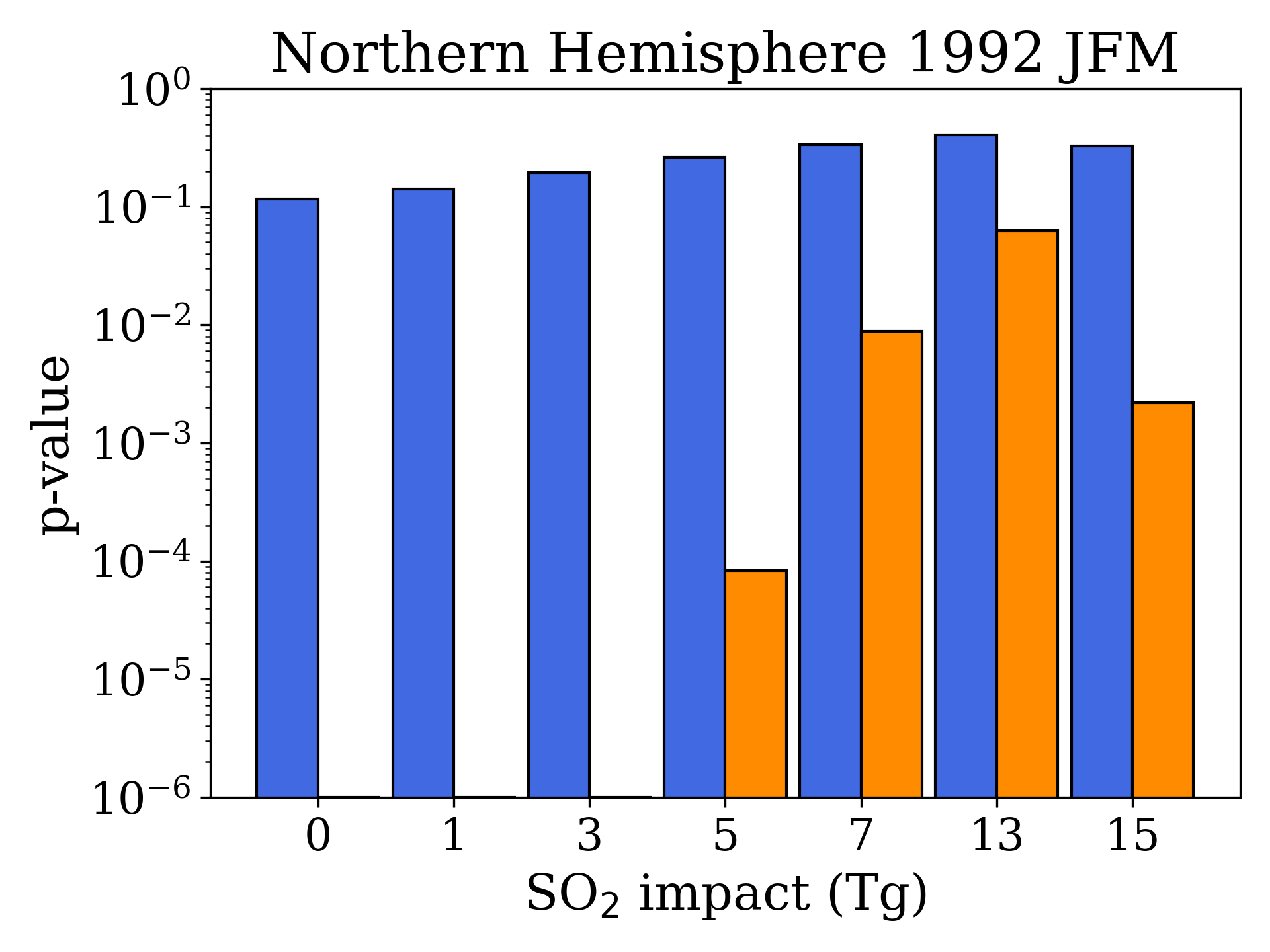}
    \end{minipage}

    \begin{minipage}{0.49\linewidth}
        \includegraphics[width=0.99\linewidth]{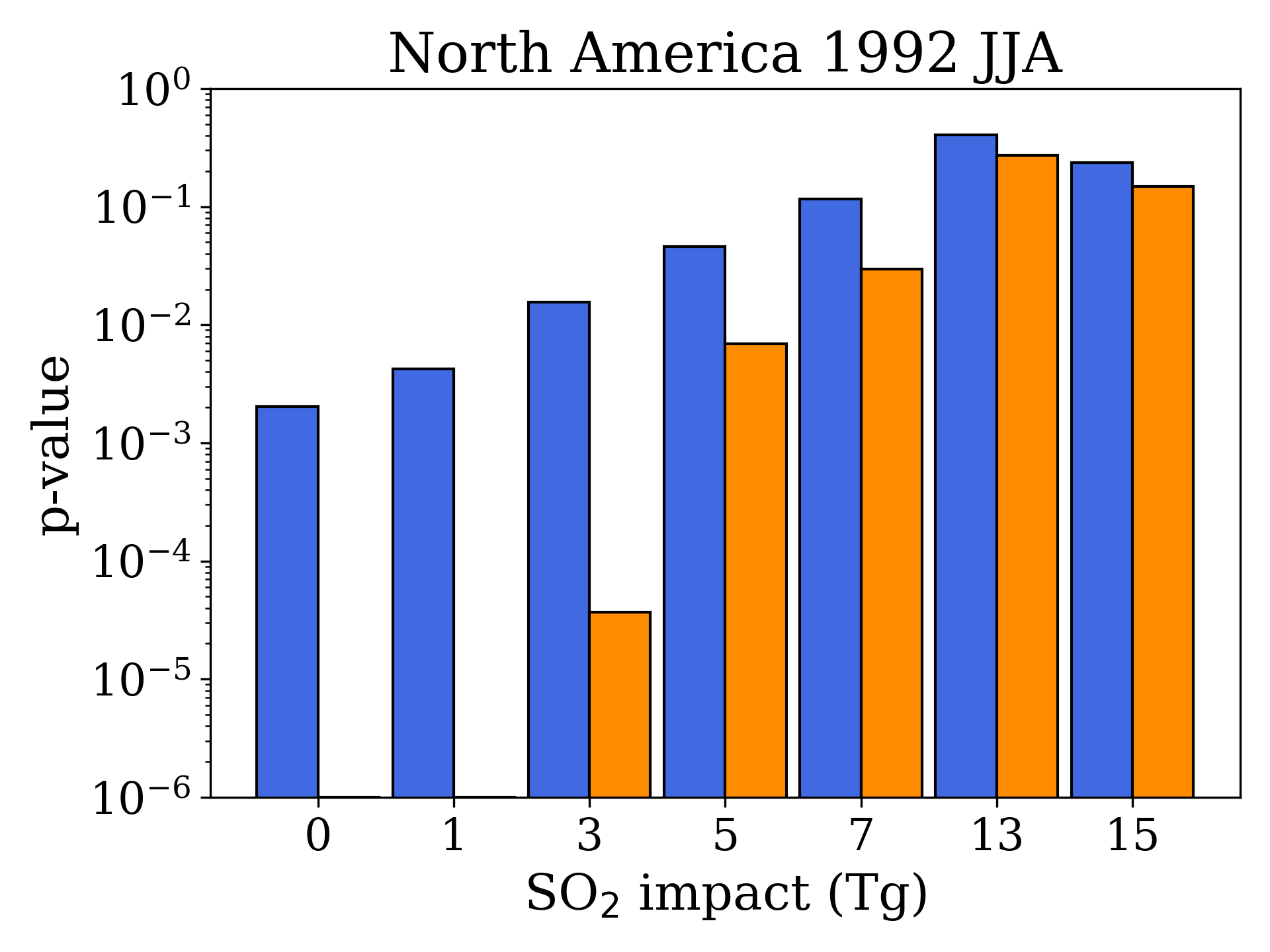}
    \end{minipage}
    \begin{minipage}{0.49\linewidth}
        \includegraphics[width=0.99\linewidth]{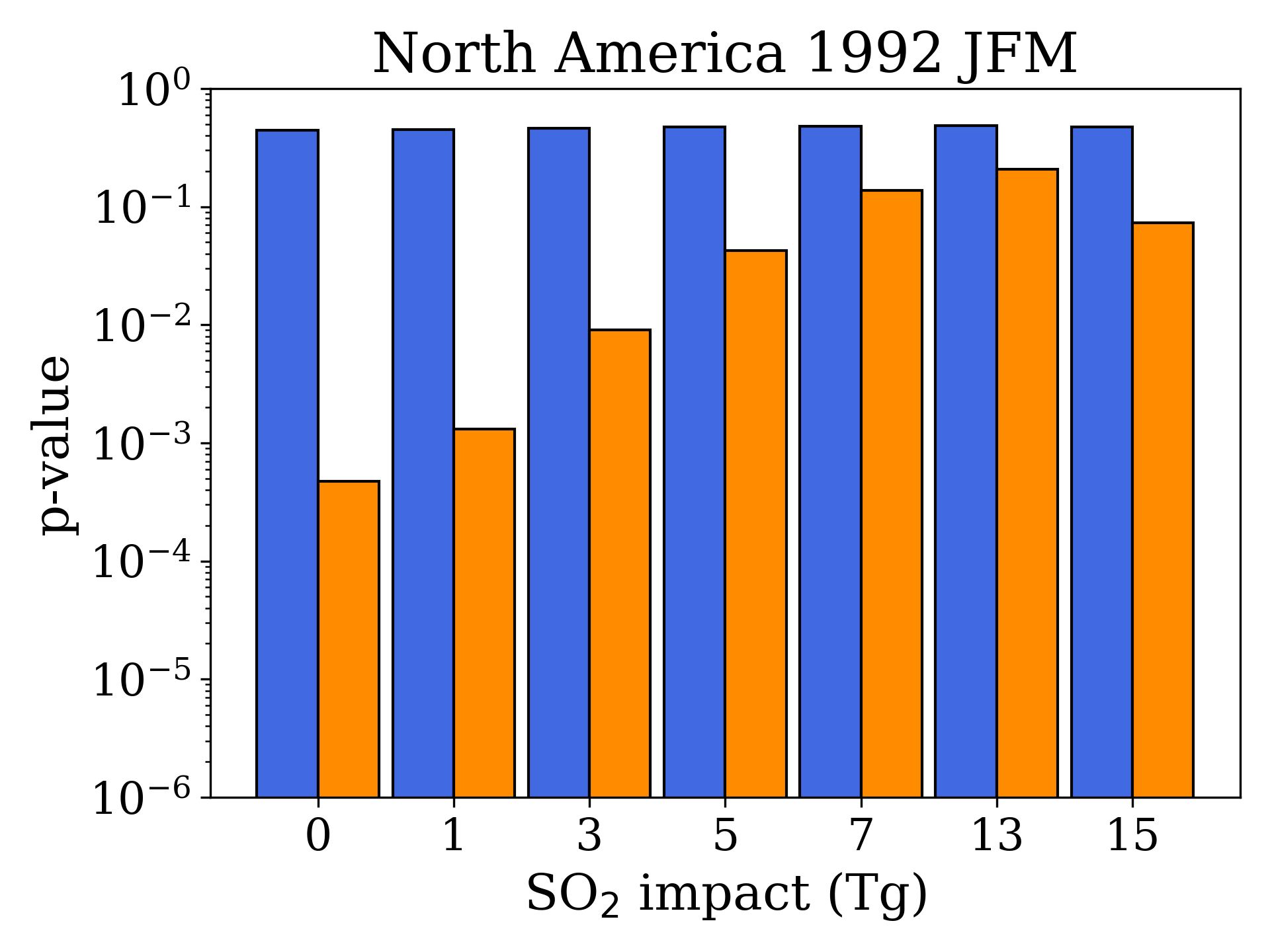}
    \end{minipage}

    \caption{Likelihood ratio test $p$-values for varied spatial regions (Northern Hemisphere and North America) and temporal periods (1992 JJA and JFM), comparing posited 10 Tg eruption against all other simulated eruption magnitudes.}
    \label{fig:pval-nh-na}
\end{figure}

In Section~\ref{subsec:multistep-results} we employed the same definition of NH and NA as~\citet{Zanchettin2019}. However, we considered the much smaller Mt.\ Pinatubo eruption with a more limited range of internal variability than ~\citet{Zanchettin2019}, as we initialized the simulations with ENSO and the QBO in historically accurate states. Figure~\ref{fig:pval-nh-na} contrasts the $p$-values from Table~\ref{tab:pValue} between the single (blue) and multi-step (orange) approaches for NH and NA in the 1992 summer (JJA) and winter (JFM) timeframes. The blue bars are mainly flat in the winter indicating a high $\linvarest_{\varIdx}$ that is confirmed in Appendix~\ref{app:variances}. This flat behavior implies a lack of interpretability from the single-step attribution. Hence, for a Mt. Pinatubo sized eruption employing single-step attribution, we can confirm~\citet{Zanchettin2019}'s distinguishability in the NH and NA summer from the 0 Tg eruption (at 99.9\%  and 99.0\% confidence levels respectively). Further, limited to a single-step, one cannot decipher a 10$\pm$5 Tg eruption at a 95\% confidence level in the NH with the magnitude range expanding for NA.

However, using the conditional pathways-based approach, with the incorporation of radiative flux, a peaked behavior is revealed in Figure~\ref{fig:pval-nh-na}. There is now a clear distinction (at a 99.9\% confidence level) from the 0 Tg eruption in both the summertime and wintertime in the NH and NA even for an eruption that is approximately $\sim\frac{1}{3}$ the size of the Mt.\ Tambora eruption. Furthermore, the conditional pathways-based attribution is able to distinguish summertime forcing within $\pm$3 Tg in the NH at a 95\% confidence level, instead of $\pm$5 Tg in the single step, and within $\pm$5 Tg in NA at a confidence level lower than 90\%. Since variability does not scale with eruption magnitude, being able to achieve distinguishability between forcings that are $\sim\frac{1}{3}$ the magnitude ($\pm$3 Tg) of those used by~\citet{Zanchettin2019} ($\sim\pm$9 Tg) is significant. Although wintertime confidence levels are available, the quality of the regression cautions against using them directly. However, the summit in $p$-values within $\pm$5 Tg in the NH and NA in Figure \ref{fig:pval-nh-na} offers limited insights into the range of distinguishability even if with lower assurances because of the quality of the regressions. 
The ability to distinguish from the unforced scenario in the summer and winter as well as quantitatively (qualitatively) tighten the range of distinguishability in the summer (winter) highlights the power of the conditional pathways-based attribution.

Geoengineering through stratospheric aerosol injection (SAI) highlights the need for attribution techniques exhibiting confidence for low magnitude forcings producing short-duration and regional responses. SAI is currently being studied as a method to limit global~\citep{Richter2022, Tilmes2018} and regional~\citep{Lee2021, Duffey2023, Wheeler2025} temperature rise. However, projected injections for global influence do not reach magnitudes equivalent to Mt.\ Pinatubo until 2067 in the community dataset ARISE-SAI~\citep{Richter2022}. As shown in Figure \ref{fig:betas-two} and Table~\ref{tab:pValue}, it is extremely difficult to attribute the change in global reference height temperature alone to a 10 Tg eruption. As such, it would be extremely difficult for the global community to use standard techniques to quantitatively assess the effectiveness of interventions with low injection magnitudes on global temperature in the face of internal variability. As highlighted by~\citet{Keys2022}, this opens the door to perceived failures of SAI and could result in scientifically misinformed decisions. New attribution formalisms, like that presented here, have the potential to incorporate mechanistic knowledge as additional conditional information to increase the confidence that initial SAI injections are, or are not, influencing the surface temperature as desired.

Finally, it could be important to determine what constitutes the pathway, i.e. which mediating mechanisms are part of a response. For instance, this conditional pathway-based attribution method could also be applied to the Australian wildfire response pathway causing the ``triple-dip'' La Ni\~na proposed by~\citet{Fasullo2023}. The expertise employed in their inference across variables, space, and time could potentially be represented in a regression model like that shown in Figure~\ref{fig:regressions-glob}. This would require simulations of scaled bushfire magnitude and the selection of appropriate features for the regression model. This would specifically query each step for its correlation with the bushfire forcing. Hence, not only would this framework contribute to testing the validity of the hypothesized pathway, but it could also directly demonstrate attribution with a similar hypothesis testing through likelihood ratios. The strength of the attribution could in turn rebuff or grant alternative pathways capable of explaining the persistent equatorial Pacific cooling.

%% file: conclusions.tex
In this paper we have formulated a novel approach to climate impact attribution which leverages strong relationships along a conditionally-dependent pathway of climate variables linking downstream effects to their source forcing. A rigorous hypothesis consistency framework built from likelihood ratio testing allows for detailed attribution of such effects to specific forcing levels, supplying a new tool for analysts to better understand the sensitivity of the climate to continuously varying forcings. The approach is demonstrated for the short-term, point-source forcing produced by the 1991 eruption of Mt.\ Pinatubo, and compared favorably against a traditional fingerprinting detection and attribution approach which is ill-equipped to analyze such forcings. The use of intermediate climate variables such as the net shortwave radiation flux greatly improves attribution power and demonstrates the benefit of expert understanding of the climate system.

It remains to be seen whether this methodology can succeed for more complex pathways like those presented in Section \ref{sec:discussion}. Linking climate forcings to human-relevant impacts (such as agricultural productivity) will necessitate more complex pathways with potentially disjoint steps. Strong linear relationships in average impact space may not exist for effects far downstream of the source forcing, and higher variance will decrease the certainty of attribution. Additionally, the method may require modification to extend to more sustained forcings such as geoengineering or climate tipping points. Scalar features and linear models may not be the most applicable, requiring careful consideration. Ultimately, this framework opens the door to a host of interesting analyses and equips climate scientists with a new rigorous, probabilistic framework for tackling attribution for a multitude of modern climate problems.

%% file: acknowledgements.tex
We thank the members of the CLDERA LDRD team for helpful discussions. In particular, we thank Thomas Ehrmann and Benjamin Wagman for assistance in characterizing the temperature response and deep expertise in climate systems, Meredith Brown and Justin Li for initial work on fingerprinting and variability, Joseph Hart, Mamikon Gulian, Indu Manickam, J. Jake Nichol, Irina Tezaur, Kara Peterson, and Lyndsay Shand for helpful discussions.

This research used resources of the National Energy Research Scientific Computing Center (NERSC), a Department of Energy Office of Science User Facility using NERSC award BER-ERCAP0026535.

%% file: declarations.tex
\bmhead{Funding}
Sandia National Laboratories is a multimission laboratory managed and operated by National Technology \& Engineering Solutions of Sandia, LLC, a wholly owned subsidiary of Honeywell International Inc., for the U.S. Department of Energy's National Nuclear Security Administration under contract DE-NA0003525. This work was funded through Sandia's Laboratory Directed Research and Development program. The views expressed in the article do not necessarily represent the views of the U.S. Department of Energy or the United States Government.

\bmhead{Competing Interests}
This article has been authored by employees of National Technology \& Engineering Solutions of Sandia, LLC under contract DE-NA0003525 with the U.S. Department of Energy (DOE). The employees own all right, title and interest in and to the article and is solely responsible for its contents. The United States Government retains and the publisher, by accepting the article for publication, acknowledges that the United States Government retains a non-exclusive, paid-up, irrevocable, world-wide license to publish or reproduce the published form of this article or allow others to do so, for United States Government purposes. The DOE will provide public access to these results of federally sponsored research in accordance with the DOE Public Access Plan.

\bmhead{Author Contributions}
Christopher R. Wentland processed simulation data, wrote calculation and visualization scripts, and assisted in manuscript drafting and editing. Michael Weylandt developed the conditional attribution approach including the use of likelihood ratios, and assisted in manuscript drafting and editing. Laura P. Swiler outlined the fingerprinting methods, provided context for related attribution work, and assisted in manuscript drafting and editing. Diana Bull conceived the research, interpreted results, and drafted and edited the manuscript. 

\bmhead{Data and Code Availability}
The global average impact time series data described in Section~\ref{subsec:simulations}, along with scripts for generating the figures presented in Section~\ref{sec:results}, can be found at \url{https://github.com/sandialabs/conditional-multistep-attribution}.

%% file: appendices.tex
\section{Normalized linear regression model variances}\label{app:variances}

Table~\ref{tab:vars} displays residual variance measurements $\linvarest_{\varIdx}$ for the linear forcing response models reported in Section~\ref{subsec:multistep-results}, for which the data has been normalized to the range [-1, 1] in order to compare the relative amounts of internal variability present in each step of the single- and multi-step surface cooling pathways. These normalized quantities are ultimately not used in computing likelihood distributions in Section~\ref{subsec:multistep-results}, and are solely for discussion purposes.

\begin{table}
    \caption{\label{tab:vars}Linear regression $\widehat{\sigma}^2_{\varIdx}$ values for single-step, intermediary, and multi-step regressions, as computed from data normalized to the range [-1, 1] for each spatio-temporal region.}
    \begin{tabular}{llrrr}
     \toprule
     Region & Time & \SOtwo $\rightarrow$ TREFHT & \SOtwo $\rightarrow$ FSNT & \SOtwo, FSNT $\rightarrow$ TREFHT \\
     \midrule
             & 3 years  & 0.0501 & 0.0067 & 0.0282 \\
    Global   & 1992 JJA & 0.0521 & 0.0238 & 0.0500 \\
             & 1992 JFM & 0.1000 & 0.0284 & 0.0879 \\
    \midrule
             & 3 years  & 0.0533 & 0.0088 & 0.0384 \\
    N. Hem.  & 1992 JJA & 0.0484 & 0.0281 & 0.0451 \\
             & 1992 JFM & 0.1540 & 0.0229 & 0.1360 \\
    \midrule
             & 3 years  & 0.0907 & 0.0294 & 0.0706 \\
    N. Amer. & 1992 JJA & 0.0772 & 0.0683 & 0.0370 \\
             & 1992 JFM & 0.1750 & 0.0653 & 0.1640 \\
    \bottomrule
    \end{tabular}
\end{table}

%% file: main.bbl
\begin{thebibliography}{82}
\providecommand{\natexlab}[1]{#1}
\providecommand{\url}[1]{{#1}}
\providecommand{\urlprefix}{URL }
\providecommand{\doi}[1]{\url{https://doi.org/#1}}
\providecommand{\eprint}[2][]{\url{#2}}
 \bibcommenthead

\bibitem[{Allen and Stott(2003)}]{Allen2003}
Allen MR, Stott PA (2003) {Estimating signal amplitudes in optimal
  fingerprinting, Part I: Theory}. Climate Dynamics 21(5):477--491.
  \doi{10.1007/s00382-003-0313-9}

\bibitem[{Allen and Tett(1999)}]{Allen1999}
Allen MR, Tett SFB (1999) Checking for model consistency in optimal
  fingerprinting. Climate Dynamics 15(6):419--434. \doi{10.1007/s003820050291}

\bibitem[{Berliner et~al(2000)Berliner, Levine, and Shea}]{Berliner2000}
Berliner LM, Levine RA, Shea DJ (2000) {Bayesian climate change assessment}.
  Journal of Climate 13(21):3805--3820.
  \doi{10.1175/1520-0442(2000)013<3805:BCCA>2.0.CO;2}

\bibitem[{Bindoff et~al(2013)Bindoff, Stott, AchutaRao, Allen, Gillett,
  Gutzler, Hansingo, Hegerl, Hu, Jain et~al}]{Bindoff2013}
Bindoff NL, Stott PA, AchutaRao KM, et~al (2013) Detection and attribution of
  climate change: from global to regional. Climate change 2013: the physical
  science basis Contribution of Working Group I to the Fifth Assessment Report
  of the Intergovernmental Panel on Climate Change

\bibitem[{Brown et~al(2024)Brown, Wagman, Bull, Peterson, Hillman, Liu, Ke, and
  Lin}]{hbrown2024}
Brown HY, Wagman B, Bull D, et~al (2024) Validating a microphysical prognostic
  stratospheric aerosol implementation in {E3SMv2} using the {M}ount {P}inatubo
  eruption. EGUsphere 2024:1--46. \doi{10.5194/egusphere-2023-3041}

\bibitem[{Cattiaux et~al(2010)Cattiaux, Vautard, Cassou, Yiou, Masson-Delmotte,
  and Codron}]{Cattiaux2010}
Cattiaux J, Vautard R, Cassou C, et~al (2010) Winter 2010 in {E}urope: A cold
  extreme in a warming climate. Geophysical Research Letters 37(20).
  \doi{https://doi.org/10.1029/2010GL044613}

\bibitem[{Chiang et~al(2021)Chiang, Greve, Mazdiyasni, Wada, and
  AghaKouchak}]{Chiang2021}
Chiang F, Greve P, Mazdiyasni O, et~al (2021) A multivariate conditional
  probability ratio framework for the detection and attribution of compound
  climate extremes. Geophysical Research Letters 48(15):e2021GL094361.
  \doi{https://doi.org/10.1029/2021GL094361}

\bibitem[{Church et~al(2005)Church, White, and Arblaster}]{church2005}
Church JA, White NJ, Arblaster JM (2005) Significant decadal-scale impact of
  volcanic eruptions on sea level and ocean heat content. Nature
  438(7064):74--77. \doi{10.1038/nature04237}

\bibitem[{Dogar et~al(2024)Dogar, Fujiwara, Zhao, Ohba, and Kosaka}]{dogar2024}
Dogar MM, Fujiwara M, Zhao M, et~al (2024) {ENSO} and {NAO} linkage to strong
  volcanism and associated post-volcanic high-latitude winter warming.
  Geophysical Research Letters 51(1):e2023GL106114.
  \doi{https://doi.org/10.1029/2023GL106114}

\bibitem[{Duffey et~al(2023)Duffey, Irvine, Tsamados, and Stroeve}]{Duffey2023}
Duffey A, Irvine P, Tsamados M, et~al (2023) Solar geoengineering in the polar
  regions: A review. Earth's Future 11(6):e2023EF003679.
  \doi{https://doi.org/10.1029/2023EF003679}

\bibitem[{Ehrmann et~al(2024)Ehrmann, Wagman, Bull, Hillman, Hollowed, Brown,
  Peterson, Swiler, Watkins, and Hart}]{ehrmann2024}
Ehrmann T, Wagman B, Bull D, et~al (2024) Identifying northern hemisphere
  stratospheric and surface temperature responses to the {M}t. {P}inatubo
  eruption within {E3SMv2-SPA}. Tech. rep., Sandia National Laboratories

\bibitem[{Ehrmann et~al(Submitted July 2025)Ehrmann, Wagman, Bull, Hillman,
  Hollowed, Brown, Peterson, Swiler, Watkins, and Hart}]{ehrmann2025}
Ehrmann T, Wagman B, Bull D, et~al (Submitted July 2025) Identifying the
  northern hemisphere winter warming response to the {M}t. {P}inatubo eruption
  through limited variability ensembles. Atmospheric Chemistry and Physics

\bibitem[{Eyring et~al(2021{\natexlab{a}})Eyring, Gillett, Achutarao,
  Barimalala, Barreiro~Parrillo, Bellouin, Cassou, Durack, Kosaka, McGregor
  et~al}]{IPCCar6}
Eyring V, Gillett N, Achutarao K, et~al (2021{\natexlab{a}}) {Human Influence
  on the Climate System: Contribution of Working Group I to the Sixth
  Assessment Report of the Intergovernmental Panel on Climate Change}. Tech.
  rep., IPCC Sixth Assessment Report

\bibitem[{Eyring et~al(2021{\natexlab{b}})Eyring, Gillett, Achutarao,
  Barimalala, Barreiro~Parrillo, Bellouin, Cassou, Durack, Kosaka, McGregor
  et~al}]{IPCCar6ch10}
Eyring V, Gillett N, Achutarao K, et~al (2021{\natexlab{b}}) {Human Influence
  on the Climate System: Contribution of Working Group I to the Sixth
  Assessment Report of the Intergovernmental Panel on Climate Change}. Tech.
  rep., IPCC Sixth Assessment Report

\bibitem[{Eyring et~al(2021{\natexlab{c}})Eyring, Gillett, Achutarao,
  Barimalala, Barreiro~Parrillo, Bellouin, Cassou, Durack, Kosaka, McGregor
  et~al}]{IPCCar6ch3}
Eyring V, Gillett N, Achutarao K, et~al (2021{\natexlab{c}}) {Human Influence
  on the Climate System: Contribution of Working Group I to the Sixth
  Assessment Report of the Intergovernmental Panel on Climate Change}. Tech.
  rep., IPCC Sixth Assessment Report

\bibitem[{Fasullo et~al(2023)Fasullo, Rosenbloom, and Buchholz}]{Fasullo2023}
Fasullo JT, Rosenbloom N, Buchholz R (2023) {A multiyear tropical Pacific
  cooling response to recent Australian wildfires in CESM2}. Science Advances
  9(19):eadg1213. \doi{https://doi.org/10.1126/sciadv.adg1213}

\bibitem[{Gillett et~al(2004)Gillett, Weaver, Zwiers, and Wehner}]{gillett2004}
Gillett N, Weaver A, Zwiers F, et~al (2004) {Detection of volcanic influence on
  global precipitation}. Geophysical Research Letters 31(12).
  \doi{10.1029/2004GL020044}

\bibitem[{Golaz et~al(2022)Golaz, Roekel, Zheng, Roberts, Wolfe, Lin, Bradley,
  Tang, Maltrud, Forsyth, Zhang, Zhou, Zhang, Zender, Wu, Wang, Turner, Singh,
  Richter, Qin, Petersen, Mametjanov, Ma, Larson, Krishna, Keen, Jeffery,
  Hunke, Hannah, Guba, Griffin, Feng, Engwirda, Vittorio, Dang, Conlon, Chen,
  Brunke, Bisht, Benedict, Asay-Davis, Zhang, Zhang, Zeng, Xie, Wolfram, Vo,
  Veneziani, Tesfa, Sreepathi, Salinger, Eyre, Prather, Mahajan, Li, Jones,
  Jacob, Huebler, Huang, Hillman, Harrop, Foucar, Fang, Comeau, Caldwell,
  Bartoletti, Balaguru, Taylor, McCoy, Leung, and Bader}]{Golaz:2022}
Golaz JC, Roekel LPV, Zheng X, et~al (2022) {The DOE E3SM Model Version 2:
  Overview of the Physical Model and Initial Model Evaluation}. Journal of
  Advances in Modeling Earth Systems 14(12). \doi{10.1029/2022MS003156}

\bibitem[{Greenwald et~al(2006)Greenwald, Bergin, Xu, Cohan, Hoogenboom, and
  Chameides}]{greenwald2006}
Greenwald R, Bergin M, Xu J, et~al (2006) {The influence of aerosols on crop
  production: A study using the CERES crop model}. Agricultural Systems
  89(2-3):390--413. \doi{10.1016/j.agsy.2005.10.004}

\bibitem[{Gu et~al(2003)Gu, Baldocchi, Wofsy, Munger, Michalsky, Urbanski, and
  Boden}]{gu2003}
Gu L, Baldocchi DD, Wofsy SC, et~al (2003) {Response of a deciduous forest to
  the Mount Pinatubo eruption: Enhanced photosynthesis}. Science
  299(5615):2035--2038. \doi{10.1126/science.1078366}

\bibitem[{Guo et~al(2004)Guo, Bluth, Rose, Watson, and Prata}]{guo2004}
Guo S, Bluth GJ, Rose WI, et~al (2004) {Re-evaluation of SO2 release of the 15
  June 1991 Pinatubo eruption using ultraviolet and infrared satellite
  sensors}. Geochemistry, Geophysics, Geosystems 5(4).
  \doi{10.1029/2003GC000654}

\bibitem[{Hasselmann(1993)}]{Hasselmann1993}
Hasselmann K (1993) {Optimal Fingerprints for the Detection of Time-dependent
  Climate Change}. Journal of Climate 6(10):1957--1971.
  \doi{10.1175/1520-0442(1993)006<1957:OFFTDO>2.0.CO;2}

\bibitem[{Hasselmann(1997)}]{hasselmann1997}
Hasselmann K (1997) {Multi-pattern fingerprint method for detection and
  attribution of climate change}. Climate Dynamics 13(9):601--611.
  \doi{10.1007/s003820050185}

\bibitem[{Hegerl and North(1997)}]{hegerl1997B}
Hegerl GC, North GR (1997) {Comparison of statistically optimal approaches to
  detecting anthropogenic climate change}. Journal of Climate 10(5):1125--1133.
  \doi{10.1175/1520-0442(1997)010<1125:COSOAT>2.0.CO;2}

\bibitem[{Hegerl et~al(1997)Hegerl, Hasselmann, Cubasch, Mitchell, Roeckner,
  Voss, and Waszkewitz}]{hegerl1997A}
Hegerl GC, Hasselmann K, Cubasch U, et~al (1997) {Multi-fingerprint detection
  and attribution analysis of greenhouse gas, greenhouse gas-plus-aerosol and
  solar forced climate change}. Climate Dynamics 13(9):613--634.
  \doi{10.1007/s003820050186}

\bibitem[{Hegerl et~al(2010)Hegerl, Hoegh-Guldberg, Casassa, Hoerling, Kovats,
  Parmesan, Pierce, and Stott}]{hegerl2010}
Hegerl GC, Hoegh-Guldberg O, Casassa G, et~al (2010) Good practice guidance
  paper on detection and attribution related to anthropogenic climate change.
  Tech. rep., Intergovernmental Panel on Climate Change Expert Meeting on
  Detection and Attribution of Anthropogenic Climate Change

\bibitem[{Kahn et~al(2023)Kahn, Andrews, Brock, Chin, Feingold, Gettelman,
  Levy, Murphy, Nenes, Pierce et~al}]{Kahn2023}
Kahn RA, Andrews E, Brock CA, et~al (2023) Reducing aerosol forcing uncertainty
  by combining models with satellite and within-the-atmosphere observations: A
  three-way street. Reviews of Geophysics 61(2):e2022RG000796.
  \doi{https://doi.org/10.1029/2022RG000796}

\bibitem[{Kapoor and Narayanan(2023)}]{Kapoor2023}
Kapoor S, Narayanan A (2023) {Leakage and the reproducibility crisis in
  machine-learning-based science}. Patterns 4(9).
  \doi{10.1016/j.patter.2023.100804}

\bibitem[{Keys et~al(2022)Keys, Barnes, Diffenbaugh, Hurrell, and
  Bell}]{Keys2022}
Keys PW, Barnes EA, Diffenbaugh NS, et~al (2022) {Potential for perceived
  failure of stratospheric aerosol injection deployment}. Proceedings of the
  National Academy of Sciences of the United States of America 119(40):1--8.
  \doi{10.1073/pnas.2210036119}

\bibitem[{Kirchner et~al(1999)Kirchner, Stenchikov, Graf, Robock, and
  Antu{\~n}a}]{kirchner1999}
Kirchner I, Stenchikov GL, Graf HF, et~al (1999) Climate model simulation of
  winter warming and summer cooling following the 1991 {M}ount {P}inatubo
  volcanic eruption. Journal of Geophysical Research: Atmospheres
  104(D16):19039--19055. \doi{https://doi.org/10.1029/1999JD900213}

\bibitem[{Kremser et~al(2016)Kremser, Thomason, von Hobe, Hermann, Deshler,
  Timmreck, Toohey, Stenke, Schwarz, Weigel, Fueglistaler, Prata, Vernier,
  Schlager, Barnes, Antuña-Marrero, Fairlie, Palm, Mahieu, Notholt, Rex,
  Bingen, Vanhellemont, Bourassa, Plane, Klocke, Carn, Clarisse, Trickl, Neely,
  James, Rieger, Wilson, and Meland}]{Kremser:2016}
Kremser S, Thomason LW, von Hobe M, et~al (2016) {Stratospheric
  Aerosol--Observations, Processes, and Impact on Climate}. Reviews of
  Geophysics 54(2):278--335. \doi{10.1002/2015RG000511}

\bibitem[{Labitzke and McCormick(1992)}]{labitzke1992}
Labitzke K, McCormick MP (1992) {Stratospheric temperature increases due to
  Pinatubo aerosols}. Geophysical Research Letters 19(2):207--210.
  \doi{10.1029/91GL02940}

\bibitem[{Lackmann(2015)}]{Lackmann2015}
Lackmann GM (2015) Hurricane {S}andy before 1900 and after 2100. Bulletin of
  the American Meteorological Society 96(4):547--560.
  \doi{https://doi.org/10.1175/BAMS-D-14-00123.1}

\bibitem[{Lee et~al(2005)Lee, Zwiers, Hegerl, Zhang, and Tsao}]{Lee2005}
Lee TCK, Zwiers FW, Hegerl GC, et~al (2005) A {B}ayesian climate change
  detection and attribution assessment. Journal of Climate 18(13):2429 -- 2440.
  \doi{10.1175/JCLI3402.1}

\bibitem[{Lee et~al(2021)Lee, MacMartin, Visioni, and Kravitz}]{Lee2021}
Lee WR, MacMartin DG, Visioni D, et~al (2021) High-latitude stratospheric
  aerosol geoengineering can be more effective if injection is limited to
  spring. Geophysical Research Letters 48(9):e2021GL092696.
  \doi{https://doi.org/10.1029/2021GL092696}

\bibitem[{Lehner et~al(2016)Lehner, Schurer, Hegerl, Deser, and
  Fr{\"o}licher}]{Lehner2016}
Lehner F, Schurer AP, Hegerl GC, et~al (2016) The importance of {ENSO} phase
  during volcanic eruptions for detection and attribution. Geophysical Research
  Letters 43(6):2851--2858. \doi{https://doi.org/10.1002/2016GL067935}

\bibitem[{Li et~al(2021)Li, Zhang, and Kondragunta}]{Li2021Aus}
Li F, Zhang X, Kondragunta S (2021) Highly anomalous fire emissions from the
  2019--2020 {A}ustralian bushfires. Environmental Research Communications
  3(10):105005. \doi{https://doi.org/10.1088/2515-7620/ac2e6f}

\bibitem[{Liu et~al(2012)Liu, Easter, Ghan, Zaveri, Rasch, Shi, Lamarque,
  Gettelman, Morrison, Vitt, Conley, Park, Neale, Hannay, Ekman, Hess,
  Mahowald, Collins, Iacono, Bretherton, Flanner, and Mitchell}]{liu2012}
Liu X, Easter RC, Ghan SJ, et~al (2012) {Toward a minimal representation of
  aerosols in climate models: description and evaluation in the Community
  Atmosphere Model CAM5}. Geoscientific Model Development 5(3):709--739.
  \doi{10.5194/gmd-5-709-2012}

\bibitem[{Liu et~al(2016)Liu, Ma, Wang, Tilmes, Singh, Easter, Ghan, and
  Rasch}]{liu2016}
Liu X, Ma PL, Wang H, et~al (2016) {Description and evaluation of a new
  four-mode version of the Modal Aerosol Module (MAM4) within version 5.3 of
  the Community Atmosphere Model}. Geoscientific Model Development
  9(2):505--522. \doi{10.5194/gmd-9-505-2016}

\bibitem[{Lloyd and Shepherd(2023)}]{LloydShepherd2023}
Lloyd EA, Shepherd TG (2023) Foundations of attribution in climate-change
  science. Environmental Research: Climate 2(3):035014.
  \doi{https://doi.org/10.1088/2752-5295/aceea1}

\bibitem[{Malchow et~al(2023)Malchow, Hartig, Reeg, K{\'e}ry, and
  Zurell}]{Malchow2023}
Malchow AK, Hartig F, Reeg J, et~al (2023) Demography--environment
  relationships improve mechanistic understanding of range dynamics under
  climate change. Philosophical Transactions of the Royal Society B
  378(1881):20220194. \doi{https://doi.org/10.1098/rstb.2022.0194}

\bibitem[{Marshall et~al(2019)Marshall, Johnson, Mann, Lee, Dhomse, Regayre,
  Yoshioka, Carslaw, and Schmidt}]{marshall2019}
Marshall L, Johnson JS, Mann GW, et~al (2019) {Exploring how eruption source
  parameters affect volcanic radiative forcing using statistical emulation}.
  Journal of Geophysical Research: Atmospheres 124(2):964--985.
  \doi{10.1029/2018JD028675}

\bibitem[{Marvel et~al(2020)Marvel, Biasutti, and Bonfils}]{Marvel2020}
Marvel K, Biasutti M, Bonfils C (2020) {Fingerprints of external forcings on
  Sahel rainfall: aerosols, greenhouse gases, and model-observation
  discrepancies}. Environmental Research Letters 15(8):084023.
  \doi{10.1088/1748-9326/ab858e}

\bibitem[{McGraw et~al(2016)McGraw, Barnes, and Deser}]{Mcgraw2016}
McGraw MC, Barnes EA, Deser C (2016) Reconciling the observed and modeled
  southern hemisphere circulation response to volcanic eruptions. Geophysical
  Research Letters 43(13):7259--7266.
  \doi{https://doi.org/10.1002/2016GL069835}

\bibitem[{Mindlin et~al(2020)Mindlin, Shepherd, Vera, Osman, Zappa, Lee, and
  Hodges}]{Mindlin2020}
Mindlin J, Shepherd TG, Vera CS, et~al (2020) Storyline description of southern
  hemisphere midlatitude circulation and precipitation response to greenhouse
  gas forcing. Climate Dynamics 54:4399--4421.
  \doi{https://doi.org/10.1007/s00382-020-05234-1}

\bibitem[{Minnis et~al(1993)Minnis, Harrison, Stowe, Gibson, Denn, Doelling,
  and Smith~Jr}]{Minnis1993}
Minnis P, Harrison E, Stowe L, et~al (1993) Radiative climate forcing by the
  {M}ount {P}inatubo eruption. Science 259(5100):1411--1415.
  \doi{10.1126/science.259.5100.1411}

\bibitem[{Mitchell et~al(2001)Mitchell, Karoly, Hegerl, Zwiers, Allen, and
  Marengo}]{IPCCar3ch12}
Mitchell J, Karoly D, Hegerl G, et~al (2001) Detection of climate change and
  attribution of causes. Tech. rep., Intergovernmental Panel on Climate Change
  (IPCC), Assessment Report 3

\bibitem[{{National Academies of Sciences, Engineering, and
  Medicine}(2016)}]{NAP21852}
{National Academies of Sciences, Engineering, and Medicine} (2016) Attribution
  of Extreme Weather Events in the Context of Climate Change. The National
  Academies Press, Washington, DC, \doi{10.17226/21852}

\bibitem[{North and Stevens(1998)}]{North1998}
North GR, Stevens MJ (1998) {Detecting climate signals in the surface
  temperature record}. Journal of Climate 11(4):563--577.
  \doi{10.1175/1520-0442(1998)011<0563:DCSITS>2.0.CO;2}

\bibitem[{Otto(2017)}]{Otto2017}
Otto FE (2017) Attribution of weather and climate events. Annual Review of
  Environment and Resources 42(1):627--646.
  \doi{https://doi.org/10.1146/annurev-environ-102016-060847}

\bibitem[{Paciorek et~al(2018)Paciorek, Stone, and Wehner}]{Paciorek2018}
Paciorek CJ, Stone DA, Wehner MF (2018) Quantifying statistical uncertainty in
  the attribution of human influence on severe weather. Weather and Climate
  Extremes 20:69--80. \doi{https://doi.org/10.1016/j.wace.2018.01.002}

\bibitem[{Parker et~al(1996)Parker, Wilson, Jones, Christy, and
  Folland}]{parker1996}
Parker D, Wilson H, Jones PD, et~al (1996) The impact of {M}ount {P}inatubo on
  world-wide temperatures. International Journal of Climatology: A Journal of
  the Royal Meteorological Society 16(5):487--497.
  \doi{https://doi.org/10.1002/(SICI)1097-0088(199605)16:5<487::AID-JOC39>3.0.CO;2-J}

\bibitem[{Polvani et~al(2019)Polvani, Banerjee, and Schmidt}]{Polvani:2019}
Polvani LM, Banerjee A, Schmidt A (2019) Northern {H}emisphere continental
  winter warming following the 1991 {M}t.\ {P}inatubo eruption: reconciling
  models and observations. Atmospheric Chemistry and Physics 19(9):6351--6366.
  \doi{10.5194/acp-19-6351-2019},
  \urlprefix\url{https://acp.copernicus.org/articles/19/6351/2019/}

\bibitem[{Proctor et~al(2018)Proctor, Hsiang, Burney, Burke, and
  Schlenker}]{proctor2018}
Proctor J, Hsiang S, Burney J, et~al (2018) {Estimating global agricultural
  effects of geoengineering using volcanic eruptions}. Nature
  560(7719):480--483. \doi{10.1038/s41586-018-0417-3}

\bibitem[{Ramachandran et~al(2000)Ramachandran, Ramaswamy, Stenchikov, and
  Robock}]{ramachandran2000}
Ramachandran S, Ramaswamy V, Stenchikov GL, et~al (2000) {Radiative impact of
  the Mount Pinatubo volcanic eruption: Lower stratospheric response}. Journal
  of Geophysical Research: Atmospheres 105(D19):24409--24429.
  \doi{10.1029/2000JD900355}

\bibitem[{Reed et~al(2022)Reed, Goldenson, Grotjahn, Gutowski, Jagannathan,
  Jones, Leung, McGinnis, Pryor, Srivastava et~al}]{Reed2022}
Reed KA, Goldenson N, Grotjahn R, et~al (2022) Metrics as tools for bridging
  climate science and applications. Wiley Interdisciplinary Reviews: Climate
  Change 13(6):e799. \doi{https://doi.org/10.1002/wcc.799}

\bibitem[{Ribes et~al(2013)Ribes, Planton, and Terray}]{ribes2013}
Ribes A, Planton S, Terray L (2013) {Application of regularised optimal
  fingerprinting to attribution. Part I: method, properties and idealised
  analysis}. Climate Dynamics 41(11):2817--2836.
  \doi{10.1007/s00382-013-1735-7}

\bibitem[{Ribes et~al(2017)Ribes, Zwiers, Aza{\"\i}s, and Naveau}]{Ribes2017}
Ribes A, Zwiers FW, Aza{\"\i}s JM, et~al (2017) A new statistical approach to
  climate change detection and attribution. Climate Dynamics 48(1):367--386.
  \doi{https://doi.org/10.1007/s00382-016-3079-6}

\bibitem[{Richter et~al(2022)Richter, Visioni, Macmartin, Bailey, Rosenbloom,
  Dobbins, Lee, Tye, and Lamarque}]{Richter2022}
Richter JH, Visioni D, Macmartin DG, et~al (2022) {Assessing Responses and
  Impacts of Solar climate intervention on the Earth system with stratospheric
  aerosol injection (ARISE-SAI): protocol and initial results from the first
  simulations}. Geoscientific Model Development 15(22):8221--8243.
  \doi{10.5194/gmd-15-8221-2022}

\bibitem[{Robock(2000)}]{Robock:2000}
Robock A (2000) {Volcanic eruptions and climate}. {Reviews of Geophysics}
  38(2):191--219. \doi{10.1029/1998RG000054}

\bibitem[{Robock and Mao(1992)}]{RobockMao1992}
Robock A, Mao J (1992) Winter warming from large volcanic eruptions.
  Geophysical Research Letters 19(24):2405--2408.
  \doi{https://doi.org/10.1029/92GL02627}

\bibitem[{Rossow and Schiffer(1999)}]{rossow1999}
Rossow WB, Schiffer RA (1999) Advances in understanding clouds from {ISCCP}.
  Bulletin of the American Meteorological Society 80(11):2261--2288.
  \doi{https://doi.org/10.1175/1520-0477(1999)080<2261:AIUCFI>2.0.CO;2}

\bibitem[{Santer et~al(1993)Santer, Wigley, and Jones}]{santer1993}
Santer B, Wigley T, Jones P (1993) {Correlation methods in fingerprint
  detection studies}. Climate Dynamics 8(6):265--276. \doi{10.1007/BF00209666}

\bibitem[{Santer et~al(2011)Santer, Mears, Doutriaux, Caldwell, Gleckler,
  Wigley, Solomon, Gillett, Ivanova, Karl et~al}]{santer2011}
Santer BD, Mears C, Doutriaux C, et~al (2011) {Separating signal and noise in
  atmospheric temperature changes: The importance of timescale}. Journal of
  Geophysical Research: Atmospheres 116(D22). \doi{10.1029/2011JD016263}

\bibitem[{Santer et~al(2014)Santer, Bonfils, Painter, Zelinka, Mears, Solomon,
  Schmidt, Fyfe, Cole, Nazarenko, Taylor, and Wentz}]{Santer2014}
Santer BD, Bonfils C, Painter JF, et~al (2014) {Volcanic contribution to
  decadal changes in tropospheric temperature}. Nature Geoscience
  7(3):185--189. \doi{10.1038/ngeo2098}

\bibitem[{Saunois et~al(2024)Saunois, Martinez, Poulter, Zhang, Raymond,
  Regnier, Canadell, Jackson, Patra, Bousquet et~al}]{Saunois2024}
Saunois M, Martinez A, Poulter B, et~al (2024) Global methane budget
  2000--2020. Earth System Science Data Discussions 2024:1--147.
  \doi{https://doi.org/10.5194/essd-2024-115}

\bibitem[{Shepherd(2016)}]{Shepherd2016}
Shepherd TG (2016) A common framework for approaches to extreme event
  attribution. Current Climate Change Reports 2:28--38.
  \doi{https://doi.org/10.1007/s40641-016-0033-y}

\bibitem[{Soden et~al(2002)Soden, Wetherald, Stenchikov, and
  Robock}]{soden2002}
Soden BJ, Wetherald RT, Stenchikov GL, et~al (2002) {Global cooling after the
  eruption of Mount Pinatubo: A test of climate feedback by water vapor}.
  Science 296(5568):727--730. \doi{10.1126/science.296.5568.727}

\bibitem[{Swain et~al(2020)Swain, Singh, Touma, and Diffenbaugh}]{Swain2020}
Swain DL, Singh D, Touma D, et~al (2020) Attributing extreme events to climate
  change: A new frontier in a warming world. One Earth 2(6):522--527.
  \doi{https://doi.org/10.1016/j.oneear.2020.05.011}

\bibitem[{Tilmes et~al(2018)Tilmes, Richter, Kravitz, MacMartin, Mills,
  Simpson, Glanville, Fasullo, Phillips, Lamarque et~al}]{Tilmes2018}
Tilmes S, Richter JH, Kravitz B, et~al (2018) {CESM1} ({WACCM}) stratospheric
  aerosol geoengineering large ensemble project. Bulletin of the American
  Meteorological Society 99(11):2361--2371.
  \doi{https://doi.org/10.1175/BAMS-D-17-0267.1}

\bibitem[{Toohey(2025)}]{tooheyPersonalCorrespondence}
Toohey M (2025) personal communication

\bibitem[{Trenberth et~al(2015)Trenberth, Fasullo, and
  Shepherd}]{Trenberth2015}
Trenberth KE, Fasullo JT, Shepherd TG (2015) Attribution of climate extreme
  events. Nature climate change 5(8):725--730.
  \doi{https://doi.org/10.1038/nclimate2657}

\bibitem[{Ukhov et~al(2023)Ukhov, Stenchikov, Osipov, Krotkov, Gorkavyi, Li,
  Dubovik, and Lopatin}]{Ukhov2023}
Ukhov A, Stenchikov G, Osipov S, et~al (2023) Inverse modeling of the initial
  stage of the 1991 pinatubo volcanic cloud accounting for radiative feedback
  of volcanic ash. Journal of Geophysical Research: Atmospheres
  128(12):e2022JD038446. \doi{https://doi.org/10.1029/2022JD038446}

\bibitem[{Watson-Parris et~al(2020)Watson-Parris, Bellouin, Deaconu, Schutgens,
  Yoshioka, Regayre, Pringle, Johnson, Smith, Carslaw et~al}]{Watson2020}
Watson-Parris D, Bellouin N, Deaconu L, et~al (2020) Constraining uncertainty
  in aerosol direct forcing. Geophysical Research Letters 47(9):e2020GL087141.
  \doi{https://doi.org/10.1029/2020GL087141}

\bibitem[{Weierbach et~al(2023)Weierbach, LeGrande, and
  Tsigaridis}]{weierbach2023}
Weierbach H, LeGrande AN, Tsigaridis K (2023) The impact of {ENSO} and {NAO}
  initial conditions and anomalies on the modeled response to pinatubo-sized
  volcanic forcing. Atmospheric Chemistry and Physics 23(24):15491--15505.
  \doi{https://doi.org/10.5194/acp-23-15491-2023}

\bibitem[{Weylandt and Swiler(2024)}]{Weylandt:2024-BeyondPCA}
Weylandt M, Swiler LP (2024) Beyond {PCA}: Additional dimension reduction
  techniques to consider in the development of climate fingerprints. Journal of
  Climate 37:1723–1735. \doi{10.1175/JCLI-D-23-0267.1}

\bibitem[{Wheeler et~al(2025)Wheeler, Wagman, Smith, Davies, Cook, Brunell,
  Glen, Hackenburg, Lien, Shand, and Zeitler}]{Wheeler2025}
Wheeler L, Wagman B, Smith W, et~al (2025) Design and simulation of a
  logistically constrained high-latitude, low-altitude stratospheric aerosol
  injection scenario in the energy exascale earth system model ({E3SM}).
  Environmental Research Letters \doi{https://doi.org/10.1088/1748-9326/adba01}

\bibitem[{Wohland(2022)}]{Wohland2022}
Wohland J (2022) Process-based climate change assessment for {E}uropean winds
  using {EURO-CORDEX} and global models. Environmental Research Letters
  17(12):124047. \doi{https://doi.org/10.1088/1748-9326/aca77f}

\bibitem[{Wu et~al(2020)Wu, Yang, Hu, and Wei}]{Wu2020}
Wu Y, Yang S, Hu X, et~al (2020) Process-based attribution of long-term surface
  warming over the tibetan plateau. International Journal of Climatology
  40(15):6410--6422. \doi{https://doi.org/10.1002/joc.6589}

\bibitem[{Zanchettin et~al(2019)Zanchettin, Timmreck, Toohey, Jungclaus,
  Bittner, Lorenz, and Rubino}]{Zanchettin2019}
Zanchettin D, Timmreck C, Toohey M, et~al (2019) Clarifying the relative role
  of forcing uncertainties and initial-condition unknowns in spreading the
  climate response to volcanic eruptions. Geophysical Research Letters
  46(3):1602--1611. \doi{https://doi.org/10.1029/2018GL081018},
  \urlprefix\url{https://agupubs.onlinelibrary.wiley.com/doi/abs/10.1029/2018GL081018},
  {\href{https://arxiv.org/abs/https://agupubs.onlinelibrary.wiley.com/doi/pdf/10.1029/2018GL081018}{{https://agupubs.onlinelibrary.wiley.com/doi/pdf/10.1029/2018GL081018}}}

\bibitem[{Zanchettin et~al(2022)Zanchettin, Timmreck, Khodri, Schmidt, Toohey,
  Abe, Bekki, Cole, Fang, Feng, Hegerl, Johnson, Lebas, LeGrande, Mann,
  Marshall, Rieger, Robock, Rubinetti, Tsigaridis, and
  Weierbach}]{Zanchettin2022}
Zanchettin D, Timmreck C, Khodri M, et~al (2022) Effects of forcing differences
  and initial conditions on inter-model agreement in the {VolMIP}
  volc-pinatubo-full experiment. Geoscientific Model Development
  15(5):2265--2292. \doi{10.5194/gmd-15-2265-2022}

\bibitem[{Zappa and Shepherd(2017)}]{ZappaShepherd2017}
Zappa G, Shepherd TG (2017) Storylines of atmospheric circulation change for
  {E}uropean regional climate impact assessment. Journal of Climate
  30(16):6561--6577. \doi{https://doi.org/10.1175/JCLI-D-16-0807.1}

\end{thebibliography}
